\documentclass[useAMS,usenatbib]{mn2e}

\usepackage{graphicx, float}

\usepackage{amsfonts}

\usepackage[toc, page]{appendix}

\usepackage[normalem]{ulem}

\usepackage{epsfig}

\usepackage{color}

\usepackage{times}

\usepackage{amsmath}

\usepackage{amssymb}

\usepackage{xspace}

\usepackage{txfonts}

\usepackage{enumitem}

\usepackage{algorithm}

\usepackage{algorithmic}

\usepackage[outercaption]{sidecap}

\usepackage{subfig}

\usepackage{rotating}

\usepackage{setspace}

\def\Mpch{h^{-1} {\rm Mpc}}
\def\Mpchk{h {\rm Mpc}^{-1}}

\newcommand{\Mspace}     {\mm{{\mathbb M}}}
\newcommand{\Rspace}     {\mm{{\mathbb R}}}

\newcommand{\Zspace}     {\mm{{\mathbb Z}}}
\newcommand{\Bspace}     {\mm{{\mathbb B}}}
\newcommand{\Tspace}     {\mm{{\mathbb T}}}
\newcommand{\Hspace}     {\mm{{\mathbb H}}}
\newcommand{\Nspace}     {\mm{{\mathbb N}}}

\newcommand{\cH}  {{\mathcal H}}

\def\smallhalf{\mbox{ $\frac{1}{2}$}}

\newcommand {\mm}[1] {\ifmmode{#1}\else{\mbox{\(#1\)}}\fi}

\newcommand{\beq}{\begin{eqnarray}}
\newcommand{\eeq}{\end{eqnarray}}
\newcommand{\beqq}{\begin{eqnarray*}}
\newcommand{\eeqq}{\end{eqnarray*}}

\newcommand{\beqn}{\begin{equation}}
\newcommand{\eeqn}{\end{equation}}
\newcommand{\beqqn}{\begin{equation*}}
\newcommand{\eeqqn}{\end{equation*}}

\newcommand{\lips}{{\cal L}}

\newcommand{\real}{R}
\newcommand{\mink}{{\mathcal M}}
\newcommand{\Min}{{\mathcal M}}

\def\cO{{\cal O}}

\def\Min{\mathcal M}

\newcommand{\sqbinom}[2]{\begin{bmatrix}#1 \\ #2 \end{bmatrix}}
\newcommand{\ssqbinom}[2]{{\mbox{\scriptsize $\sqbinom{#1}{#2}$ }}}

\def\LK{Lipschitz-Killing}
\def\LKC{Lipschitz-Killing curvature}
\def\LKCs{Lipschitz-Killing curvatures}
\newcommand{\bc}{\begin{center}}
\newcommand{\ec}{\end{center}}

\begin{document}

\title[Topology and geometry of Gaussian random fields]{Topology and Geometry of Gaussian random fields I: on  Betti Numbers, Euler characteristic and Minkowski functionals}
\author[Pratyush Pranav, Rien van de Weygaert, Gert Vegter, Bernard J.T. Jones]
	{Pratyush Pranav$^{1,2,3}$\thanks{pratyush.pranav@ens-lyon.fr},
	Rien van de Weygaert$^2$,
	Gert Vegter$^4$,
	Bernard J. T. Jones$^2$,
	\and Robert J. Adler$^3$,
	Job Feldbrugge$^{2,5}$,
	Changbom Park$^6$,
	Thomas Buchert$^1$,
        \and Michael Kerber$^7$\\
$^1$Univ Lyon, ENS de Lyon, Univ Lyon1, CNRS, Centre de Recherche Astrophysique de Lyon UMR5574, FV69007, Lyon, France\\
$^2$Kapteyn Astronomical Institute, Univ. of Groningen, PO Box 800, 9700 AV Groningen, The Netherlands\\
$^3$Technion -- Israel Institute of Technology, Haifa, 32000, Israel\\
$^4$Johann Bernoulli Inst. for Mathematics and Computer Science, Univ. of Groningen, P.O. Box 407, 9700 AK Groningen, The Netherlands\\
$^5$Perimeter Institute for Theoretical Physics, University of Waterloo,
Waterloo, Canada\\
$^6$Korean Institute of Advanced Studies, Hoegiro 87, Dongdaemun-gu, Seoul 130-722, Korea\\
$^7$Institut f\"ur Geometrie, TU Graz, Kopernikusgasse 24A 8010 Graz  \\
}

\maketitle


\begin{abstract}
This study presents a numerical analysis of the topology of a set of cosmologically interesting three-dimensional Gaussian random fields in terms of their Betti numbers $\beta_0$, $\beta_1$ and $\beta_2$. We show that Betti numbers entail a considerably richer characterization of the topology of the primordial density field. Of particular interest is that Betti numbers specify which topological features - islands, cavities or tunnels - define its spatial structure. 

A principal characteristic of Gaussian fields is that the three Betti numbers dominate the topology at different density ranges. At extreme density levels, the topology is dominated by a single class of features. At low levels this is a \emph{Swiss-cheeselike} topology, dominated by isolated cavities, at high levels 
a predominantly \emph{Meatball-like} topology of isolated objects.
At moderate density levels, two Betti number define a more 
\emph{Sponge-like} topology. At mean density, the topology even needs three Betti numbers, quantifying a  field consisting of several disconnected complexes, not of one connected and percolating overdensity.

A {\it second} important aspect of Betti number statistics is that they are sensitive to the power spectrum. It reveals a monotonic trend in which at a moderate density range a lower spectral index corresponds to a considerably higher (relative) population of cavities and islands. 

We also assess the level of complementary information that Betti numbers represent, in addition to conventional measures such as  Minkowski functionals. To this end, we include an extensive description of the Gaussian Kinematic Formula (GKF), which represents a major theoretical underpinning for this discussion. 
\end{abstract}

\begin{keywords}
 {large-scale structure of the universe, Gaussian random fields -- 
 cosmology: theory; topology, Betti numbers, genus -- topology: 
 theory}
\end{keywords}

\section{Introduction}
\label{sec:intro}

The richness of the big data samples emerging from astronomical experiments and simulations demands increasingly complex algorithms in order
to derive maximal benefit from their existence.  Generally speaking, most current analyses express inter-relationships between
quantitative properties of the datasets, rather than geometric or topological, ie. structural, properties.

Here we introduce a new technique that successfully attacks the problem of characterising the structural nature of data.  This exercise 
involves an excursion into the relatively complex and unfamiliar domain of \textit{homology}, which we attempt to present in a
straightforward manner, that should enable others to use and extend this aspect of data analysis. On the application side, we demonstrate the power of the formalism through a systematic study of Gaussian random fields using this novel methodology.

A Gaussian random field is a stochastic process, $X$, defined over some parameter space of $S$, and characterised by the fact  that the
vector $(X(s_1),\dots,X(s_k))$ has a $k$-dimensional, multivariate normal distribution for any collection of points $(s_1,\dots,s_k)$ in $S$.
Gaussian random fields play a key role in cosmology: in the standard cosmological view, the primordial density and velocity fields have a
Gaussian character, making Gaussian fields the initial conditions for the formation of all structure in the Universe. A Gaussian random field is fully specified
by its power spectrum, or in the real space, its correlation function. 
As a result, the determination and characterization of the power spectrum of the theoretical models as well as observational data has been one of the main
focal points in the analysis of the primordial cosmic fluctuation field as well as the Megaparsec - large scale - matter and galaxy distribution at low
redshifts. 

A substantial body of theoretical and observational evidence underpins the assumption of the Gaussianity of the
primordial cosmic density and velocity fields. These have established Gaussian random fields as a prominent aspect
of the current standard cosmological worldview. The primary evidence for this is the near-perfect Gaussian nature of
the Cosmic Microwave Background radiation (CMB) temperature fluctuations. These directly reflect
the density and velocity perturbations on the surface of last scattering, and thus the mass distribution at the
recombination and decoupling epoch 379,000 years after the Big Bang, at a redshift of $z \approx 1090$ \citep[see e.g.][]{Pee80,jones2017}.
In particular the measurements by the COBE, WMAP and Planck satellites established that to high accuracy, the CMB temperature fluctuations
define a homogeneous and isotropic Gaussian random field \citep{smoot1992,bennett2003,spergel2007,komatsu2011,planck2016,buchert2017,PlanckCosmoparams2018}. 
Second, that the primordial fluctuations have a Gaussian nature, narrowly follows from the theoretical predictions of the inflationary
scenario, at least in its simplest forms. According to this fundamental cosmological theory, the early universe underwent a phase
transition at around $t \approx 10^{-35}$ sec after the Big Bang \citep{guth1981,linde1981,kolb1990,liddle2000}. As a result, 
the Universe underwent a rapid exponential expansion over at least 60 e-foldings. The inflationary expansion of quantum fluctuations
in the generating inflaton (field) leads to a key implication of this process, the generation of cosmic density and velocity
fluctuations. It involves the prediction of the resulting density fluctuation field being adiabatic and a homogeneous Gaussian
random field, with a near scale-free Harrison-Zeldovich spectrum,
$P(k) \propto k^1$ \citep{harrison1970,zeldovich1972,mukhanov1981,guthpi1982,starobinsky1982,bardeen1983}. Third, the
{\it Central Limit Theorem} states that the statistical distribution of a sum of many independent and identically distributed random
variables will tend to assume a Gaussian distribution. Given that when the Fourier components of a primordial density and velocity field are statistically independent, each having the same Gaussian
distribution, then the joint probability of the density evaluated at a finite number of points will be Gaussian \citep{bbks}. 

On the basis of these facts, Gaussian random fields have played a central role in describing a multitude of fields of interest
that arise in cosmology, making their characterization an important focal point in cosmological studies
\citep{Dor70,bbks,HGW86,bertschinger1987,SV91,Mecke1991,Mecke94,weyedb1996,schmalzing1997,matsubara2010}. When assessing the
structure and patterns of the temperature fluctuations in the CMB, the interest is that of Gaussian fields
on the two-dimensional surface of a sphere, ie. on two-dimensional space $\mathbb S^2$. When studying the cosmic galaxy
and matter distribution, the parameter space is that of a large, but essentially fine, subset of three-dimensional space
$\mathbb R^3$ (i.e. assuming curvature of space is almost perfectly flat, as has been inferred from the WMAP and Planck
CMB measurements \citep{spergel2007,planck2016}). 

In this study, we address the topological characteristics of three-dimensional Gaussian fields, specifically in terms of 
the topological concepts and language of homology \citep[][]{munkres1984elements,vegter2004,robins2006,zomorodian2009,EdHa10,robinsa2015,robinsb2015}. These
concepts are new to cosmology (see below), and will enrich the analysis of cosmological datasets considerably \citep[see eg.][]{RST,elbers2018}. 
The principal rationale for this study of Gaussian field homology is the definition and development of a reference base line. In most
cosmological scenarios Gaussian fields represent the primordial mass distribution out of which 13.8 Gigayears of gravitational evolution has
morphed the current cosmic mass distribution. Hence, for a proper understanding of the rich (persistent) homology of the cosmic web, a full
assessment of Gaussian field homology as reference point is imperative. 

Topology is the branch of mathematics that is concerned with the properties of space that are preserved under 
continuous deformations of manifolds, including their stretching (compression) and bending, but excluding tearing or
glueing. It also includes invariance of properties such as connectedness and boundary. As such it addresses key aspects of the
structure of spatial patterns, the ones concerning the organization, i.e. shape, and connectivity \citep[see e.g.][]{robins2006,robinsa2015,patania2017}. 
The topological characterization of the models of cosmic mass distribution has been a focal point
of many studies \citep{Dor70,Adl81,bbks,HGW86,GDM86,CSO98,CE04,pogosyan2009,CPKG10,PK10}. Such topological studies
provide insight into the global structure, organization and connectivity of cosmic density fields. These aspects
provide key insight into how these structures emerged, and subsequently interacted and merged with neighbouring
features. Particularly helpful in this context is that topological measures are relatively insensitive to systematic effects such 
as non-linear gravitational evolution, galaxy biasing, and redshift-space distortion \citep{PK10}. 

The vast majority of studies of the topological characteristics of the cosmic mass distribution concentrate on the measurement of the genus and
the Euler characteristic \citep{GDM86,HGW86,GMTS89}.
The notion of genus is, technically, only well defined for 2-dimensional surfaces, where it is  a simple linear function of the Euler characteristic.
For 3-dimensional manifolds with smooth boundaries, there is also a simple relationship between the Euler characteristic of a set and the genus of
its boundary. Beyond these examples, however, these relationships break down and, in higher dimensions,  only the Euler characteristic is well defined.
We will therefore typically work with the Euler characteristic, rather than the  genus, even when both are defined. 

While the genus, the Euler characteristic - and the Minkowski functionals discussed below - have been extremely instructive in gaining an
understanding of the topology of the mass distribution in the Universe, there is a substantial scope for an enhancement of the topological
characterization in terms of a richer and more informative description. In this study we present a topological analysis of Gaussian random
fields through \emph{homology} \citep{munkres1984elements,EdHa10,adler2010persistent,weygaert2011,feldbrugge2012,PPC13,BobK,KahleR,PEW16,wasserman2017}.
Homology is a mathematical formalism for specifying in a quantitative and unambiguous manner about how a space is connected
\footnote{There is a notion of k-connectedness, $k = 0,\ldots, d$, where $d$ is the dimension of the 
  manifold. Within this, 0-connectedness is the same as the 'usual' notion of connectedness.}, through assessing the boundaries of
a manifold \citep{munkres1984elements}. To this end, we evaluate the topology of a manifold in terms of the holes that it contains, by assessing their boundaries. 

A $d$-manifold can be composed of topological holes of $0$ up to ($d-1$) dimensions. For $d < 3$, the holes have an intuitive interpretation. 
A $0$-dimensional hole is a \emph{gap} between two isolated independent objects. A $1$-dimensional hole is a \emph{tunnel} through which 
one can pass in any one direction without encountering a boundary. A $2$-dimensional hole is a \emph{cavity} or \emph{void} fully enclosed within a 
2-dimensional surface. This intuitive interpretation in terms of \emph{gaps} and \emph{tunnels} is only valid for surfaces embedded in $\Rspace^3$, 
$\mathbb{S}^3$ or $\mathbb{T}^3$. Following the realization that the identity, shape and outline of these entities is more straightforward to
describe in terms of their boundaries, homology turns to the definition of holes via \emph{cycles}. A $0$-cycle is a connected object (and hence, 
a $0$-hole is the gap between two independent objects). A $1$-cycle is a \emph{loop} that surrounds a tunnel. A $2$-cycle is a \emph{shell} 
enclosing a void.

The statistics of the holes in a manifold, and their boundaries, are captured by its \emph{Betti numbers}. Formally, the Betti numbers are the ranks
of the homology groups. The $p$-th homology group is the assembly of all $p$-dimensional cycles of the manifold, and the rank of the group is
the number of \emph{independent} cycles. In all, there are $d+1$ Betti numbers $\beta_p$, where $p = 0, \ldots , d$ \citep{Bet71,vegter1997,vegter2004,robins2006,EdHa10,weygaert2011,robinsa2015,PEW16}. The first three Betti numbers have intuitive meanings: $\beta_0$ counts the number of independent components, $\beta_1$ counts
the number of loops enclosing the independent tunnels and $\beta_2$ counts the number of shells enclosing the isolated voids.

There is a profound relationship between the homology characterization in terms of Betti numbers and the Euler characteristic.
The \emph{Euler-Poincar\'{e}} formula \citep{Eul58}  states that the Euler characteristic is the alternating sum of the Betti numbers (see equation 35 below).
One immediate implication of this is that the set of Betti numbers contain more topological information than is expressed by the Euler characteristic
(and hence the genus used in cosmological applications). Visually imagining the 3D situation as the projection of three Betti numbers on to
a one-dimensional line, we may directly appreciate that two manifolds that are branded as topologically equivalent in terms of their Euler characteristic 
may actually turn out to possess intrinsically different topologies when described in the richer language of homology. Evidently, in a cosmological
context this will lead to a significant increase of the ability of topological analyses to discriminate between different cosmic structure formation
scenarios. 

The Euler characteristic of a set is an essentially topological quantity. For example, the Euler
characteristic of a three dimensional set is the number of its connected components, minus the number of its holes, plus the number of voids it contains (where
each of terms requires careful definition).  Numbers are important here, but the sizes and shapes of the various objects are not. Nevertheless, it is a deep
result, known as the Gauss-Bonnet-Chern-Alexandrov Theorem, going back to \citep{Eul58}, requiring both Differential and Algebraic Topology
\footnote{Algebraic Topology is a branch of mathematics that uses concepts from abstract algebra to study topological spaces . Differential Topology is the
field of mathematics dealing with differentiable functions on differentiable manifolds.} to prove that - at least for smooth, stratified manifolds \footnote{A
  topologically stratified manifold $\Mspace$ is a space that has been decomposed into pieces called strata; these strata are topological submanifolds and
  are required to fit together in a certain way. Technically, $\Mspace$ needs to be a `$C^2$ Whitney stratified manifold' satisfying mild side conditions
  \citep{Adl10}.} - the Euler characteristics can  actually be computed from  geometric quantities. That is, the Euler characteristic also has a
geometric interpretation, and is actually associated with the integrated Gaussian curvature of a manifold. In fact, together with other quantities related
to volume, area and length, the Euler characteristic forms a part of a more extensive geometrical description via the \emph{Minkowski functionals},
or \emph{Lipschitz-Killing curvatures} of a set. 

There are $d+1$ Minkowski functionals, $\{Q_k\}_{k = 0, \ldots, d}$, defined over nice subsets of $\mathbb R^d$ \citep{Mecke94,schmalzing1997,sahni1998,schmalzing1999,Kersher2000}. All are predominantly geometric in nature. For compact subsets of $\mathbb R^3$, the four Minkowski functionals, in increasing order,  are 
proportional to volume, surface area, integrated mean curvature or total contour length, and integrated Gaussian curvature, itself proportional to the
Euler characteristic. Analyses based on Minkowski functionals, genus and Euler characteristic have played  key roles in understanding
and testing models and observational data of the cosmic mass distribution \citep{GDM86,HGW86,Mecke94,schmalzing1997,SigIRAS1997,sahni1998,IRASMF1998,CSO98,J-stat1999,PSCzMF2001,SDSSMF2003,CE04,pogosyan2009,CPKG10,weygaert2011,PK10,codis2013,PPC13,SDSSMFCode2014}. Generalizations of Minkowski functionals for vector and tensor fields have also been applied in cosmology, and have been useful in quantifying substructures in galaxy clusters \citep{Morphometry2001}. Tensor-valued Minkowski functionals allow to probe directional information and to characterize preferred directions, e.g., to measure the anisotropic signal of redshift space distortions \citep{3DMFTensors2018}, or to characterize anisotropies and departures from Gaussianity in the CMB \citep{PravaTensorMF2016,TensorMF2D}.

The topological analysis of Gaussian fields using genus, Euler characteristic and Minkowski functionals has occupied a place of key importance 
within the methods and formalisms enumerated above. Of fundamental importance, in this respect, has been the realization that the expected value
of the genus in the case of a 2D manifold, and the Euler characteristic in the case of a 3D manifold, as a function of density threshold has an
analytic closed form expression for Gaussian random fields \citep{Adl81,bbks,Adl10}. Amongst others, this makes them an ideal tool for validating
the hypothesis of initial Gaussian conditions through a comparison with the observational data. Important to note is that the functional form of
the genus, the Euler characteristic and the Minkowski functionals is independent of the specification of the power spectrum for Gaussian fields, and is a 
function only of the dimensionless density threshold $\nu$. The contribution from power spectrum is restricted to the amplitude of the 
genus curve through the variance of the distribution, or equivalently, the amplitude of the power spectrum. This indicates that the shape of these
quantities is invariant with respect to the choice of the power spectrum. While this makes them highly suitable measures for testing fundamental
cosmological questions such as the Gaussian nature of primordial perturbations, they are less suited when testing for differences between
different structure formation scenarios is the primary focus.

Given the evident importance of being able to refer to solid analytical expressions, in this study we will report on the fundamental developments of the
past decade which have demonstrated that the analytic expressions for the genus, Euler characteristic and
Minkowski functionals of Gaussian fields belong to an extensive family of such formulae, all emanating from the so called {\it Gaussian kinematic formula} or GKF \citep{Adl10,ATSF,ARF}. The GKF, in one compact formula, gives the expected values of the Euler characteristic (and so genus), all
the \LKCs, and so Minkowski functionals, as well as their extensions, for the superlevel sets (and their generalisations in vector valued cases) of
a wide class of random fields, both Gaussian and only related somehow to Gaussian. This is for both homogeneous and non-homogeneous cases, and cover all examples required
in cosmology. Even though hardly known in the cosmological and physics literature, its relevance and application potential for the study of cosmological
matter and galaxy distributions, as well as other general scenarios, is self-evident \citep[see e.g.][]{codis2013}.

Because of its central role for understanding a range of relevant topological characteristics of Gaussian and other random fields, we discuss the
Gaussian Kinematic Formula extensively in Section~\ref{sec:gkf}. Of conclusive importance for the present study, the interesting observation is
that homology and the associated quantifiers such as Betti numbers are not covered by the Gaussian Kinematic Formula. In fact, a detailed and complete
statistical theory parallel to the GKF for them does not exist. In this respect, it is
good to realize that the Gaussian Kinematic Formula is mainly about geometric quantifiers. The exception to this is the Euler characteristic. Nonetheless,
in a sense the latter may also be seen as a geometric quantity via the Gauss-Bonnet Theorem. To date there is no indication that - along the
lines of the GKF - an analytical description for Betti numbers and other homological concepts is feasible \citep[also see][]{wintraecken2013}.
Nonetheless, this does not exclude the possibility of analytical expressions obtained via alternative routes. One example is analytical
expressions for asymptotic situations, such as those for Gaussian field excursion sets at very high levels. For this situation, the seminal study
by \cite{bbks} obtained the statistical distribution for Betti numbers, ie. for the \emph{islands} and \emph{cavities} in the cosmic matter distribution.
Even more generic is the approach followed by the recent study of \cite{feldbrugge2012} (\cite{feldbrugge2018}). They derived path integral expressions for
Betti numbers and additional homology measures, such as persistence diagrams. While it is not trivial to convert these into concise formulae, the
numerically evaluated approximate expression for 2D Betti numbers turns out to be remarkably accurate.

\bigskip
The current paper presents a numerical investigation of the topological properties of Gaussian random fields through homology and
Betti numbers. Given the observation that generic analytical expressions for their statistical distribution are not available, this study is mainly
computational and numerical. It numerically infers and analyzes the statistical properties of Betti numbers, as well as those of the corresponding Euler
characteristic and Minkowski functionals. The extensive analysis concerns a large set of 3-D Gaussian field realizations, for a range of different power spectra, generated in cubic volumes with periodic boundary conditions. 

In an earlier preliminary paper \citep{PPC13}, we presented a brief but important aspect of the analysis of the homology of three-dimensional Gaussian
random fields via Betti numbers. It illustrated the thesis forwarded in \cite{weygaert2011} that Betti numbers represent a richer source of topological information
than the Euler characteristic. For example, while the latter is insensitive to the power spectrum, Betti numbers reveal a systematic dependence on
power spectrum. It confirms the impression of homology and Betti numbers as providing the next level of topological information. The current paper extends
this study to a more elaborate exploration of the property of Gaussian random fields as measured by the Betti numbers, paying particular attention to the
statistical aspects. Together with the information contained in Minkowski functionals, it shows that homology establishes a more comprehensive and detailed
picture of the topology and morphology of the cosmological theories and structure formation scenarios. 

A powerful extension of homology is its hierarchical
variant called \emph{persistent homology}. The related numerical analysis of the persistent homology of the set of Gaussian field realizations presented in
this paper is the subject of the upcoming related article \citep{pranavb2018}. Our work follows up on early explorations of Gaussian field
homology by Adler \& Bobrowski \citep{adler2010persistent,bobrowskiphd2012,bobrowski2012}. These studies address fundamental and generic aspects and are
strongly analytically inclined, but also give numerical results on Gaussian field homology. Particularly insightful were the presented results on their persistent homology in terms of bar diagrams. 

In addition to the topological analysis of Gaussian fields by means of genus, Minkowski functionals and Betti numbers, we also include a thorough
discussion of the computational procedure that was used for evaluating Betti numbers. The homology computational procedure detailed in \cite{PEW16} is
for a \emph{discrete particle distribution}. On the other hand, in this paper, we detail the homology procedure for evaluating the Betti numbers for random fields whose values have been
sampled on a \emph{regular cubical grid}. The procedure is generic and can be used for the full Betti number and persistence analysis of arbitrary random
fields. In the case of Gaussian fields, one may exploit the inherent symmetries of Gaussian fields to compute only 2 Betti numbers, from which one may
then seek to determine the third one via the analytical expectation value for the Euler characteristic. Indeed, this is the shortcut that was followed
in our preliminary study \citep{PPC13}. 

This study, along with earlier articles \citep{PEW16,pranavThesis,WVEJ11}, gives the fundamental framework and so forms the basis of a planned series of articles aimed at introducing the topological concepts and language of homology - new to cosmology - for the analysis and
description of the cosmic mass distribution. They define a program for an elaborate topological data analysis
of cosmological data \citep[see][for an up-to-date review of topological data analysis in a range of scientific applications]{wasserman2017}. 
The basic framework, early results and program are described and reviewed in \cite{weygaert2011}, which introduced the concepts of homology to the cosmological
community. Following this, in \cite{pranavThesis} and \cite{PEW16} we described in formal detail the mathematical foundations and computational aspects of topology, homology and
persistence. These provide the basis for our program to analyse and distinguish between models of cosmic structure formation in terms of their topological
characteristics, working from the expectation that they offer a considerably richer, more profound and insightful characterization of their
topological structure.

Our program follows the steadily increasing realization in the cosmological community that homology and persistent topology  offer a range of innovative
tools towards the description and analysis of the complex spatial patterns that have emerged from the gravitational evolution of the cosmic matter distribution
from its primordial Gaussian conditions to the intricate spatial network of the cosmic web seen in the current Universe on Megaparsec scales. In this respect,
we may refer to the seminal contribution by \cite{sousbie2011a,sousbie2011b}, and the recent studies applying these topological measures to various cosmological and astronomical scenarios \citep{pranavDE,PPC13,chna2015,shivashankar2016,RST,irinaMagnet,codis2018,xu2018,cole2018,MakarenkoMHD2018}. 
\bigskip

The remainder of the paper is structured as follows: We begin in Section~\ref{sec:grf} by 
providing an introduction to Gaussian random fields, and the presentation of the set
of Gaussian field realizations that forms the basis of this study's numerical investigation. A 
description of the topological background follows in Section~\ref{sec:topology_ch2}. Gaussian fields and
topology are then combined in Section~\ref{sec:gkf} with a discussion of the Gaussian kinematic formula,
which gives a rigorous formulation of what is known about mean Euler characteristic and Minkowski Functionals for Gaussian level sets. This
section also explains why, with topological
quantifiers such as Betti numbers, analytic results at least appear far from trivial to obtain. These sections all describe pre-existing
material, but it is their combination which represents a novel approach towards characterizing the rich topology of cosmological
density fields. The novel computational aspects of this study are outlined in detail in Section~\ref{sec:computation}. This is followed by a description of the model realizations used for the computational studies in Section~\ref{sec:models}. Section~\ref{sec:betti_and_genus_result} describes the Betti number
analysis of our sample of Gaussian random field realizations. Subsequently, the relationship and differences between the distribution
of ``islands'' and ``peaks'' in a Gaussian random field is investigated in Section~\ref{sec:peaks_vs_islands}. This is followed
in Section~\ref{sec:minkowski_result} by an assessment of the comparative information content of Minkowski functionals and
Betti numbers. The homology characteristics of the LCDM Gaussian field are discussed in Section~\ref{sec:lcdm_result}.
Finally, we conclude the paper with some general discussion in Section~\ref{sec:discussion_ch2}.

\section{Gaussian random fields:\ \ \ \ \ definitions }
\label{sec:grf}

In this section, we define the basic concepts of Gaussian random fields, along with definitions and a description of the models analyzed in this paper.
Standard references for the material in this section are \cite{Adl81} and \cite{bbks}.

\subsection{Definitions}

Recall that, at the most basic level,  a \emph{random field} is simply a collection of random variables, $f(x)$, where the values of $x$ run over some parameter space $\mathcal X$. This space might be finite or infinite, countable or not. The probabilistic properties of  random fields are determined by their   $m$-point, 
joint, distribution functions, 
\begin{equation}
P[f_1, \ldots , f_m]\,df_1 \ldots df_m,
\end{equation}
where the $f_1,\dots,f_m$ are the values of the random field at  $m$ points $x_1,\dots,x_m$.
 
A random field is called   \emph{zero mean, Gaussian}, if the $m$-point distributions are all multivariate Gaussian, so that 

\begin{equation}
\begin{aligned}
&P\left[f_1, \ldots , f_m\right]df_1 \ldots df_m  \\
&= \frac{1}{(2 \pi)^N (\text{det} M)^{1/2}}
\times \exp\left(-\frac12
\sum f_i(M^{-1})_{ij}  f_j
\right)\,df_1 \ldots df_m, 
\label{eqn:prob_grf_ch2}
\end{aligned}
\end{equation}
\noindent where $M$ is the  $m \times m$ \emph{covariance 
matrix} of the $f_i$, determined by the \emph{covariance} or \emph{autocovariance function} 
\begin{equation}\xi(x_1,x_2) \, =\,  \left\langle f(x_1) f(x_2)\right\rangle
\label{eq:rob:xi}
\end{equation}
via the correspondence

\begin{equation}
M_{ij} \,=\, \xi(x_i, x_j).
\end{equation}
The angle bracket in \eqref{eq:rob:xi} denotes  ensemble averaging.

It follows from  \eqref{eqn:prob_grf_ch2} that the distribution of zero mean Gaussian random fields is fully specified by 
 second order moments, as expressed via the autocovariance  function. (From now on we shall always assume zero mean.) If we now specialise to random fields defined over  $\mathbb R^D$, $D\geq 1$, so that the points in the parameter set are vectors, we can introduce the notions of \emph{homogeneity} (or \emph{stationarity}) and \emph{isotropy}. A Gaussian random field is called homogeneous if  $\xi (\vec x, \vec y)$ can be written as a function of the difference $ \vec x - \vec y$, and isotropic if it is also a function only of the (absolute)  distance $\| \vec x - \vec y\|$.  In the homogeneous, isotropic, case we write, with some abuse of notation,

\begin{equation}
\xi(r)\,=\,\xi(\|{{\vec r}}\|)\,\equiv\,\langle f({\vec x}) f({\vec x}+{{\vec r}}) \rangle \,.
\label{eq:xi_ch2}
\end{equation}

An immediate consequence of homogeneity is that the variance 
\begin{equation}
\label{eq:rob:sigma2}
\sigma^2=\xi(0)=\langle f^2(\vec x)\rangle
\end{equation}
 of $f$ is constant. Normalising the autocovariance function by $\sigma^2$ gives  the \emph{autocorrelation function}.

In many situations, and generally for cosmological applications of homogeneous random fields, it is more natural to work with the Fourier transform 
\begin{eqnarray}
\label{eq:rob:fhat}
\hat f(\vec k) &\, =  \,& \int_{\mathbb R^D} d^D\vec x\, \,f(\vec x)\,\exp(i\vec k\cdot \vec x) ,\nonumber\\
\ \\
f(\vec x) &\, = \,& \int_{\mathbb R^D} \frac{d^D\vec k}{(2\pi)^D}\, \,{\hat f}(\vec k) \exp(-i\vec k\cdot \vec x) \,\nonumber
\end{eqnarray}
of  both $f$ and, particularly, its autocovariance function $\xi$. The Fourier transform of $\xi$  is known as the \emph{power spectrum}
$P({\vec k})$. Here, and
throughout our study, we follow the Fourier convention of \cite{bbks}\footnote{also known as ``Kaiser convention'', personal communication.}.
For a random field to be strictly homogeneous and Gaussian, its Fourier modes ${\hat f}(\vec k)$ must be mutually independent, and 
the real and imaginary parts ${\hat f}_r(\vec k)$ and ${\hat f}_i(\vec k)$,
\begin{equation}
  {\hat f}(\vec k)\ = \ {\hat f}_r(\vec k) \,+\,i {\hat f}_i(\vec k)\,,
  \end{equation}
 each have a Gaussian distribution, whose dispersion is given by the value of the power spectrum for the corresponding wavenumber ${\vec k}$,
\begin{eqnarray}
  P({\hat f}_r(\vec k))\, &=& \, \frac{1}{\sqrt{2\pi\,P(k)}}\ \exp{\left(-\frac{{\hat f}_r^2(\vec k)}{2P(k)} \right)}\,,\nonumber\\
  P({\hat f}_i(\vec k))\, &=& \, \frac{1}{\sqrt{2\pi\,P(k)}}\ \exp{\left(-\frac{{\hat f}_i^2(\vec k)}{2P(k)} \right)}\,. 
\end{eqnarray}
This means that the Fourier phases $\hat \phi(\vec k)$, 
\begin{equation}
  {\hat f}(\vec k)\,=\,\|{\hat f}(\vec k)\|\,e^{i \phi(\vec k)}\,,
  \end{equation} 
of the field are random, ie. it the phases $\hat \phi(\vec k)$ have a uniform distribution, $U[0,2\pi]$. The moduli $|{\hat f}(\vec k)|$ have a
Rayleigh distribution \citep{bbks}. 

Under an assumption of ergodicity, which we will assume throughout, the power spectrum, denoted by $P({\vec k})$, is continuous.
For $\vec k\in \mathbb R^D$ this leads to 
\begin{equation}
\label{eq:rob:PF}
(2\pi)^D\,P(\vec k)\,\delta_D({\vec k}-{\vec k}')\, = \, 
\left\langle {\hat f}({\vec k}) {\hat f}^*({\vec k}')\right\rangle,
\end{equation}
\noindent where $\delta_D$ is the Dirac delta function. 

In the case of isotropic $f$, $P$ is spherically symmetric, and, once again abusing notation, we write
\begin{equation}
P(k) \, =\, P(\|\vec k\|) \, = \, P(\vec k)\,.
\end{equation}
The power spectrum  breaks down the total variance of $f$ into components at different frequencies, in the sense that 
\begin{eqnarray}
\label{eq:rob:sigmaP}
\sigma^2 \ =\ \int_{\mathbb R^D}  \frac{d^D\vec k}{(2\pi)^D}\, P(\vec k)  &=& \frac{2}{(4\pi)^{D/2} \Gamma(D/2)} \int_0^{\infty}dk\,k^{D-1}P(k)\nonumber\\
&=& \frac{2}{(4\pi)^{D/2} \Gamma(D/2)} \int_0^{\infty}d(\mathrm{ln}\,k)\,k^DP(k).\nonumber\\
\end{eqnarray}
\noindent where $\Gamma(x)$ is the Gamma function. From this, one can interpret $k^D P(k)$ - with the
addition as the contribution of the power spectrum, on a logarithmic scale,  to the total variance of the density field.
The numerical prefactors can be computed with the help of the recurrence relation $\Gamma(1+x)=x\Gamma(x)$, and the values
$\Gamma(1)=1$ and $\Gamma(1/2)=\sqrt{\pi}$ for the Gamma function. For two-dimensional space, $D=2$, the
field variance $\sigma^2$ is given by
\begin{equation}
\sigma^2 \ = \ \frac{1}{2\pi} \int_0^{\infty}d(\mathrm{ln}\,k)\,k^2 P(k)\,,
\end{equation}
\noindent while for three-dimensional space, $D=3$, we have 
\begin{equation}
\sigma^2 \ = \ \frac{1}{2\pi^2} \int_0^{\infty}d(\mathrm{ln}\,k)\,k^3 P(k)\,.
\end{equation}
Finally, we make the observation that since the distribution of a homogeneous Gaussian random field is completely determined by its
covariance function. Hence, the distribution of isotropic Gaussian fields is determined purely and fully by the spectral density $P(k)$. 

\subsection{Filtered fields}
When assessing the mass distribution of a continuous density field, $f({\vec x})$, a common practice in
cosmology is to identify structures of a particular scale $R_s$ by studying the field smoothed at
that scale. This is accomplished by means of a convolution of the field $f({\vec x})$ with a particular smoothing
kernel function $W_s({\vec r}; R_s)$, 
  \begin{equation}
  f_s (\vec x)  \,=\, \int f(\vec y)\, W_s(\vec y -\vec x; R_s)\, d\vec y \,.
  \end{equation}
Following Parseval's theorem, this can be written in terms of the Fourier integral, 
\begin{equation}
  f_s (\vec x)  \,=\,  \int_{\mathbb R^3} \frac{d^3\vec k}{(2\pi)^3}\, \,{\hat f}(\vec k)\,{\hat W}(kR_s)\, \exp(-i\vec k\cdot \vec x)\,, 
\end{equation}
in which ${\hat W}(kR_s)$ is the Fourier transform of the filter kernel. From this, it is straightforward to see that
the corresponding power spectrum $P_s(k)$ of the filtered field is the product of the unfiltered power spectrum $P(k)$
and the square of the filter kernel ${\hat W}(kR_s)$
\begin{equation}
  P_s(k; R_s)\,=\,P(k)\,{\hat W}^2(kR_f)\,.
\end{equation}

\subsection{Excursion sets}
The \emph{superlevel sets} of the smoothed field $f_s(\vec x)$
define a manifold $\Mspace_\nu$ and consists of the regions
\begin{eqnarray}
  \Mspace_\nu  &=&  \left\{ \vec x \in \Mspace \mid f_s (\vec x) \in [f_{\nu}, \infty) \right\} \nonumber\\
                 &=&  f_s^{-1} [f_{\nu}, \infty) . 
\label{eq:excursionset}
\end{eqnarray}
In other words, they are the regions where the smoothed density is less than or equal to the threshold value $f_{\nu}$, 
\begin{equation}
\nu\ =\ \frac{f_{\nu}}{\sigma}\,,
  \end{equation}
with $\sigma$ the dispersion of the smoothed density field.

\medskip
Our analysis of the Betti numbers, Euler characteristic and Minkowski functionals of Gaussian random fields
consists of a systematic study of the variation of these topological and geometric quantities as a function
of excursion manifolds $\Mspace_\nu$, ie. as a function of density field threshold $\nu$. In other words, we
investigate topological and geometric quantities as function of density parameter $\nu$.

\section{Topology and geometry:\\
  \ \ \ \ \ \ Betti numbers, Euler characteristic and \\
  \ \ \ \ \ \ Minkowski functionals}
\label{sec:topology_ch2}

In this section, we first define the cosmologically familiar genus, Euler characteristic, and the Minkowski functionals. Subsequently, we give an informal presentation and a summary on 
the theory of homology, and the concepts 
essential to its formulation. For a more detailed description, in a cosmological framework, we refer the reader to \cite{weygaert2010}, \cite{weygaert2011}, \cite{pranavThesis},
and \cite{PEW16}.

\subsection{Euler characteristic and genus}

The Euler characteristic (or Euler number, or Euler-Poincar\'e characteristic) is a topological invariant, an integer  that describes aspects of a topological space's shape or structure regardless of the way it is bent. It was originally defined for polyhedra but, as we will see in the following subsection, has deep ties with homological algebra. 

Despite this generality, for the moment we will concentrate on the two and three dimensional settings, since these are the most relevant to cosmology. Suppose $\Mspace$ is a solid body in $\mathbb R^3$,  and we triangulate it, and  its boundary $\partial \Mspace$ using $v$ vertices, $e$ edges, and $t$ triangles and $T$ tetrahedra, all of which are examples of  \emph{simplices}.  A vertex is a 0-dimensional simplex, an edge is a 1-dimensional simplex, a triangle is a 2-dimensional simplex, and a tetrahedron is a 3-dimensional simplex \citep{okabe2000,vegter1997,vegter2004,zomorodian2009,EdHa10,PEW16}. The triangulation of $\partial \Mspace$ is made up of a subset of the vertices, edges, and triangles used to triangulate $\Mspace$, and we denote the numbers of these by
 $v_\partial$, $e_\partial$ and $t_\partial$.
 
 Formulae going back, essentially, to \cite{Eul58}, define the  \emph{Euler characteristics} of $\Mspace$ and $\partial\Mspace$ - traditionally denoted as $\chi(\Mspace)$ and $\chi(\partial\Mspace)$ - as the alternating sums 
 \begin{equation} 
\chi(\Mspace)\,=\, v-e+t-T,\qquad
\chi(\partial\Mspace)\,  = \,  v_\partial - e_\partial + t_\partial,
\label{eq:eulerpolyh} 
\end{equation} 
with similar alternating sums appearing in higher dimensions. It is an important and deep result that the Euler characteristic does not depend on the triangulation.

A more global, but equivalent, definition of the Euler characteristic would be to take $\chi(\Mspace)$ to be the number of its connected components, minus the number of its `holes' (also known as  `handles' or `tunnels'; regions through which one can poke a finger) plus the number of its enclosed voids (connected, empty regions).
For $\partial\Mspace$, or, indeed, any general, connected, closed two-dimensional surface, the Euler characteristic is equal to twice the  number of components minus twice the number of tunnels. If the surface is not closed, but has $b$ boundary components, then the number of such components needs to be subtracted from this difference.

The number of holes  of a connected, closed surface $S$ can be formalized in terms of its \emph{genus}, $g(S)$.
For a connected, orientable surface, the genus
is defined , up to a constant factor,
as the maximum number of disjoint closed curves that can be drawn on $S$ so that cutting along them does not leave the surface disconnected.  It
thus follows that the genus of a surface is closely related to its Euler 
characteristic, via:

\begin{equation}
\label{eq:rob:chi}
\chi(S)\, =\, 2-2g(S) .   
\end{equation}

Another result linking the Euler characteristic with the genus is that  three dimensional regions $\Mspace$ which have smooth, closed manifolds $\partial \Mspace$ as boundary, $\chi(\Mspace) = \smallhalf\chi(\partial\Mspace)$. It thus follows from \eqref{eq:rob:chi} that 
\begin{equation}
\chi(\Mspace)\, = \, \smallhalf \chi(\partial \Mspace) \, =\,   1 - g(\partial  \Mspace)\,.
\end{equation}

Both the genus and the Euler characteristic  have been an important focal point of topological studies in cosmology since their introduction in the cosmological setting \citep{GDM86,HGW86}.  Both have been used extensively in the study of models as well as observational data, with a strong emphasis on the test of the
assumption of Gaussianity of the initial phases of matter distribution in the Universe, as well as the large scale structure at the later epochs.
One reason for this is because of the existence of a closed analytical expression for the mean genus and the Euler characteristic of the excursion
sets of Gaussian random fields. For excursion sets $\Mspace_\nu$ of a Gaussian field at normalized level $\nu=f/\sigma$ (Equation~\ref{eq:excursionset}),
the mean Euler characteristic $\langle\chi(\nu)\rangle$ in a unit volume is given by \citep{Dor70,Adl81,bbks,HGW86}
\begin{equation}
\langle\chi(\nu)\rangle~=~-\frac{\lambda^3}{2\pi^2}\,(1-\nu^2)e^{-\nu^2/2},
\label{eqn:genusgauss}
\end{equation}
\noindent where $\lambda$ is proportional to the second order moment $\langle k^2\rangle$ of the power spectrum $P(k)$, and thus
proportional to the second order gradient of the autocorrelation function,
\begin{equation}
  \lambda^2 \,=\,\frac{\langle \|k^2\|\rangle}{3}\,=\,\frac{\sigma_1^2}{\sigma^2}\,=\,
  \frac{\int_0^{\infty} {d^3\vec k}\ k^2 P(k)}{\int_0^{\infty} {d^3\vec k}\ P(k)}\,,
  \label{eq:lambdak}
\end{equation}
or, in other words, proportional to the second order gradient of the correlation function,
\begin{equation}
  \lambda^2\,=\,-\frac{\xi''(0)}{\xi(0)}\,.
  \label{eq:lambdaxi}
\end{equation}
From this expression we may immediately observe that the Euler characteristics has only a weak sensitivity on the power spectrum
of a Gaussian field. It is limited to the overall amplitude, via its $2^{nd}$ order moment, while the variation as  a function of
threshold level $\nu$ does not bear any dependence on power spectrum. For the purpose of evaluating the Gaussianity of a field,
the Euler characteristic - and related genus - therefore provide a solid testbed. It is one of the reasons why the analytical expression
of Equation~\ref{eqn:genusgauss} plays a central role in topological studies of the Megaparsec scale cosmic mass distribution. Nonetheless, the
principal reason is that it establishes the reference point for the assessment and comparison of the majority of topological measurements.

Nonetheless, some care should be taken. As we will argue below, when discussing in Section~\ref{sec:gkf} the general context for such geometric measures in terms
of the Gaussian Kinematic Formula, this expression is valid only under strict conditions on the nature of the manifold $\Mspace_\nu$. The expression is only valid in the case where the superlevel
set is a smooth, closed manifold. Additional terms would appear when the boundary $\partial\Mspace_\nu$ of the manifold has edges or
corners. For the idealized configurations of the cubic boxes with periodic boundary conditions, such additional terms are not relevant. However,
in the real-world setting of cosmological galaxy surveys, selection effects may yield effective survey volumes that suffer a range of artefacts.

The Euler characteristic and Genus have been used extensively in the study of models as well as observational data, with a strong emphasis on the test of
the assumption of Gaussianity of the initial phases of matter distribution in the Universe, as well as the large scale structure at the later epochs.

\subsection{Minkowski functionals}
Although, as we emphasised in the previous subsection, the Euler characteristic is an essentially topological concept, it also has a role to play in geometry, as one of a number of geometric quantifiers,  which include the notions of volume and surface area.
There are $D+1$ such quantifiers for $D$-dimensional sets, and they go under a number of names, orderings, and normalisations, including, Minkowski functionals, 
 quermassintegrales,  Dehn and Steiner functionals, curvature integrals,  intrinsic volumes,  and  \LKCs. 
Most of the mathematical literature treating them is integral geometric in nature (e.g.\ \citep{Mecke94,schmalzing1997,schmalzing1999,sahni1998}) but they also often computable via differential geometric techniques (for which \cite{Adl10} is a useful reference for what we need).
We need only Minkowski functions $Q_j$ and \LKCs\ $\lips_j$, which, when both are defined, are related by the fact that
\beqn
\label{eq:rob:LipMink}
Q_j(\Mspace) \ =\  j! \omega_j  \lips_{D-j}(\Mspace),   \qquad j=0,\dots,D,
\eeqn
and $\omega_k = \pi^{k/2}\Gamma(1+k)/2)$ is the volume of a $k$-dimensional unit ball.  ($\omega_0=1,$   $\omega_1=2$, $\omega_2=\pi$, $\omega_3=4\pi/3$.) We will invest a little more space on these quantities than actually necessary for this paper, exploiting the opportunity to clarify some inconsistencies  in the ways these terms are used in the cosmological and mathematical literatures.

A useful way to define these quantities is via what is known as Steiner's formula  (which is generally quoted in the integral geometric setting of convex sets) or Weyl's tube formula (in the differential geometric setting of regions bounded by pieces of smooth manifolds, glued together in a `reasonable' fashion). Writing $V_D$ to denote $D$-dimensional volume, this reads as  
\begin{eqnarray}
\label{geometry:steiner:formula}
V_D
\left(\left\{x\in\mathbb R^D: \min_{y\in\Mspace} \|x-y\|\right\} \leq \rho \right) &=&
 \sum_{j=0}^D  \frac{\rho^{j}}{j!}  Q_j(\Mspace)\nonumber \\
&=& \sum_{j=0}^D \omega_{D-j} \rho^{D-j} \lips_j(\Mspace).
\end{eqnarray}
where $\rho$ is small, and the set in the left hand side is know as the tube around $\Mspace$ of radius $\rho$.

In any dimension, it is trivial (set $\rho=0$) to check from the definition \eqref{geometry:steiner:formula} that $Q_0$ and $\lips_D$ measure $D$-dimensional volume. It is not a lot harder to see that $Q_1$ and $2\lips_2$ measure  surface area. The other functionals are somewhat harder to define, but it is always true, and a deep result,  that 
\begin{equation}
\chi(\Mspace) \, =\,    \lips_0(\Mspace)   \, = \, \frac{1}{D!\omega_D} Q_D(\Mspace)\,. 
\end{equation}
\noindent In the 3-dimensional case of most interest to us, this leaves only $Q_2$ and $\lips_1$ to be defined. Integral geometrically, if the manifold
$\Mspace$ is convex,  $\lips_1(\Mspace)=Q_2(\Mspace)/2\pi$  is twice the {\it caliper diameter} of $\Mspace$. The latter is defined as follows: 
place $\Mspace$ between two parallel planes (calipers), measure the distance between the planes, and  average over all rotations of $\Mspace$.

A property that will actually be important for us later is the scaling property that, for any $\lambda>0$,
\beqn
\label{rob:eq:scaling}
\lips_j(\lambda\Mspace) \ = \ \lips_j(\lambda x: x\in\Mspace) \ = \ \lambda^j  \lips_j(\Mspace) .
\eeqn

As we already noted, in general all the LKCs\  can also be calculated via differential geometry and curvature integrals, at least when $\partial\Mspace$ is a smooth \emph{stratified manifold}. These include, for example, cubes, for which the interior of the sides, edges, along with the corners, are all submanifolds of the cube, along with cubes which have been deformed in a smooth manner.  In the future, we will assume that $\Mspace$ is a nice stratified manifold. The simplest situation for describing the differential geometric approach to Minkowski functionals occurs when $\partial\Mspace$ is actually a smooth closed, manifold. (i.e.\ non-stratified, and without a boundary). The formulae, for $D=3$, are then
\beq
 {\tilde Q}_0(\Mspace) &=&  \int_\Mspace d^3x, 
\label{rob:eqn:Q_0} \\
 {\tilde Q}_1(\Mspace) &=& \int_{\partial \Mspace} d^2 S({x}),\\
\label{rob:eqn:Q_1}
 {\tilde Q}_2(\Mspace) &=& 
\int_{\partial \Mspace} d^2 S({x}) \left(\kappa_1  + \kappa_2\right),\\
\label{rob:eqn:Q_2}
 {\tilde Q}_3(\Mspace) &=& 
2\int_{\partial \Mspace} d^2 S({x}) \, \kappa_1\kappa_2 ,
\label{rob:eqn:Q_3}
\eeq
where $\kappa_1(x)$ and $\kappa_2(x)$ are the principal curvatures of $\partial\Mspace$ at the point $x  \in \partial\Mspace$,  and $S$ is surface measure. Equation \eqref{rob:eqn:Q_3}, known as the Gauss-Bonnet theorem, 
encapsulates the remarkable fact that a topological characteristic such as the Euler characteristic of a set, which is invariant to bending and stretching, is accessible as the integral of the curvature of its boundary. In Section~\ref{rob:sec:CandA}, we will relate these formulae to the standard formulae used in cosmology to compute the Minkowski functionals. 

There are two very important facts to always remember when using the above four formulae. The first is that different authors often define the $Q_j$ slightly
differently, so that factors of 2 and $\pi$ may appear in front of the integrals. As long as there is consistency within a particular paper, this is of little
consequence. Our own choice of constants is dictated by the tube formula of \eqref{geometry:steiner:formula} and the simple connection \eqref{eq:rob:LipMink}
between the
\LK\ and Minkowski functionals. More important, however, is the fact that the simple expressions in 
\eqref{rob:eqn:Q_0}--\eqref{rob:eqn:Q_3} hold only because of the assumption that the space $\Mspace$
is a smooth, closed, manifold. As we will argue in the discussion in Section~\ref{sec:gkf} on the Gaussian Kinematic Formula in less idealistic circumstances
the situation is less straightforward. If the boundary $\partial\Mspace$ has edges or corners then there are additional terms, involving curvature integrals
along the edges and angle calculations at the corners. These terms have  typically been ignored in the cosmological literature when discussing the mean values
of excursion sets, leading to results which are actually approximations, rather than  exact formulae, as they are often presented. This point will be taken up
again below, in section \ref{sec:gkf}, where, while giving exact results, we shall also show why the approximations are well justified.

\subsection{Homology and Betti numbers}

We now return to purely topological descriptions of sets, in essence breaking up the information encoded in the Euler characteristic to component,
and more informative, pieces.

A stratified manifold, which need not be connected,  can be composed of a number of objects of different topological natures. For example,
in three dimensions, each of these might be topological balls, or might have tunnels and voids in them. These 
independent objects, tunnels and voids are different topological 
components of a manifold, and  have direct relevance to some familiar 
properties of the cosmic mass distribution. For example, the 
distribution and statistics of independent components as a function of 
scale or density threshold is a direct measure of the clustering 
properties of the mass distribution. The number of tunnels 
as well as the changes in their connectivity, as a function of scale 
or density threshold, can be an indicator of percolation properties of 
the cosmic mass distribution. Similarly, the topological voids have a 
direct correspondence with the vast near empty regions of cosmic mass 
distribution called the cosmic voids.

The notions of connectedness, tunnels, and voids, along with their extensions to higher dimension, have  formal definitions 
through the notion of homology \citep[see e.g.][]{munkres1984elements}. They are associated with the $p$-dimensional 
\emph{cycles} of a $d$-dimensional manifold ($p = 0 \ldots d$). In dimension 3,  a
$0$-cycle corresponds to  a connected object, a $1$-cycle to  a loop enclosing a 
tunnel, and a $2$-cycle to  a shell enclosing a void. In general, when properly formulated, a $k$-cycle in an object of dimension
greater than $k$ corresponds to the $k$-dimensional boundary of a $(k+1)$-dimensional void. 

Not all these 
cycles are independent. For example, one can draw many loops around 
a cylinder, all of which are topologically equivalent. The collection 
of all $p$-dimensional cycles is the $p$-th homology {group} 
$\Hspace_p$ of the 
manifold, and the rank of this group is the collection of all 
\emph{linearly independent} cycles. The rank is denoted by the \emph{Betti 
numbers} $\beta_p$, where $p = 0, \ldots , d$ \citep{Bet71}.  In dimension 3, the 
three Betti numbers have simple, intuitive 
meanings: $\beta_0$ counts the number of independent components, $\beta_1$ 
counts the number of loops enclosing the independent tunnels, and 
$\beta_2$ counts the number of shells enclosing the independent voids. 

A more mathematically rigorous definition of these concepts can be found in  the
traditional literature of homology; e.g.\ \cite{munkres1984elements} and \cite{EdHa10}. For  more details,
in an intuitive and cosmological setting, see \cite{PEW16} and \cite{weygaert2011}. 

\begin{figure*}
\centering
  \includegraphics[width=0.98\textwidth]{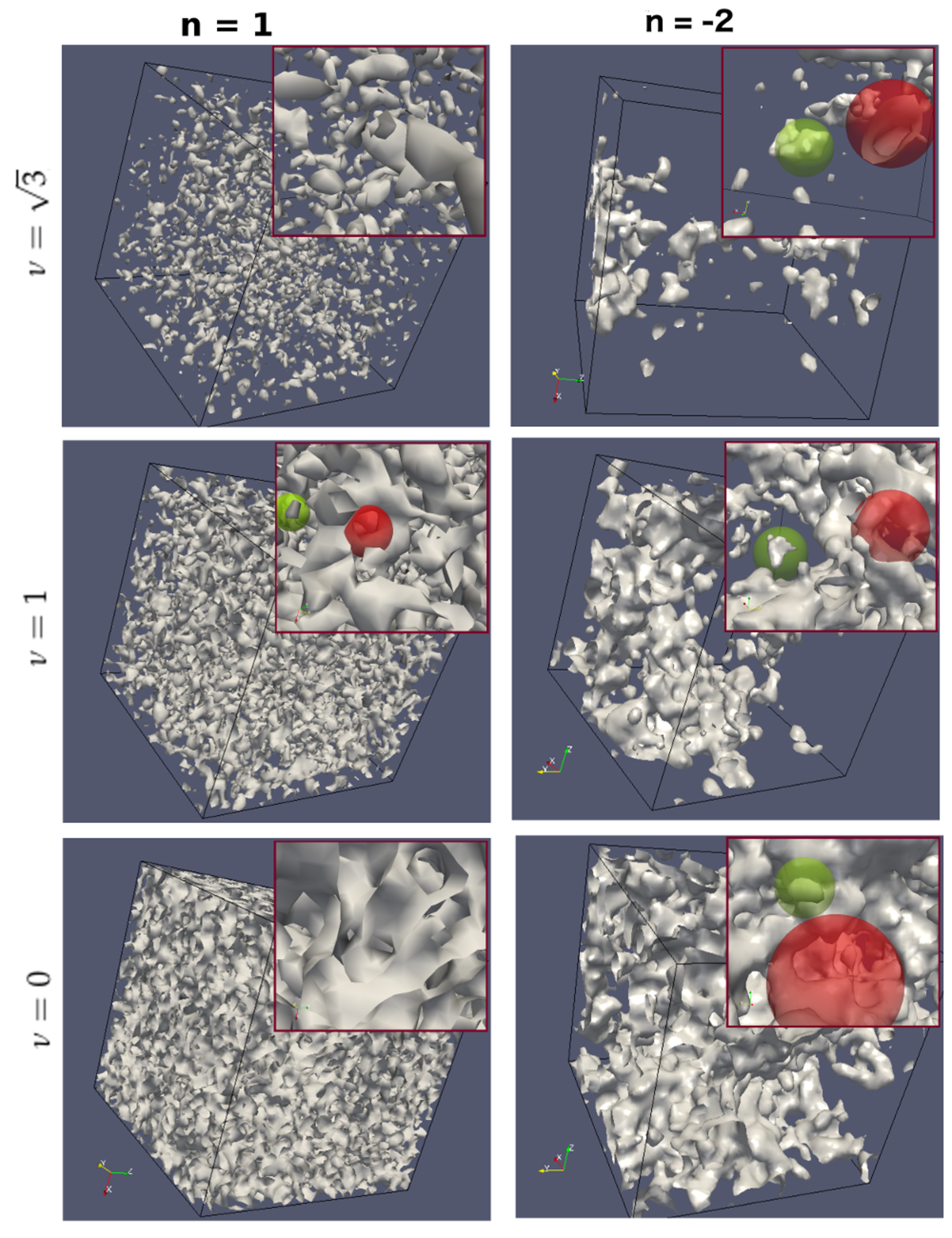}\\  
  \caption{  Iso-density surfaces denoting the structure of the field for 
  three different density thresholds $\nu = \sqrt{3}, 1, \text{and } 
  0$, for the $n = 1$ and the $n = -2$ models. The left column presents 
  the iso-density surfaces for the $n = 1$ model and the right column 
  presents the contour surfaces for the $n = -2$ model. Examples of typical tunnels 
  are enclosed in translucent 
  red spheres; examples of typical isolated islands are enclosed in green spheres.
  The topology of the contour 
  surfaces shows a dependence on the choice of the power spectrum, as 
  well as the density threshold.} 
  \label{fig:contour_surfaces}
\end{figure*}

\subsubsection{Betti numbers and Euler characteristic}
Like the Euler characteristic, the Betti numbers are topological 
invariants of a manifold, meaning that they do not change under 
systematic transformations under rotation, translation and deformation. 
Their relationship to the Euler 
characteristic is given by the following formula, which is a algebraic topological version of the original Euler-Poincar\'{e} Formula,
in which the summands were numbers of simplices of varying dimension in a triangulation.
\begin{equation}
  \chi ~~=~~ \beta_0 - \beta_1 + \beta_2 - \cdots + (-1)^d \beta_d .
\label{eq:eulerpoincare_intro}
\end{equation}
Yet, even the Betti numbers don't determine a manifold completely. Two topologically inequivalent manifolds my have equal
Betti numbers. 
One implication of this is that the set of $d$ Betti numbers contain more topological information than is contained in the
Euler characteristic. Hence, two manifolds may have the same Euler characteristic, yet be topologically distinctly different
in terms of its Betti numbers. In the context of Gaussian random fields we will see that this finds its expression of power
spectrum sensitivity: while the variation of the Euler characteristic as a function of density threshold of a superlevel set is
independent of power spectrum, we find distinct sensitivities of Betti numbers on the power spectrum (see Section~\ref{sec:betti_and_genus_result}
and \cite{PPC13}).

\subsubsection{\emph{Meatball-like}, \emph{Swiss-cheeselike} and \emph{Sponge-like} topologies}
\label{sec:spongetopology}
The description of topology through connected components, tunnels and voids has parallels in the earlier works related to the topological studies of cosmic mass
distribution. \cite{GDM86} introduced the terms \emph{Meatball-like} and \emph{Swiss-cheeselike} topologies to describe the dominance of either \emph{islands} - connected
components - and \emph{voids}. As is apparent from the terms, \emph{Meatball-like} topology refers to sets dominated by mainly isolated objects. Opposite to this are
the \emph{Swiss-cheeselike} topologies, denoting a  manifold composed of a single or a few components with the presence of fully enclosed cavities much like the inside of
cheese. In other words, while a pattern with \emph{Meatball-like} topology resembles that of black polka dots on a white background, the \emph{Swiss-cheeselike} topology is
that of white polka dots on a dark background \citep[see][]{GDM86}. These terminologies are intuitively meaningful, and present a clear picture in the mind of the reader.
Formally, however, they are no more than a colourful way of indicating the dominant Betti number. Nevertheless, we will borrow these terms from \cite{GDM86} to
augment intuitive understanding for the reader.

The topological \emph{Meatball-like} and \emph{Swiss-cheeselike} configurations are characteristic for two extreme outcomes of different cosmological structure formation scenarios. The
\emph{Meatball-like} topology would involve the formation of high-density islands - dependent on scale galaxy halos, clusters, or superclusters - in a low-density ocean. It was
supposed to be the typical outcome of bottom-up hierarchical formation scenarios such as Cold Dark Matter cosmologies. The \emph{Swiss-cheeselike} topologies were more characteristic
of the top-down formation scenarios, which produce a texture in which low-density or empty void regions appear to be carved out an otherwise higher density background. 
This would be the result of a formation scenario in which primordial perturbations over a narrow range of scales would assume a dominant role, manifesting itself
with voids would occupy most of space \citep[see e.g.][]{weygaert2002}. 

\cite{GDM86} and subsequent studies of the genus or Euler characteristic of the cosmic matter and galaxy distribution claimed that its topology is only
manifestly \emph{Meatball-like} at high density thresholds, and \emph{Swiss-cheeselike} at very low density thresholds, while it is characteristically
\emph{Sponge-like} at the median density level. A \emph{Sponge-like} topology points to a set with
a percolating structure, which signifies the presence of a single or a few connected components, each marked by the presence of tunnels that percolate the
structure. In this phase, tunnels are the dominant topological features. Strictly speaking, and usually interpreted as such in cosmology
\citep[see e.g.][]{GDM86,GHVP08}, a sponge-like topology means that at median density level (which for the symmetric Gaussian fields corresponds to the
mean density level $\nu=0$), at which high and low density regions each take up
50\% of the volume, the high density regions form one multiply connected region while the low density regions also form one connected region that
is interlocking with the high-density region \citep{GHVP08}. In other words, in a pure \emph{Sponge-like} topology there is only one underdense void region and only
one overdense region, each of these evidently characterized by an irregular and indented surface and by numerous percolating alleys or tunnels. In other words,
these claims  suggest that \emph{Sponge-like} topologies correspond to one where the Betti numbers $\beta_0=1$ and $\beta_1=1$ at the median density. We will soon
see that the reality is slightly more complex. 

For a visual appreciation of the different topological identities, Figure~\ref{fig:contour_surfaces} presents the iso-density surfaces 
of a simulated Gaussian random field  over a cubic region for three different density thresholds $\nu = \sqrt{3},\ 1,\ 0$, and for
two different Gaussian fields with a power-law power spectrum, namely the $n = 1$ and the $n = -2$ models. The left column presents 
the contour surfaces for the $n = 1$ model, the right column the contour surfaces for the $n = -2$ model. By means of enclosing translucent
spheres we show a typical tunnel, and we show isolated objects by means of an enclosing green translucent sphere.
The visualizations in Figure~\ref{fig:contour_surfaces} immediately reveal the considerable contrast in topology between the different
Gaussian field realizations, most evidently when assessed at around the mean density level $\nu=0$. While both are \emph{Sponge-like} at around
this threshold, we do note some stark differences. For the $n = 1$ model, the topology is predominantly sponge-like, with a dominant presence of
short loops, most of which are like indentations of a single, large connected surface. By contrast the topology of the $n = -2$ model is a visible
mixture of loops and  as well as isolated islands. In general, the overall topology consists of a mixture of the various topological
components, with different mixing fractions for Gaussian fields with different power spectra.

It is at this point that we may appreciate the increased information content of Betti numbers, as opposed to the more limited topological
characterization by the Euler characteristic or genus only. In the context of homology, we can directly relate terms like \emph{Meatball-like}, \emph{Swiss-cheeselike}
or \emph{Sponge-like} topology to a more quantitative characterization in terms of the relative values of $\beta_0$ and $\beta_2$. The situation where
the $\beta_0$ assumes the vast share of the topological signal is the \emph{Meatball-like} topology of \cite{GDM86}. The opposite situation of a dominant
$\beta_2$ signal is that of the \emph{Swiss-cheeselike} topology, while a \emph{Sponge-like} topology corresponds to the entire field divided into a low number
of overdense and underdense regions, and thus low values for $\beta_0$ and $\beta_2$, always in combination with a large value for $\beta_1$,
corresponding to the tunnels and loops that form indentations of these connected regions. 

We refer to Section~\ref{sec:betti_and_genus_result} for a considerably more quantitative evaluation of the relative contributions of topological features in
terms of the corresponding Betti numbers $\beta_0$, $\beta_1$ and $\beta_2$. 

\section{The Gaussian kinematic formula}
\label{sec:gkf}

As mentioned above, one of the main reasons that the  Euler characteristic, genus and the Minkowski functionals have played such a useful role in cosmology is that there are exact, analytic, formula for their expected values, when the characteristics that are being computed are generated by the superlevel sets of Gaussian random fields. These formula are old, going back to \cite{Dor70} for a simple 2-dimensional case, with the cosmological literature generally relying mainly on \cite{Adl81} and \cite{bbks} for full results. Over the last decade or so, major extensions of these formulae have been developed, going under the name of the \emph{Gaussian kinematic formula}, or, hereafter, \emph{GKF}. The GKF, in one compact formulae, gives the expected values of the Euler characteristic (and so genus), all the \LKCs\ (and so Minkowski functionals) described earlier as well as extensions of them, for the superlevel sets (and their generalisations in vector valued cases) of a wide class of random fields, both Gaussian and only related somehow to Gaussian, and both homogeneous and non-homogeneous. The parameter sets of these random fields are also very general, and cover all examples required in cosmology, without  any need to ignore boundary effects.

We do not actually use the GKF in the current paper, since later on we shall be more concerned with Betti numbers than Euler characteristics or Minkowski functionals, and, unfortunately, these are not covered by the GKF. In fact, for reasons we shall explain later, there is no detailed statistical theory for them, which is why this paper is mainly computational. Nevertheless, since most of the literature around the GKF is highly technical differential topology, we take this opportunity to discuss the GKF in a language that should be more natural for cosmology. Our basic references are \cite{Adl10} for all the details,
and \cite{ATSF} and \cite{ARF} for less detailed, but  more user-friendly, treatments.

\subsection{The GKF}

The first component of the GKF is a $D$-dimensional parameter space $\Mspace$, which is taken to be a $C^2$ Whitney stratified manifold. As mentioned earlier, this is a set made out of glued together pieces, each one of which is a sub-manifold of $\Mspace$, along with rules about how to glue the pieces together. We group all the $k$-dimensional  submanifolds together, and write the collection as $\partial\Mspace_k$, $k=0,\dots,D$. For example, if $\Mspace$ is a 3-dimensional cube, then $\partial\Mspace_3$ is the interior of the cube,  $\partial\Mspace_2$ contains the interiors of its six sides, $\partial\Mspace_1$ collects the interiors of the eight edges, and $\partial\Mspace_0$ is the collection of the eight vertices. In general, we write 
\beqn
\label{rob:eq:decompositon}
\Mspace \ = \ \bigsqcup_{k=0}^D \partial\Mspace_k,
\eeqn
where the union is of disjoint sets. The parameter space $\Mspace$ could be a subset of a Euclidean space, or a general, abstract, stratified manifold. To the best of our knowledge, the Euclidean setting is (so far) the only one used in cosmology.

The second component of the GKF is a twice differentiable, constant mean, Gaussian random field, $f:\Mspace\to\mathbb R$, with constant variance. There is no requirement of stationarity or isotropy, only of constant mean and variance. For convenience, we take these to be 0 and 1, respectively. Changing them in the formulae to follow involves nothing more than  addition, or multiplication, by constants. An extension of the second component, which is crucial for getting away from the purely Gaussian setting, is to take $d\geq 1$ independent copies, $f_1,\dots,f_d$ of $f$, and we write $\vec f = (f_1,\dots,f_d)$ for the vector valued random field made up of these as components.

The third, and final, component is a set $\cH\subset\mathbb R^d$, called a \emph{hitting set}. In most of the cases of interest to cosmology, $d=1$ and $\cH=[\nu,\infty)$ for some $\nu$.

The aim of the GKF is to give a formula for the expectations of geometric and topological measures of the  \emph{excursion sets}
\beqn
\label{rob:eq:AH}
A_\cH \ \equiv \ A_\cH\left(\vec f,\Mspace\right) \ = \ \left\{x\in \Mspace: \vec f(x)\in \cH\right\}.
\eeqn
In the particular case that  $d=1$, so that $f$ is real-valued, and $\cH$ is the set $[\nu,\infty)$, we are looking at super level sets of $f$, and write
\beqn
\label{intro:Au:equn}
A_\nu \ \equiv\  A_u\left(\vec f,\Mspace\right)\ =\  \{x\in \Mspace: f(x)\geq \nu\}.
\eeqn

In order to formulate the GKF, we need to revisit one definition and add an additional one. Recall the \LKCs\ of \eqref{geometry:steiner:formula}, which, together with the Minkowski functionals, we chose to define via a tube volume formula. This definition is adequate for a Euclidean set, but the most general version of the GKF works on abstract stratified manifolds. In that case the most natural definition of the \LKCs\ is not via a tube formula, but rather via curvature integrals akin to \eqref{rob:eqn:Q_0}-\eqref{rob:eqn:Q_3}. These curvatures will now involve the Riemannian curvatures and second fundamental forms of all the submanifolds in all the $\partial \Mspace_k$, and the Riemannian metric underlying all these turns out to be one related to the covariance function of the random field. All of this is  beyond the scope of this paper.  Nevertheless, although we shall concentrate on stationary random fields on subsets of Euclidean spaces (for which the decomposition  \eqref{rob:eq:decompositon} will still be relevant)  for the remainder of this paper, it is worthwhile remembering that this is but a small part of a much larger theory.

The remaining definition is of a Minkowski-like functional which, instead of measuring the size of objects, measures their (Gaussian)
probability content. To define it,  let $\vec X$ be a vector of $d$ 
independent, identically distributed, standard Gaussian random variables, and,
for a nice subset (e.g.\ locally convex, stratified manifold)  $\cH\subset\mathbb R^d$, and sufficiently small $\rho>0$, consider the Taylor series expansion
\beqn
\label{rob:eq:GMdef}
\text{Pr}\left\{X\in \left\{x\in\mathbb R^d: \min_{y\in\cH} \|y-x\| \right\} \leq \rho \right\}
\ = \ \sum_{j=0}^{\infty} 
\frac{\rho^j}{j!} \Min^d_j(\cH).
\eeqn
The coefficients, $\Min^d_j(\cH)$,  in this expansion,   due to  \cite{Taylor1}, are known as the \emph{Gaussian Minkowski functionals} of 
$\cH$, and play a similar role to the usual Minkowski functionals, with the exception that all measurements of size are now weighted with respect to probability content. 

In dimension $d=1$, with $\cH=[\nu,\infty)$, the $\mink^1_j(\cH)$ take a particularly simple form, and it is easy to check from a Taylor expansion of the Gaussian density that 
\beqn
\label{tubes:minkequalshermite:equation}
\Min^{1}_j([\nu,\infty)) \
=  \  H_{j-1}(\nu) \frac{e^{-\nu^2/2}}{\sqrt{2\pi}}.
\eeqn
where, for $n\geq 0$, $H_n$ is the $n$-th Hermite polynomial, 
\beqqn
H_n(x) =  n!\, \sum_{j=0}^{\lfloor n/2\rfloor} 
\frac{(-1)^j x^{n-2j}}{j!\, (n-2j)!\, 2^j},
\eeqqn
 and, for $n=-1$, we set
\beqn
H_{-1}(x)   \ = \ \sqrt{2\pi} e^{x^2/2}\Psi(x).
\eeqn
where
\beqn
\Psi(x) \ = \   \frac{1}{\sqrt{2\pi}} \int_u^\infty e^{-x^2/2}\,dx 
\eeqn
is the Gaussian tail probability.

We now have all we need to define the GKF, which  is the result that, under all the conditions above, and some minor technical conditions for which \cite{Adl10} is the best reference,
\beqn
  \label{eq:isotropic:kff:1d}
\left\langle\lips_i\left(A_\cH(f,\Mspace)\right)\right\rangle
\ =\
 \sum_{j=0}^{D-i} \sqbinom{i+j}{j} (2\pi)^{-j/2}\lips_{i+j}(\Mspace) 
\, \mink_j^{d}(\cH),
\eeqn
where the  combinatorial `flag coefficients' are defined by
\beqn
\label{geometry:square-binom:def}
\ssqbinom{n}{j} = \binom{n}{j} \frac{\omega_n}{\omega_{n-j} \; \omega_j}\,,
\eeqn
where $\omega_m$ is the volume of the unit ball in $\Rspace^m$:
\begin{equation}
  \omega_m\,=\,\frac{\pi^{m/2}}{\Gamma(\frac{n}{2}+1)}\,,
  \end{equation}
ie. $\omega_1=2$, $\omega_2=\pi$ and $\omega_3=4\pi/3$. 
(Note that all $\lips_j$ for $j>D$ are defined to be identically zero, so that the highest order \LKC\ in \eqref{eq:isotropic:kff:1d} is always $\lips_D(\Mspace$.)

All this is very general. The parameter space $\Mspace$ might be an abstract stratified manifold, and the \LKCs\ on both sides of the GKF might be  Riemannian curvature integrals.  On the other hand, the Gaussian Minkowski functionals are independent of the structure of the random field, and dependent only on the structure of the hitting set $\cH$. To see how this result works in simpler cases, we look at some more concrete examples.

\subsection{Examples: Rectangles, cubes and spheres}

To start,  we will take $f$ to be a mean zero Gaussian random field on $D$-dimensional Euclidean space, and allow a little more generality, with possibly general variance 
\beqn
\langle f^2(x) \rangle \ = \ \sigma^2.
\eeqn
To make the formulae tidier, we will also assume that $f$ has a mild form of isotropy, in that the covariance between two partial derivatives of $f$, in directions $\vec v_1$ and $\vec v_2$, is equal to $\lambda^2 \langle\vec v_1,\vec v_2\rangle$; viz.\ it is proportional to the usual Euclidean product of the directions. This will be the case, for example, if $f$ is homogeneous and covariance function has a  Taylor series expansion at the origin of the form
\begin{equation}
\xi (x) = \sigma^2\,-\,\smallhalf \lambda^2 \sigma^2 \|x\|^2\, +\, o(\|x\|^2)\,.
\end{equation}
\noindent Isotropy implies this, but we are actually assuming far less.  
This requirement implies that  $\lambda^2$ is the variance of any partial derivative of $f$, and that this variance is independent of the direction in which the derivative is taken.  In the homogeneous, isotropic case (see Equation~\ref{eq:lambdak} for the specific 3-D case), 
\beqn
\lambda^2 \sigma^2\, =\,  - \frac{1}{D}\,\sum_{j=1}^{D} \frac{\partial^2\xi(x)}{\partial x_j^2}\Big|_{x=0}
\, = \, \frac{\langle \|\vec k \|^2\rangle}{D}\,\xi(0)\,,
\label{rob:eq:lambdadef}
\eeqn
 where the partial derivative can be taken in any of the $D$ directions. Thus $\lambda^2$ can be found directly from the covariance function or, equivalently, as the second spectral moment.

For our first example, let $\Mspace$ be the $D$-dimensional rectangle $\Mspace_{Rec}= \prod_{j=1}^D[0,m_i]$. The usual, Euclidean, \LKCs\ of $\Mspace$ will then be 
\beqn
\label{rob:eq:lkcrectangle}
\lips_j^E\left(\Mspace_{Rec}\right) \ = \ \sum m_{i_1}\cdots m_{i_j},
\eeqn
where the sum is taken over the ${D \choose j}$ different choices of subscripts  $i_1,\dots,i_j$, and the additional superscript $E$ is to emphasise the Euclidean nature of the \LKCs. The corresponding Minkowski functionals are just products of reordered \LKCs, as in \eqref{eq:rob:LipMink}. The \emph{Riemannian} \LKCs\ needed for substitution in the GKF are then given by
\beqn
\label{rob:eq:RLKC}
\lips_j\left(\Mspace_{Rec}\right)\ = \  \lambda^{j}\lips_j^E\left(\Mspace_{Rec}\right).
\eeqn 

Let  $\cO_k$ denote the collection of all  ${D\choose k}$  $k$-dimensional faces of $\Mspace_{Rec}$  which include the origin.  The $k$-dimensional volume of a face $J\in\cO_k$ is written as $|J|$.
Then replacing the Riemannian \LKCs\ in the GKF by the Euclidean ones, for this case the GKF   reads as follows.

\beqn
\label{rgeometry:ELKC:equation} 
\langle\lips_i^E \left(A_\nu\right)\rangle \ =\   e^{-\nu^2/2\sigma^2}
 \sum_{j=0}^{D-i} \sqbinom{i+j}{j} \frac{\lambda^{j}}{(2\pi)^{(j+1)/2}}\, H_{j-1}\left(\frac{\nu}{\sigma}\right)\lips^E_{i+j}(\Mspace) .
\eeqn
It is easy to rewrite this in  terms of Minkowski functionals, when it  becomes the slightly less elegant formula
\beqn
\label{rgeometry:EQ:equation} 
\langle Q_i \left(A_\nu\right)\rangle  = e^{-\nu^2/2\sigma^2}
 \sum_{j=0}^{i} \sqbinom{D+j-i}{j}  \sqbinom{i}{j}
 \frac{ \omega_j j! \lambda^{j}    }{
  (2\pi)^{(j+1)/2}}  H_{j-1}\left(\frac{\nu}{\sigma}\right)     Q_{i-j}(\Mspace) .
\eeqn

\begin{figure}
\centering   
\includegraphics[width=0.95\columnwidth]{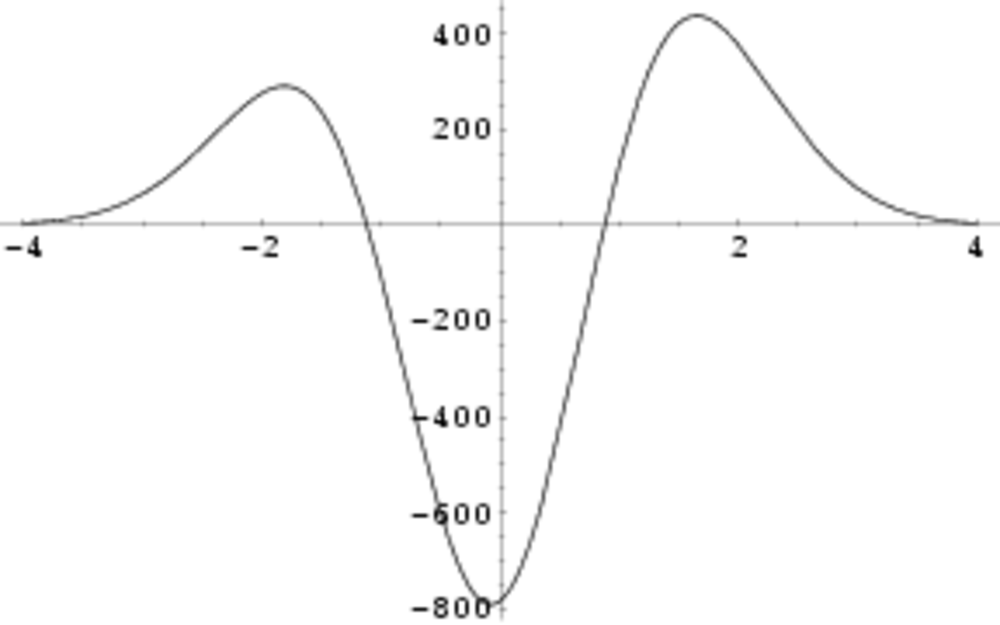}
\caption{A mean Euler characteristic curve for a Gaussian field over a three dimensional cube of limited size.
  Notice the substantial difference with the conventionally known and expected symmetric curve (see Equation~29). The latter
  forms the asymptotic situation for a very large sample size $T$ and a relatively ``quiet'' field $f$ within that volume. In a cosmological
  context this means that the symmetric curve can only be used as reference for a cosmic volume that is sufficiently large and represents
  a fair sample of the cosmic mass distribution. See text for details.}
  \label{EEC3-Gauss}
\end{figure}

To get a better feel for this equation, let us look the mean value of the Euler characteristic $\langle \chi(\Mspace) \rangle$, ie. of
the zeroth Lipschitz-Killing curvature $\lips_0(\Mspace)$R, in the cases $D=2$ and $D=3$, taking
$\Mspace$ to be a square or cube of side length $T$, and setting $\sigma^2=1$ for simplicity. In the two  dimensional case, we obtain
\beqn
\label{rgeometry:EEC-isotropic2:equation} 
 \left\langle\chi \left(A_\nu\right)\right\rangle =
\left[
\frac{T^2\lambda^2}{(2\pi)^{3/2}}\, \nu 
\ +\ \frac{2T\lambda}{2\pi} 
\right] e^{\frac{-\nu^2}{2}}  +  \Psi(\nu).\
\eeqn
In three dimensions, for the mean Euler characteristic \eqref{rgeometry:ELKC:equation} yields, again for $\sigma^2=1$,
\beqn
 \left\langle\chi \left(A_\nu\right)\right\rangle =
\left[\frac{T^3\lambda^{3}}{(2\pi)^{2}}\, (\nu^2  -1) + 
\frac{3T^2\lambda^2}{(2\pi)^{3/2}}\, \nu 
 + \frac{3T\lambda}{2\pi}  
\right] e^{\frac{-\nu^2}{2}}  +  \Psi(\nu) .
\label{eq:chigausscompl}
\eeqn
Figure~\ref{EEC3-Gauss} gives an example, over the unit cube, with $\lambda=880$ (see Equation~\ref{rob:eq:lambdadef}). 
It is clear that the Euler characteristic curve in Figure~\ref{EEC3-Gauss} differs substantially from the more conventionally
known symmetric curve specified by Equation~\ref{eqn:genusgauss}. As may be inferred from Equation~\ref{eq:chigausscompl} the symmetric curve only
represents a sufficiently valid asymptotic limit if the sample size $T$ is large and the field within this volume is relatively ``quiet''.
In a cosmological context this means that the symmetric curve can only be used as reference for a cosmic volume that is sufficiently large and represents
a fair sample of the cosmic mass distribution. This is still a relatively unknown fact in cosmological applications. 

Similar expressions as Equation~\ref{eq:chigausscompl} hold for the mean values of all the \LKCs\ and Minkowski functionals of excursion sets. 
   
There  is an important point that one should note about these formulae, which, while obvious in the simplest cases such as in \eqref{rgeometry:ELKC:equation} are actually general phenomena.    All of these formula contain obvious, or sometimes hidden, power series expansions. In the simple case of \eqref{rgeometry:ELKC:equation} there are three  such series. The most obvious one is in the size of the cube, as expressed through the side length $T$. If $T$ is large, then the first term, in $T^3$, is dominant. The opposite is true if $T$ is small. Overall, one can, correctly, relate to the coefficients of the powers of $T$ as expressions affected by the behaviour of $f$ in the interior of $\Mspace$, then on its boundary, an so forth. There is also an expansion in the second spectral moment, $\lambda^2$. The larger this moment, the rougher will be the field $f$, and this will lead to large \LKCs\ and Minkowski functionals. In the case of the general GKF \eqref{eq:isotropic:kff:1d}  in which the \LKCs\ are all Riemannian quantities, the measurements of the various `sizes' of $\Mspace$  involve a delicate combination of both the `physical' size  and shape of $\Mspace$, along with the roughness of $f$ in different regions. Nevertheless, the same general interpretation of these expansions still holds.    The final expansion is in the height parameter $\nu$. Clearly, as $\nu$ becomes large, the first term in the GKF - the one associated with the volume $\lips_D$ - is the dominant one. 

  The last two paragraphs are important for applications of the GKF. For example, 
 while the formulae of this subsection will look vaguely familiar to integral geometers, they probably look unusually complicated to a reader familiar only with the cosmology literature. We will explain the differences in section \ref{rob:sec:CandA}, but firstly briefly mention how to use the GKF for non-Gaussian random fields.

\subsection{Gaussian related random fields and the GKF}

Although the GKF is about Gaussian random fields, the way it is formulated, in terms of vector values fields and the general hitting set
$\cH$, allows it to also treat a certain class of non-Gaussian random fields as well. The class, while somewhat limited, turns out to be  broad enough
to cover many, if not most, statistical applications of random fields. 

To be more precise,
we shall call a random field $g:\Mspace\to \real^d$ a {\it Gaussian related}, $\mathbb R^d$-valued, random
 field if we can find 
a vector  valued Gaussian random field, $\vec f:\Mspace\to\mathbb R^D$, satisfying all the conditions of the GKF, and 
a function $F\:\real^D\to\real^d$,
such that $\vec g$ has the same multivariate distributions as $F(\vec f)$. 

In the trivial case that $D=1$, or, in general $D=d$ and $F$ is invertible, then the corresponding
Gaussian related process is not much harder to study than the original Gaussian
one, since  what happens at the level $u$ for $\vec f$ is precisely what happens at
the uniquely defined level $F^{-1}(u)$ for $\vec g$. In the more interesting cases in which  $F$ is not invertible, $\vec g=F(\vec f)$ can
provide a process that is qualitatively different to $\vec f$.
Three useful examples are given by 
the following three choices for $F$, where in the third we set $D=n+m$.
\beqn
\label{nongauss:threeexamples:equation}
\sum_1^D x_i^2,\qquad
\frac{x_1\sqrt{D-1}}{(\sum_2^D  x_i^2)^{1/2}},\qquad
 \frac{m\sum_1^n  x_i^2}{n\sum_{n+1}^{n+m}  x_i^2}.
\eeqn
The corresponding random fields are  known as  $\chi^2$ fields with 
$D$ degrees of freedom, the $T$ field with 
$D-1$ degrees of freedom, and the  $F$
  field with $n$ and $m$ degrees of freedom. These three random fields all have very different
spatial behaviour, and each is 
as fundamental to the statistical applications of random field theory.   In note of these three cases, as
in general for a Gaussian related random field, there is no simple point-wise transformation
which will transform   it to a real valued Gaussian field.

Note that for a Gaussian related field $\vec g$ the excursion sets $A_\cH$ 
can be rewritten as
\begin{equation}
\label{intro:nongaussexc:equation}
 \qquad\qquad  A_\cH(\vec g, \Mspace) \ =\  A_\cH(F(\vec f),\Mspace)  
\ =  A_{  F^{-1}(\cH)}(\vec f,\Mspace).        \nonumber 
\end{equation}
Thus, for example, the excursion set of a {\it real valued}
  non-Gaussian $g=F(\vec f)$ above a level
$u$ is equivalent to the excursion set for a {\it vector valued} 
 {Gaussian} $\vec f$ in $F^{-1}([u,\infty))\in \mathbb R^D$.  Consequently, as long as $F$ is smooth enough, expressions for the mean \LKCs\ of $\vec g$ follow immediately from the GKF, once one knows how to compute the corresponding Gaussian Minkowski functionals. This can be easy or hard, depending on the form of $F$. Examples  are given in  \cite{Adl10,ATSF} and \cite{ARF}  .

\subsection{Non-homogeneity}
\label{rob:sec:nonh}

Before turning to the connections between the GKF and related geometric results in the cosmology literature, we add a brief comment about computing the \LKCs\ in the non-homogeneous setting.  As mentioned above, the \LKCs, in general, implicitly incorporate information on the variance structure of the random field $f$. To see how this works, take $\Mspace$ to be a subset of $\mathbb R^D$, retain the assumptions of zero mean and constant unit variance, and write the (two-parameter) covariance function  of $f$ as
\begin{equation}
\xi(x,y)\, =\, \langle f(x)f(y)\rangle.
\end{equation}
Define a matrix valued function   $\Lambda (x) =(\lambda_{ij}(x))_{i,j=1,\dots,D}$ of second order spectral moments by 
\begin{equation}
\lambda_{ij}(x) \, = \, \left\langle k_ik_j \right\rangle \, = \, -\frac{1}{\xi(0)}\,\frac{\partial^2 \xi  (x)}{\partial x_i \partial x_j}\Big|_{x=0}\,.
\end{equation}
In terms of the previous notation for the isotropic case, with second spectral moment $\lambda^2$, we have $\lambda^2=\lambda_{ii}(x)$, independent of $i$ and $x$, and $\lambda_{ij}\equiv 0$ when $i\neq j$. In the homogeneous, but non-isotropic, case, the matrices  $\Lambda (x)$ may be a general covariance matrix, but will be independent of $x$.

It turns out that this is all that one needs to compute the leading \LKC\ in the general case, where we have 
\begin{equation}
\lips_D \, =\, \int_\Mspace \sqrt{  \text{det} (\Lambda(x)) }\, dx.
\end{equation}
If $\Mspace$ has a smooth boundary, then the next \LKC\ can be be calculated as a surface integral of a (Riemannian) curvature function, although the integrating measure is a little complicated. An easy case is that of 2-dimensional $\Mspace$, in which case if we  
first parametrize  $\partial \Mspace$  by a $C^2$ function $\gamma:[0,1]\to\mathbb R^2$, we have
\begin{eqnarray}
\lips_1 &=& \frac12 \int_0^1 \sqrt{\left(\frac{d\gamma(t)}{dt}\right)^T      
\Lambda(\gamma(t))\,   \frac{d\gamma(t)}{dt}  }\,dt.
\end{eqnarray}
For full details of the general case see \cite{Adl10}, and some specific worked cases in \cite{ARF,Eliran}. 

In many cases, it is possible to avoid analytic computation of the \LKCs, and simply estimate them from data. Differing approaches to this can be found in \cite{Adler-Bartz-Kou,Arminetal,TWJASA}.

\subsection{Cosmology: approximations and boundaries}
\label{rob:sec:CandA}

For the reader familiar with the cosmological literature on mean Euler characteristics and mean Minkowski functionals, much of the discussion will
probably seem unfamiliar and perhaps unnecessarily complicated. There are three reasons for this. The first is that cosmology has typically worked under
assumptions of homogeneity and isotropy, and we have already seen that in this case the \LKCs are considerably simpler than in the general case. The second
reason lies in the fact that there are only two main examples in cosmology: the two-dimensional sphere, for CMB studies, and subsets of $\mathbb R^3$, for
the Megaparsec galaxy and matter density studies. Under the restrictions of homogeneity and isotropy for these two cases, a general theory seems superfluous.

The third reason, however, is not so obvious, and is relevant to both of these parameter spaces. The fact is that the CMB is \emph{not} observed over the full sky, typically as a result of interference from bright foreground objects, such as our own galactic disk and bright point sources. Thus the parameter space $\Mspace$ in these cases is an, often complicated, subset of the sphere, with a convoluted boundary. Similarly, the data on the large-scale galaxy and matter
density are estimable only over sectors of the 3-dimensional universe that have been covered by observational surveys. Nearly without exception these are 
limited in terms of sky coverage and include objects only out to a certain distance. Also, they tend to suffer from incompleteness,
and usually involve similar foreground issues such as the obscuration by the gas and dust in the disk of our own Galaxy along the zone of avoidance.
In this case, $\Mspace$ is a compact 3-dimensional region with a complicated boundary and which, in fact, may not even be connected.

In other words, the boundary terms, which even in the homogeneous, isotropic case,  make the GKF so complicated, cannot be ignored in exact computations. 
A simple way out of this conundrum is to replace all the measures described above with  dimensionless, `normalised' measures. For example, rather than computing the total Euler characteristic $\chi (A_\nu(f,\Mspace))$ of a superlevel set, one works with 
 $|\Mspace |^{-1}\chi (A_\nu(f,\Mspace))$, where $\lips_D(\Mspace)=|\Mspace |$ is the volume, or surface area, of $\Mspace$, giving a `per unit volume' notion of Euler characteristic. The effect of this normalisation on the GKF is minimal. All terms, on both the right and left of the GKF, are similarly normalised. Working then on the implicit assumption that 
 ${\lips_D(\Mspace)}/{\lips_j(\Mspace)}$
 is small for large $\Mspace$, the GKF of  \eqref{eq:isotropic:kff:1d} leads to the approximation 
 \begin{equation}
 \left\langle\frac{1}{|\Mspace|} \lips_i\left(A_\cH(f,\Mspace)\right)\right\rangle
\ \approx \
 \sqbinom{D}{i} (2\pi)^{-(D-i)/2}
\, \mink_{D-i}^{d}(\cH), 
 \end{equation}
 while the simpler, Euclidean examples  \eqref{rgeometry:ELKC:equation}  and \eqref{rgeometry:EQ:equation},    in which $f$ is real valued, become
 \begin{equation}
\left\langle\frac{1}{|\Mspace|}  \lips_i^E \left(A_\nu\right)\right\rangle \ \approx \   e^{-\nu^2/2\sigma^2}
  \sqbinom{D}{i} \frac{\lambda^{D-i }}{(2\pi)^{(D-i+1)/2}}\, H_{D-i-1}\left(\frac{\nu}{\sigma}\right)
\end{equation}
and
\begin{equation}
\left\langle \frac{1}{|\Mspace|}  Q_i \left(A_\nu\right)\right\rangle  \ \approx \  e^{-\nu^2/2\sigma^2}
 \sqbinom{D}{i}  
 \frac{ \omega_i i! \lambda^{i}    }{
  (2\pi)^{(i+1)/2}}  H_{i-1}\left(\frac{\nu}{\sigma}\right)     .
\end{equation}
 Up to unimportant factors of 2 and $\pi$, due to slightly different definitions of the Minkowski functionals, 
 the last of these approximations is equivalent to the formulae given as exact equations in, for example
 \cite{tomita1993} and \cite{schmalzing1997}, following a tradition of ignoring the contributions of boundary 
effects going back at least half a century,  to \cite{Dor70}.

Under the - key - assumption that the space $\Mspace$ on which the Minkowski functionals are measured is a smooth, closed, manifold, their expected
values for Gaussian random fields, obtained from the evaluation of \eqref{rob:eqn:Q_0}--\eqref{rob:eqn:Q_3}, are given by rather straightforward
analytical expressions. These then coincide with the expected values of the Minkowski functionals ${\tilde Q}_m$ \emph{per unit volume} for 3D manifolds $\Mspace$ defined as
the excursion sets at normalized field levels $\nu=f/\sigma$, found by by \cite{tomita1993} and \cite{schmalzing1997}:
\begin{eqnarray}
\langle{\tilde Q}_0(\nu)\rangle &=& \frac{1}{2}-\frac{1}{2}\Phi\left( \frac{1}{\sqrt{2}}\nu\right)\,, \\
\langle{\tilde Q}_1(\nu)\rangle &=& \frac{\lambda}{3\pi}\exp\left(-\frac{1}{2}\nu^{2}\right)\,,\\
\langle{\tilde Q}_2(\nu) \rangle&=& \frac{2}{3}\frac{\lambda^{2}}{(2\pi)^{3/2}}\nu\exp\left(-\frac{1}{2}\nu^{2}\right)\,,\\
\langle{\tilde Q}_3(\nu) \rangle&=& \frac{\lambda^{3}}{4 \pi^2}(\nu^{2}-1)\exp\left(-\frac{1}{2}\nu^{2}\right).
\label{eqn:minkowski_ch2}
\end{eqnarray}
where $\lambda^2 = -\xi"(0)/\xi(0)$, as defined in Equation~\ref{eq:lambdaxi}, and
\begin{equation}
\Phi(x) \,=\,\int_{0}^{x}\mathrm{dt} e^{-t^{2}}\,.
\end{equation}
\noindent is the standard error function. These equations are equivalent to Equations \eqref{rob:eqn:Q_0}--\eqref{rob:eqn:Q_3}.

\subsection{On mean Betti numbers}
Returning now to the main theme of this paper, which revolves around purely topological concepts such as homology and associated quantifiers such as Betti numbers, the question that arises naturally is whether or not there is a parallel to the GKF, which, with the exception of the Euler characteristic, is about geometric quantifiers, for Betti numbers.

Unfortunately, to date the answer is mainly negative, and all indications are that it will remain that way for while \citep[see e.g.][]{wintraecken2013}.
While there are some high level, asymptotic as $\nu\to\infty$ results about the Betti numbers of excursion sets of Gaussian excursion sets in the mathematical
literature,  these are a consequence of the simple structure of Gaussian fields at these levels, and so the information on Betti numbers is minimal and
indirect (see e.g. Section~\ref{sec:peaks_vs_islands}). Perhaps most promising is the alternative approach forwarded in the recent study by \cite{feldbrugge2012}
and \cite{feldbrugge2018}. On the basis of a graph theoretical approach to Morse theory they derived path integral expressions for
Betti numbers and additional homology measures, such as persistence diagrams. While it is not trivial to convert these into concise
formulae such as entailed in the GKF, the numerically evaluated approximate expression for 2D Betti numbers turns out to be remarkably accurate. 

From a mathematical point of view, the underlying problem is that while geometric quantities, such as the \LKCs\ and Minkowski functionals, can be expressed as integrals of local functionals, the same is not true for purely topological quantities. However, even the briefest review of the derivation of the GKF in \cite{Adl10}, or any of its simpler variants over the past half century, shows that this localisation is crucial to the calculations. 
The Euler characteristic is the exception that proves the rule here, since, while topological, Gauss's Theorema Egregium expresses it via local
characteristics.

Consequently, a study of the systematics and characteristics of Betti numbers of Gaussian fields cannot be based on insightful and
versatile analytical formulae. Hence, we turn to a numerical study of their properties, assessing these on the basis of the measurements
and statistical processing of Betti numbers inferred from realizations of Gaussian fields. This involves the generation of a statistical sample of
discrete realizations of Gaussian fields in finite computational (cubic) volumes, described in section~\ref{sec:grf}.
Our investigations also involve the use of an efficient and sophisticated numerical machinery to extract the homology characteristics, and in particular
Betti numbers. It is to this computational issue that we turn in the next section. 

\section{Computation}
\label{sec:computation}

The formal definition of Betti numbers relate to a continuous field $f(\Rspace^3) \to \Rspace$. In most practical situations, 
including ours, the field or image is represented on a regular cubic grid. It results in a grid representation of the field by 
arrays of {\it voxels}, the cubic cells centered at the field sample points. In our study, we have generated the Gaussian field 
realizations on a 128$^3$ grid. For exploring the systematics of Betti numbers as a function of the Gaussian field properties, 
we need a large set of 3-D Gaussian field realizations. To facilitate the computation of the topological characteristics of these, 
we have defined a procedure consisting of two complementary algorithms. The first algorithm, detailed in \cite{BEK10}, defines a formal procedure for computing all Betti numbers of a discretely sampled image on a cubic grid. 
While optimal and exact, it is computationally expensive. We use it to infer rigorously correct results. In addition, we 
use it to assure ourselves of the validity of the results obtained by our {\it Gaussian field optimization method} \citep{PPC13}, a 
considerably faster computational procedure which is strictly valid only for Gaussian fields and limits the 
calculation to two Betti numbers. The latter exploits the intrinsic symmetries of Gaussian fields, in conjunction with the unique 
circumstance of knowing the analytical expression for the genus of isodensity surfaces of a Gaussian field \citep{Adl81,bbks,WGM87}. 

Specifically, first we compute the number of isolated islands by counting the number of isolated hot-spots (areas with positive contours as the boundary). Subsequently, we compute the Euler characteristic by evaluating the local curvature and invoking the Gauss-Bonnet theorem to relate it to the Euler characteristic. Finally, we note that the distribution of $\beta_2$ is symmetric to $\beta_0$. Using this, we arrive at the value of the first Betti number by invoking the Euler-Poincar\'{e} formula, which states that the Euler characteristic is the alternating sum of Betti numbers. Since the zeroth and the second Betti number, and the Euler characteristic are known a-priori, finding the first Betti number reduces to a simple exercise of addition (subtraction). 
In contrast, in the current work, we  provide for methods to compute them in a more generic situation from first principles. This will be particularly useful
in scenarios where the fields are not symmetric, and little known about their distribution. 

\subsection{The algorithm}

\subsubsection{Regular grids and triangulation}
The central idea of the algorithm for the homology computation of Betti numbers of a field $\rho$ sampled 
on a regular cubical grid is to construct a triangulation on the sample voxels. The geometric components of a triangulation - vertices, 
edges, faces and tetrahedra - define a simplicial complex whose topological characteristics are equivalent to that of the sampled field. 
For homology calculations on the basis of such simplicial complexes, one has access to a range of efficient algorithms 
\citep[see e.g.][]{Morozov05}.

It is not possible to construct a unique triangulation $K$ from a regular cubical grid of sample voxels. This is because such a 
cubical grid suffers from various degeneracies: the corners are shared by eight voxels, the edges by four and faces by two. The algorithm solves this by slightly perturbing the regularly spaced grid points along the space diagonal. It 
leads to a deformed grid where the corners are shared by four voxels, the edges  by three and the faces by two. This transformation defines 
the elements of the dual triangulation uniquely -- the vertices of this triangulation are defined by voxel centers, the edges defined by 
the centers of the voxels which share a common face, the triangles by the centers of the voxels which share a common edge, and the tetrahedra 
by the centers of the voxels which share a common corner.

\begin{table*}
	\centering
	\begin{tabular}{llrclll} 
		\hline
		\hline
		\ \\
		Number & power & index $n$ & $\#$ grid & $\#$ field & normal. & $k_c $ \\
		& spectrum & & point & realizations & & ($\Mpchk$) \\
		\ \\
		\hline
		\ \\
		1 & LCDM & & 128$^3$ & 100 & $\sigma_8=1.0$ & \\
		\ \\
		2 & power law & -2.0 & 128$^3$ & 100 & $\sigma_8=1.0$ & 0.785 \\
		3 & power law & -1.0 & 128$^3$ & 100 & $\sigma_8=1.0$ & 0.785 \\
		4 & power law & -0.0 & 128$^3$ & 100 & $\sigma_8=1.0$ & 0.785 \\
		5 & power law & 1.0 & 128$^3$ & 100 & $\sigma_8=1.0$ & 0.785 \\
		\ \\
		\hline
		\hline
	\end{tabular}
	\caption{Parameters Gaussian field realization dataset. \\
		The columns specify: (1) class number, (2) name power spectrum, (3) index power spectrum, 
		(4) number of grid-points, (5) number of field realizations, (6) normalization power spectrum and (7) normalization wavenumber $k_c$}
	\label{tab:gaussdata}
\end{table*}

\subsubsection{Piece-wise linear interpolation}
In a second step of the algorithm, the field values at the vertices in the triangulation are used to interpolate the values on the higher 
dimensional simplices, much akin to that used in the DTFE formalism developed by van de Weygaert \& Schaap \citep{schaapwey2000,weyschaap2009,cautun2011}.
This results in a continuous simplicial field - ie. a field defined on the edges, faces and tetrahedra of the 
resulting simplicial complex - that preserves the topology of the original density field. Of crucial importance is the fact that the choice 
of interpolation - linear, or constant - has no effect on topology. In this paper we use a piece wise constant interpolation : 
if $\tau_k (k = 1,\ldots,i)$ are the simplices on the boundary of $\sigma$, then $\rho(\sigma) = \max[\rho(\tau_1), \rho(\tau_2),\ldots,\rho(\tau_i)]$. 
This yields field values on the edges, faces and throughout the tetrahedral volumes of the triangulation $K$.

\subsubsection{(Upper star)-Filtration}
For the topology calculation, we assess the homology of a filtration of superlevel sets of the piecewise linear density field. A filtration 
is a nested sequence of subspaces $S^i$ of the field sample volume $S\subseteq{\mathbb{R}}^3$, such that $i\le j$ implies $S^i \subseteq S^j$. 
This leads to a nested sequence of subspaces
\begin{equation}
\emptyset\,=\,S^0 \subseteq S^1 \subseteq S^2 \subseteq \ldots \subseteq S^m\,=\,S\,.
\end{equation} 
While $m$ can take infinitely many values, we constrain it to a finite number of values by noting that, according to Morse theory, the 
superlevel set does not change topology as long as the density level $\nu$ does not pass a critical point. The critical points of the 
density function are the minima, maxima and saddle points. It suffices to compute homology of any one value of the level set between two 
critical points. This is equally true for a smooth  field as for the simplicial linear piecewise field that approximates it, except that 
one need to slightly adjust the concept of a critical point in the latter case. In the study reported here, the subspaces correspond to 
the regions where the density value is in excess of the corresponding density threshold $\nu_m$. 

The Betti numbers are computed for a range of superlevel sets. Computationally, this is achieved by constructing the \emph{upper star} filtration 
of the simplicial complex. Note that the upper star filtration, defined for piecewise continuous fields, is the discrete version of superlevel 
sets of the corresponding smooth continuous field. Consider a vertex $v_i$, and the simplices $\sigma_k(i = 1,\ldots,p)$ incident to it. The incident 
simplices $\sigma_k$ define the \emph{star} of the vertex. The \emph{upper star} of this vertex $v_i$ consists of all the simplices that have $v_i$ as their lowest vertex
\begin{equation}
\textrm{St}^{+}(v_i) = \{\sigma \in \textrm{St}{\_v_i} | x \in \sigma \Rightarrow \rho(x) \geq \rho(v_i)\}. 
\end{equation}
Computing homology of the superlevel set defined by a particular field value $\nu$ corresponds to 
computing the homology of the union of upper stars of all vertices whose field value is greater than or equal to $\nu$.

\subsubsection{Boundary matrix and its reduction}
To compute the Betti numbers corresponding to a superlevel set, the  algorithm subsequently proceeds by 
constructing the boundary matrix of the union of the upper-star filtration of the vertices whose value is higher or equal to the superlevel 
set value $\nu$. A p-dimensional boundary matrix is a representation where the columns correspond to p-dimensional simplices and the rows 
to $(p-1)$ dimensional simplices. The $(i,j)th$ element of the matrix is $1$ if the $i^{th}$ simplex belongs to the boundary of the $j^{th}$ 
simplex. All other entries are uniformly $0$. The boundary matrix is reduced to its Smith normal form, with a part of the matrix in 
diagonalized form, and the rest of it with empty elements. 

\subsubsection{Betti numbers: rank of the reduced boundary matrix}
The $p^{th}$ Betti number is then given by 
\begin{equation}
 \beta_p = {\rm rank}(\Zspace_p)-{\rm rank}(\Bspace_p)
\end{equation}
where $\Zspace_p$ is the null part of the $p^{th}$ boundary matrix and $\Bspace_p$ is the non-zero diagonalized part of the $(p+1)^{th}$ 
boundary matrix. The Betti numbers for subsequent superlevel sets are computed by incremental addition of simplices in the upper star of 
the newly introduced vertices, and updating the boundary matrix as we lower the density threshold. Finally, a reader inclined to gain a deeper understanding of these concepts, may refer to \cite{PEW16}, where we present the concepts in a greater detail with examples.

\section{Models }
\label{sec:models}

From now on we concentrate on Gaussian random fields over $\mathbb R^3$, since these are the ones important for the cosmological mass distribution. In order to see how different spectra impact on the topological behaviour of these fields, we consider a number of specific spectra common in cosmological modelling. The first class of examples are those with power law spectra.

\begin{figure*}
	\centering
	\includegraphics[width=0.65\textwidth]{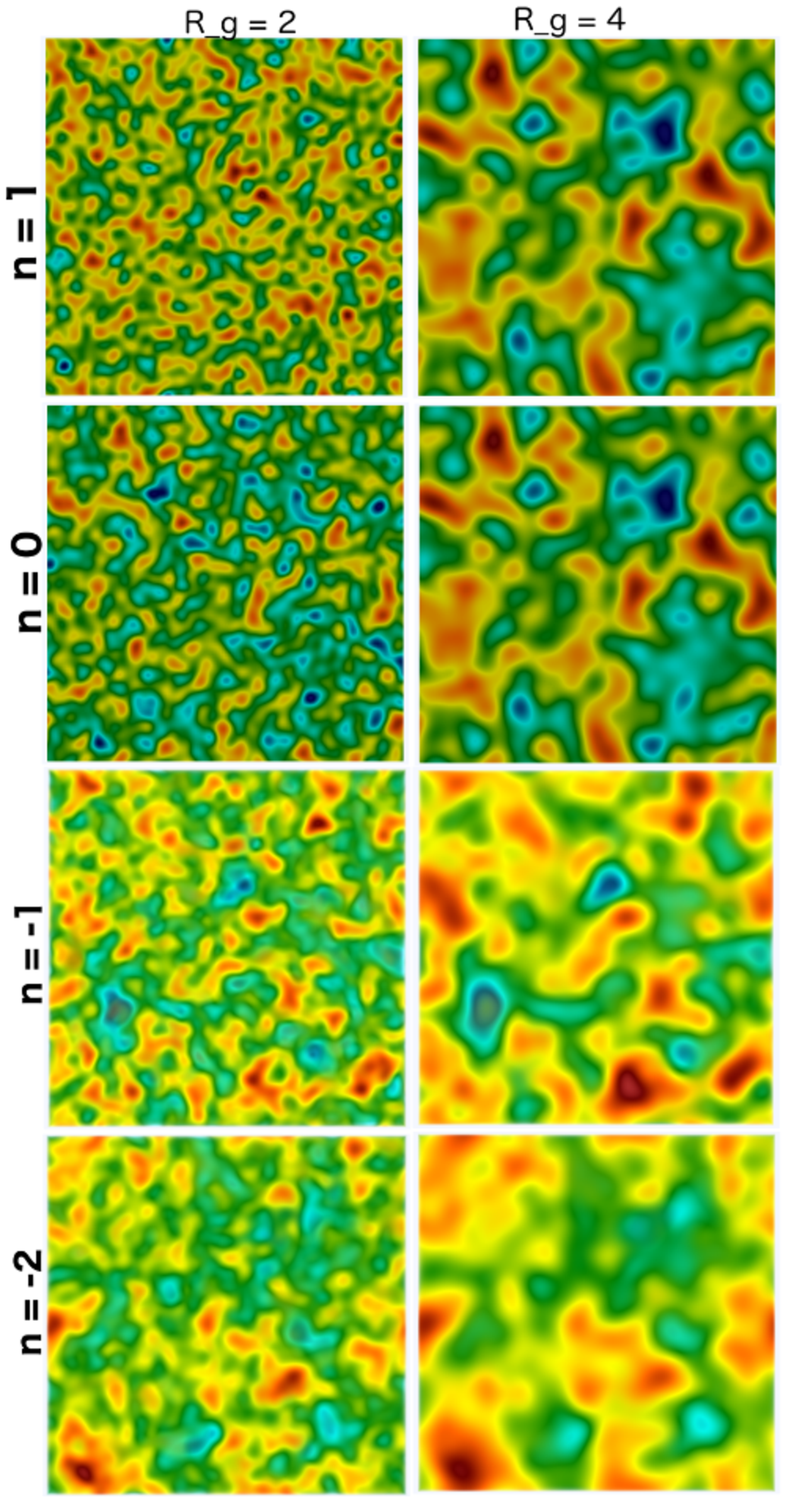}\\
\caption{  2D slices of a single realization of 3D 
	Gaussian random field 
	models investigated in this study. The models are constructed in a 
	simulation box of side $128\Mpch$ with a grid resolution of $1\Mpch$. 
	Subsequently, it is smoothed with a Gaussian kernel of scale 
	$R_f=2\Mpch$. The panels in the left column show 
	realizations of power law power spectra with spectral 
	indices $n=1, 0, -1$, and $-2$, smoothed at $2\Mpch$. As we go from positive to 
	progressively negative spectral indices, the amplitude of large-scale
	flucutations grows, resulting in structures of larger spatial coherence. The panels in the right column show the same realizations smoothed at $4\Mpch$.}
\label{fig:grf_LCDM_plaw_snaps}
\end{figure*}

\begin{figure*}
	\begin{center}
		\rotatebox{-90}{\includegraphics[height=0.9\textwidth]{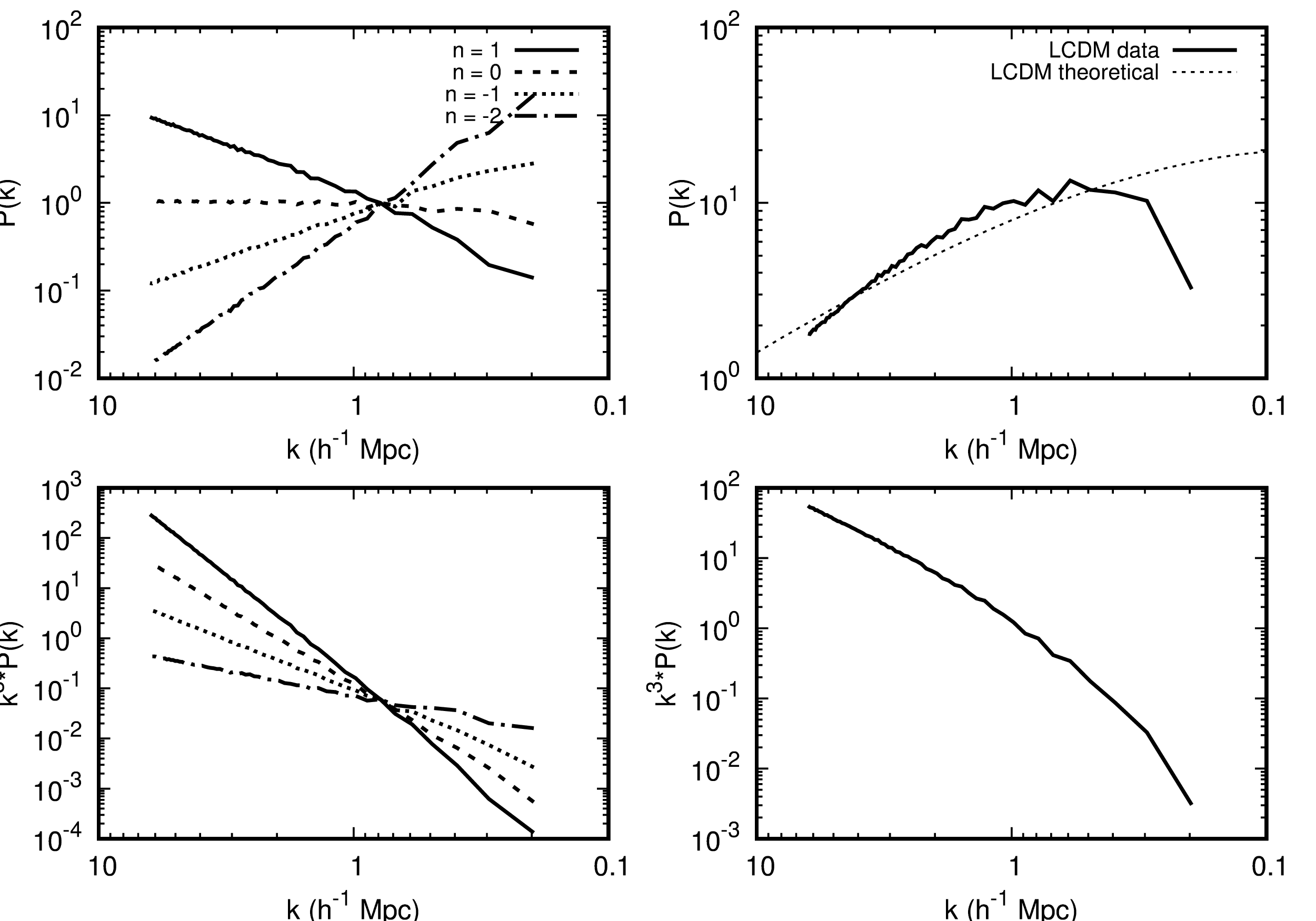}}\\
		\caption{The power spectrum $P(k)$, as well as power spectrum per unit logarithmic bin $k^3\,P(k)$. Graphs are presented for the different spectral indices of the power law model (left column), as well as the LCDM model (right column). The power law
			spectra are scaled such that different models have the same variance of the density fluctuations, when  
			filtered with a top hat filter of radius $8\Mpch$.}
		\label{fig:plaw_spectrum}
	\end{center}
\end{figure*}

\subsection{Power law spectra}

The power law power spectra are  a generic class of spectra, specified 
by the spectral index $n$, and given by
\begin{equation}
P(k) = A_n\,k^n.
\label{eqn:powerlaw_spectrum_ch2}
\end{equation}
We will treat the cases $n = 1, 0, -1$ and $-2$. In case when $n=1$, the Harrison-Zel'dovich spectrum, is the
conventionally expected spectrum for the primordial density perturbations \citep{harrison1970,PeeYu1970,zeldovich1972}. 
The measured spectrum of the primordial perturbations is very close to this,  with $n \sim 0.96$ 
\citep{DKNS09,komatsu2011,planck2016b}. In a cosmological context, we assume that the amplitude of fluctuations
at high frequencies is higher than that at low frequencies, which means that $n>-3$. This implies 
hierarchical evolution of the subsequently evolving mass distribution, with small-scale perturbations growing
faster than the large scale ones. 

To facilitate comparison between the field realizations we have normalized our spectra by equating the spectral
amplitude at one particular scale of $8 \Mpch$, corresponding to a frequency of $k_c \approx 0.785 \Mpchk$. Hence,
all spectra are set such that all power law spectra realizations have 
\begin{equation}
P(k_c)\,=\,A_n k_c^n\,=\,A_0\,=\,1.
\end{equation}

\medskip
For a visual appreciation of the impact of these spectra on the behaviour of the corresponding random field, we
turn to the visualizations in Figure~\ref{fig:grf_LCDM_plaw_snaps}. The Figure displays panels showing Gaussian field realizations in thin 2D slices,
for power-law spectra with power law index ranging from $n=1$ to $n=-2$. Each of the random field realization has been
constructed in a simulation box of side $128\Mpch$ with a grid resolution of $1\Mpch$. The panels in the left column are  smoothed with a
Gaussian kernel of scale $R_f = 2 \Mpch$, and in the right column with a Gaussian kernel of scale $R_f = 4 \Mpch$. The panels clearly show the increasing dominance of small-scale fluctuations for
realizations for higher spectral indices, while the amplitude of large-scale features increases towards progressively negative
spectral indices. As the images nicely illustrate, this results in a growing spatial coherence for fields with a more negative
spectral index. 

The power spectra themselves are shown in Figure~\ref{fig:plaw_spectrum}. Note that the spectra shown are the
ones measured from the field realizations. The top left panel shows a realization of the LCDM power spectrum. The remaining panels show realizations of power
law power spectra. For these models, there is relatively more power at the small scales 
for a higher spectral index, in comparison to a lower spectral index.  
As a result, the field fluctuates rapidly for high spectral indices. As the spectral index decreases, the power shifts 
towards larger scales. This results in a smoother  field with structures at larger scales.

\subsection{LCDM spectrum}
\label{sec:model_lcdm}

The LCDM power spectrum stems from the standard concordance model of cosmology. It fits the 
measured power spectrum of the cosmic microwave background as well as the power 
spectrum measured in the nearby large scale Universe to high accuracy. The shape of the
power spectrum can be inferred by evaluating the evolving processes through the
epoch of recombination, through the Boltzmann equation \citep{CMBfast}. A good numerical
fit is given by \citep{bbks,EH99,HE99}

\begin{align}
&P_{CDM}(k) \propto \nonumber \\ 
&\frac{k^n}{\left[1+3.89q+(16.1q)^2+(5.46q)^3+(6.71q)^4\right]^{1/2}}
\times\frac{\left[\ln(1+2.34q)\right]^2}{(2.34q)^2}, \nonumber \\
\end{align}
\noindent where
\begin{equation}
q = k/\Gamma,\,\quad\quad\,\Gamma = \Omega_m h \exp\left\{-\Omega_b-\frac{\Omega_b}{\Omega_m}\right\},
\end{equation}
$\Omega_m$ and $\Omega_b$ are the total matter density 
and baryonic matter density respectively and $\Gamma$ is referred to as the shape 
parameter. We have used the value $\Gamma \sim 0.21$, which forms a reasonable approximation
for the currently best estimates for $\Omega_b$ and $\Omega_m$ as obtained from
the Planck CMB observations \citep{planck2016}. 
In our study, the power spectrum of the LCDM Gaussian field realizations is normalized by
means of $\sigma_8=1.0$. 

Locally, the spectrum resembles a power law, with  spectral index $n_{eff}(k)$,  showing a dependence on the
scale $k$, through the relation
\begin{equation}
n_{eff}(k) = \frac{\text{d\,ln} P(k)}{\text{d ln}k}.
\end{equation}
\noindent In the asymptotic limit of  
small and large $k$, the limits of $n_{eff}(k)$ are well defined. At very large scales, 
its behavior tends towards a power law with index $n = 1$, as can be 
seen in the plot. At small scales, the LCDM power spectrum behaves like a 
power law power spectrum with index $n = -3$. The effective index 
of the model varies steeply between $n_{eff} \sim -0.5$ to $n_{eff} \sim -2.5$ for our models. At the lower limit, the Nyquist mode of the
box corresponds to the scale of  galaxies of the size of the Milky Way. At the other end the fundamental mode of the box corresponds to
wavelengths well beyond the scales at which the Universe appears homogeneous.

\subsection{Model realizations and Data sets}
\label{sec:model_real}
\bigskip
The samples of Gaussian field realizations are generated in a cubic volume on a finite grid, with periodic boundaries, achieved by identifying and gluing opposite sides, transforming a finite $\Rspace^3$ domain to $\Tspace^3$. It concerns
field realizations on a grid with $N=128^3$ grid-points. The fields are generated by our (constrained) initial conditions
code \citep{weyedb1996}. It involves the generation of $128^3$ independent Gaussian distributed Fourier field
components ${\hat f}(\vec k_i)$, and the subsequent inverse FFT transform to yield the corresponding
density field. The FFT automatically assures a cubic volume field realization with periodic boundary conditions.
Table~\ref{tab:gaussdata} lists the relevant parameters of the sample of Gaussian field realizations used in our study. 

In effect, the field realizations have the specified power spectrum amplitude (Equation~\ref{eqn:powerlaw_spectrum_ch2})
between the fundamental mode and Nyquist mode of the grid, while they are zero for lower and higher frequencies.
Effectively, the realized spectrum is therefore a block spectrum. For the power law spectra this circumvents the
divergences that beset pure power law spectra. The cubic sample volumes have a side of $128\Mpch$ with a grid resolution
of $1\Mpch$, corresponding to a fundamental mode of $k_{fund} =2\pi/128 \Mpchk \approx 0.049 \Mpchk$ and a Nyquist frequency
$k_{Nyq} = 2\pi /2 \Mpchk \approx 3.14 \Mpchk$. 
\bigskip

\bigskip

The statistical results that we obtain in our numerical study of homology and Betti numbers are based on
100 different field realizations for each tested power spectrum. For each realization we evaluate
Betti numbers, Euler characteristic and Minkowski functionals. Subsequently, we average over these 100
realizations. It is these averages which form the dataset which we will subsequently analyse in
Sections~\ref{sec:betti_and_genus_result} until \ref{sec:lcdm_result}. 

\bigskip

\section{Betti number analysis: \\ \ \ \ \ \ \ 3D Gaussian random fields}
\label{sec:betti_and_genus_result}

In this section, we analyse the topological characteristics of the models in terms of the 
Betti numbers on the basis of our numerical study of the homology of our sample of
Gaussian field realizations. The discussion is based on the statistical evaluation of
these results. 

The intention of the analysis is an evaluation of the generic properties of
Betti numbers as a function of field power spectrum, and to compare the properties
of Betti numbers with the topological behaviour in terms of the Euler characteristic. 
To this end, the three Betti numbers $\beta_0$, $\beta_1$ and $\beta_2$ are computed
for superlevel sets of the (filtered) Gaussian fields, defined by dimensionless
density threshold $\nu = f / \sigma$. The variation of the Betti numbers as a function
of the threshold $\nu$ forms the principal resource for our investigation of
the topological properties of Gaussian random fields. An important thing to note is that we perform our analyses on periodic cubes, or equivalently, the manifold is $\Tspace^3$, which is without boundary. An important consequence is that the Euler characteristic curve (and hence the Betti numbers also), are symmetrized due to the absence of boundary terms; see Section~\ref{rob:sec:CandA} for an explanation on how the boundary terms affect the Euler characteristic computation.

In an earlier article, we presented a brief investigation of Betti numbers of Gaussian 
fields, focusing on their important features with respect to a comparison with genus
statistics \citep{PPC13}.   

\begin{figure*}
\centering   
  {\includegraphics[width=\textwidth]{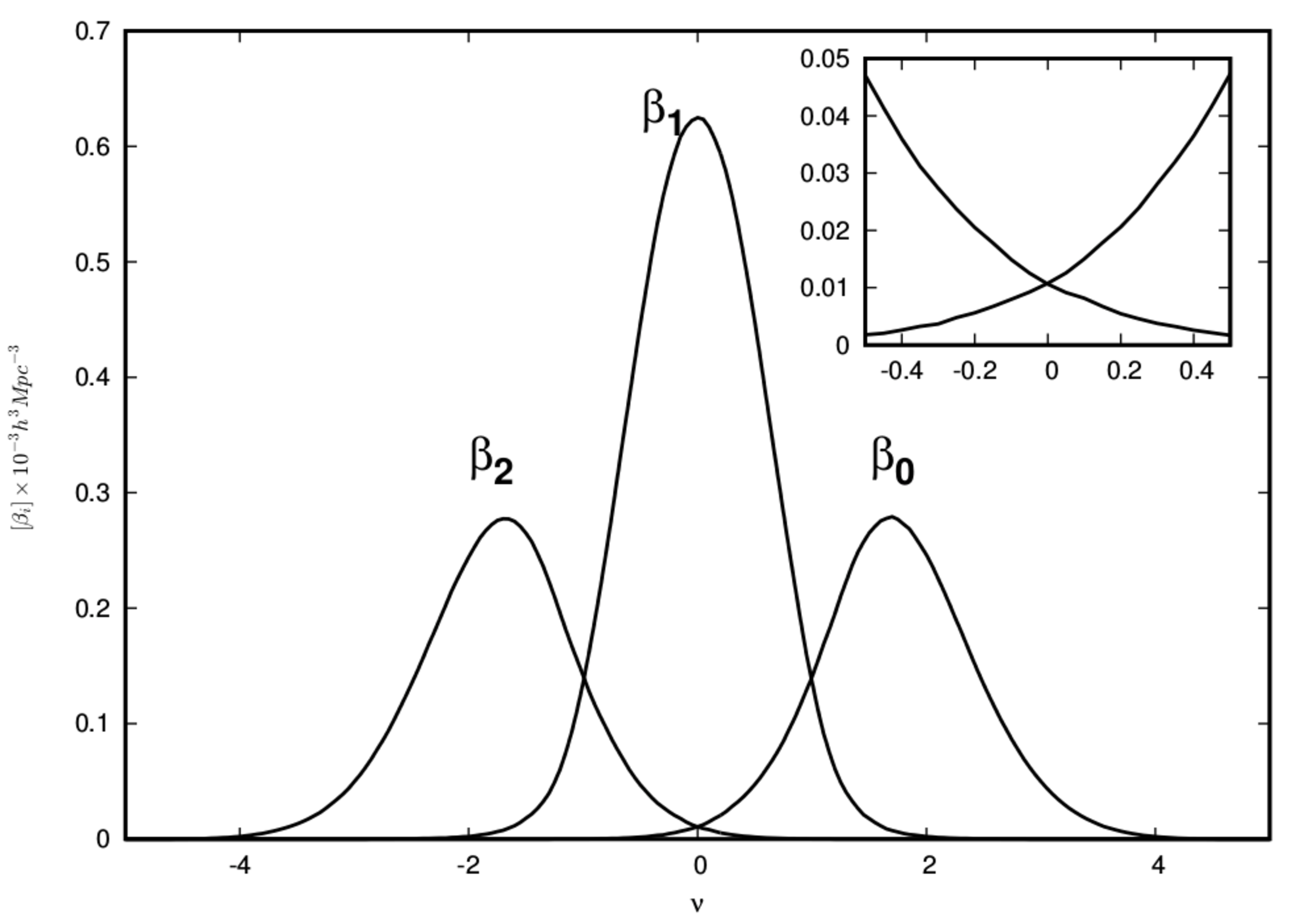}}\\
  \caption{The Betti numbers for the $n = -1$ model. It is evident that the different Betti numbers dominate the topology in the different density thresholds regions, and also the Euler characteristic curve. For high thresholds, $\beta_0$ is the dominant topological feature. For intermediate thresholds, $\beta_1$ dominates, while for low thresholds $\beta_2$ is the dominant topological entity. Inset: The zone of overlap between $\beta_0$ and $\beta_1$ curves, at the median density threshold. Both the quantities are non-unity, indicating that the manifold is not a single connected surface, and hence does not exhibit pure \emph{Sponge-like} topology.}
  \label{fig:betti_genus_PLAW_single}
 \end{figure*}
 
 \begin{figure*}
 \centering   
   {\includegraphics[width=\textwidth]{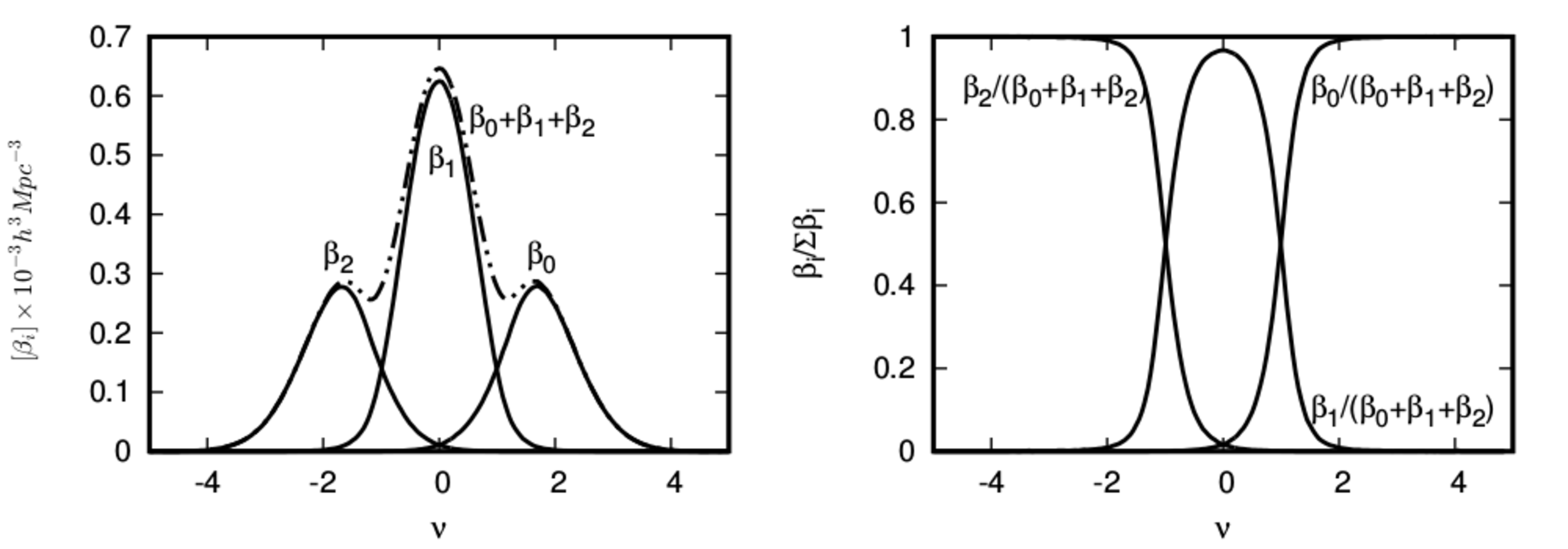}}\\
   \caption{Figure illustrating the dominance of different Betti numbers in the different density threshold regions. The left panel plots the sum of Betti numbers for the different thresholds, along side the individual Betti numbers. The right panel presents the ratio of the individual Betti numbers to their sum, as a function of the density threshold.}
   \label{fig:betti_plaw_sum}
  \end{figure*}
 
\subsection{Gaussian Betti characteristics: general properties}
\label{sec:betti_and_genus_signle}

Figure~\ref{fig:betti_genus_PLAW_single} presents the Betti number curves for a typical realization of a Gaussian field. For Gaussian fields, $\beta_0$ and $\beta_2$ appear to mirror each other about $\nu=0$.  The number of independent tunnels, in terms of $\beta_1$, appears symmetric to itself under reflection about $\nu = 0$.
The symmetries observed in the Betti number curves are a reflection of the underlying symmetry in the field itself. Because of their symmetry, an analysis with respect to the islands  
is also indicative of the properties of voids. 

At $\nu = \sim \pm\sqrt{3}$, the number of isolated islands and voids attain their maximum. At $\nu = 0$,
the number of isolated islands equals the  number of isolated voids, i.e. $\beta_0=\beta_2$. It is an interesting
observation that these numbers are \emph{not} equal to unity, as should have been the case for a pure 
\emph{Sponge-like} topology. This is evident from the inset where we zoom into the median density threshold region where $\beta_0$ and $\beta_2$ overlap, resulting in small but non-unity numbers for both islands and voids. At the same threshold, we see that the number of tunnels/loops reaches a maximum. It signifies
a morphology in which several large interconnected overdense island and void regions are interspersed with a complex 
anatomy of percolating tunnels, a result of the complex mutual intertwining of these manifolds. 

To appreciate the topological characterisation, the Betti numbers clearly provide crucial new insights.
By evaluating $\beta_0$ and $\beta_2$ at a given density level, we may assess what the contributions of the
overdense islands and underdense cavities are to the the genus characterization, and in how far tunnels
through, and in these features contribute to the definition of a complex topology. The symmetric and relatively
simple nature of Gaussian fields is helpful in identifying the connection between Betti numbers, genus and overall topological character.

One of the principal topological characteristics of Gaussian fields is their relative simplicity, in that the three
Betti numbers dominate the topology at different density ranges. In this sense, Gaussian field are rather unique.
In the case of most complex patterns encountered in nature, the density range over which all three Betti numbers have significant
non-zero values is far larger than that in Gaussian fields. In this sense, we may consider Gaussian fields to have a comparatively
simple topological structure. Both Figure~\ref{fig:betti_genus_PLAW_single} and Figure~\ref{fig:betti_plaw_sum} reveal this circumstance.

The left-hand panel of Figure~\ref{fig:betti_plaw_sum} plots the sum of the Betti numbers in dot-dashed curves, alongside the different Betti numbers
themselves (solid curves). It is clear that for $|\nu| \gtrsim 2\sigma$, $\beta_0$ and $\beta_2$ dominate the topology for the positive and the negative
thresholds respectively. In between $|\nu| \lesssim 1\sigma$, $\beta_1$ is the dominating component. Similar information can also be gleaned from the
right-hand panel, where we present the fractional Betti numbers, denoting the ratio of a particular Betti number
to the sum of all Betti numbers as a function of the density threshold. It is evident that the three Betti numbers dominate different density
regimes, albeit with a substantial range of overlap for $|\nu| \lesssim 2\sigma$.

Characteristic therefore for Gaussian fields is that the topology at extreme density values is dominated by a single class of features, cavities or islands.
At very high density levels the topology is entirely dominated by the islands and thus fully specified by $\beta_0$. The topology is predominantly 
\emph{Meatball-like}, marked by the presence of isolated components (or islands). The same is true at very low density levels,
where the topology is entirely specified by $\beta_2$ and thus exclusively dominated by cavities, ie. the central region of voids.
Here the topology is distinctly \emph{Swiss-cheeselike}. 

At more moderate levels, for $|\nu| \lesssim 2$, the topology attains an increasingly \emph{Sponge-like} character. In these regimes at least two Betti numbers are
needed to describe the topology of the superlevel manifolds. On the lower density side, the topology is dominated by $\beta_2$ and $\beta_1$. It reflects
a pattern of isolated cavities indicating agglomerates of density troughs, interspersed by tunnels and loops. On the higher density side, the topological
signal consists mostly of $\beta_0$ and $\beta_1$. The corresponding spatial pattern is that of an  agglomerate of isolated islands, infused and punctured by numerous tunnels.
In a relatively narrow density range around the mean density, for $|\nu| \lesssim 0.1-0.2$, we even observe the simultaneous existence of all three
topological features, cavities, islands, and tunnels. In that regime, all three Betti numbers are needed to quantify the Gaussian field topology.

An interesting aspect of the Gaussian field topology concerns the role of tunnels.  Starting at high density values, and proceeding towards lower
density values, the number of disconnected islands reaches a maximum at
threshold $\nu=\sqrt{3}$. At a slightly higher threshold, $\nu \approx 2$, we start to see the rapid increase of the $\beta_1$ curve. This is the
result of the formation of ever larger island complexes by the merger of higher density objects, along with the
emergence of tunnels that permeate their interior and surface. At intermediate thresholds, $|\nu| \lesssim 1$, tunnels become the most
populous topological feature. They gradually attain their maximum presence at the median density threshold, as nearly all high-density components have merged
into one huge percolating and irregularly shaped complex. The surface and interior of the complex is marked by the presence of a large number
of these permeating tunnels. To follow the entire process in detail, we refer to our description and discussion of
the corresponding persistence diagrams in the upcoming accompanying paper \citep{pranavb2018}.

A similar role of tunnels and loops is seen on the low density side of the Gaussian field. Starting at the very low densities, 
from the $\beta_2$ curve we note the dominant presence of under-dense troughs and enclosed cavities. The number of independent cavities
reaches a maximum at threshold $\nu=-\sqrt{3}$. Proceeding towards higher density values, an increasing number of cavities that were isolated, start to connect and merge, forming
ever larger ``oceans''. By density level $\nu=-2$, the merged ocean complexes are accompanied by a strongly growing presence of loops, signifying a complex topology.
The steep rise of the $\beta_1$ curve reflects this quantitatively. Approaching median density
levels, nearly all cavities have been absorbed in one large ocean, whose irregular shape and surface is reflected in the $\beta_1$ curve reaching
its maximum value. 

It is also interesting to assess the topological identity at the median density level, $\nu=0$. At that level, we see the presence
of an equal number of islands and cavities. That is, $\beta_0$ and $\beta_2$ are equal at $\nu=0$. Conventionally in the literature, for Gaussian fields, it is assumed
that all over-dense regions have merged into one percolating complex at the median density threshold, interlocked with one equivalent under-dense ocean.  
This would define a pure \emph{Sponge-like} topology. This has been assumed on the basis of the analyses of genus curves, e.g. \cite{GDM86,GMTS89}. However, it remains to be seen whether indeed such an ideal sponge-like topology exists, even
in the case of Gaussian random fields. Below we will find that, in general, this
it not true, with the topology at median levels determined by a few - disconnected - over-dense complexes, intertwined with a few
under-dense ones. The dissection of the genus curve into the contributing Betti curves reveals this phenomenon, shedding new insights on to this issue.

\subsection{Betti number characteristics: \\ \ \ \ \ \ \ \ \ \ dependence on power spectrum}
\label{sec:betti_plaw}


\begin{figure*}
	\centering   
	{\includegraphics[width=0.75\textwidth]{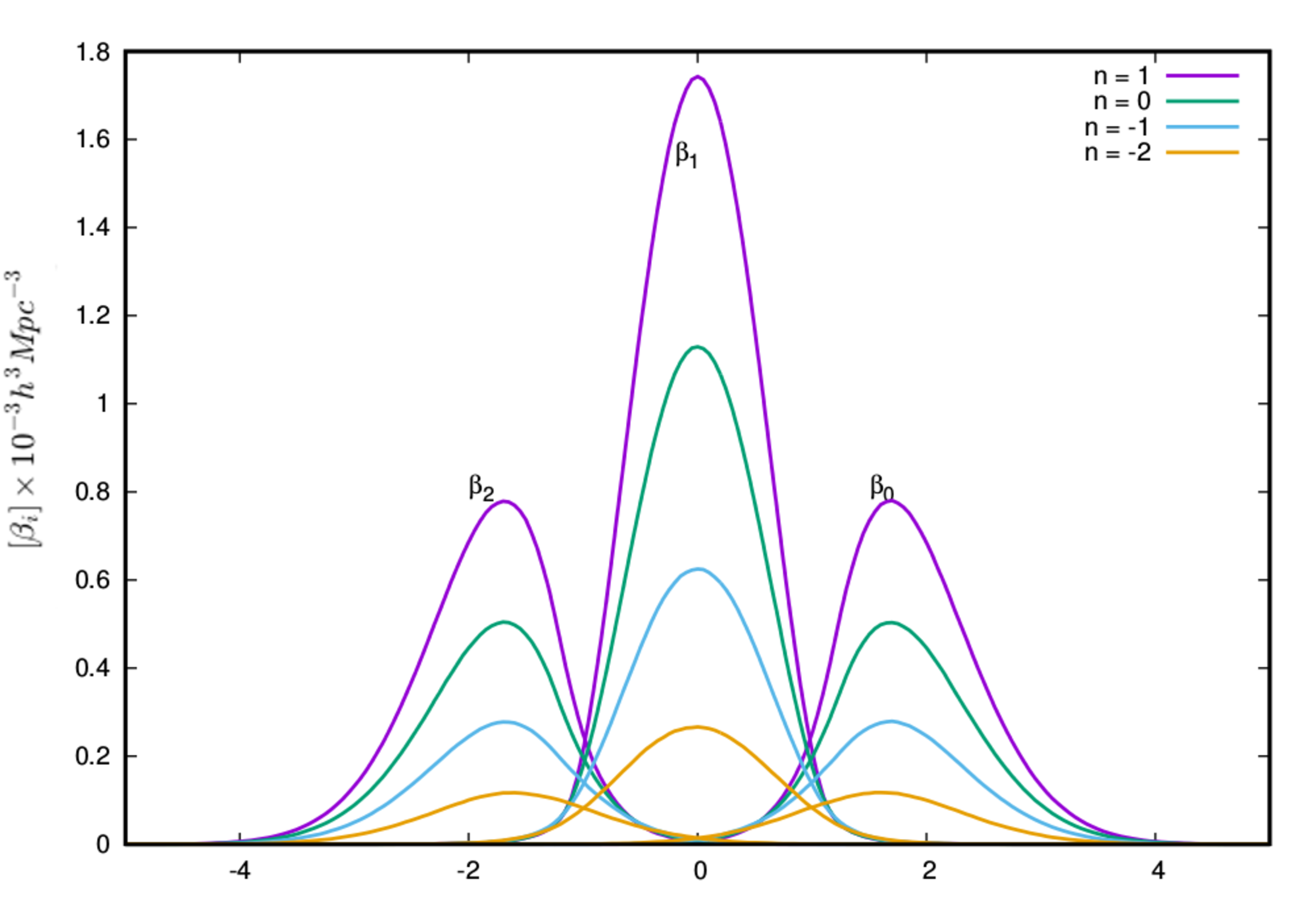}}\\
	{\includegraphics[width=0.45\textwidth]{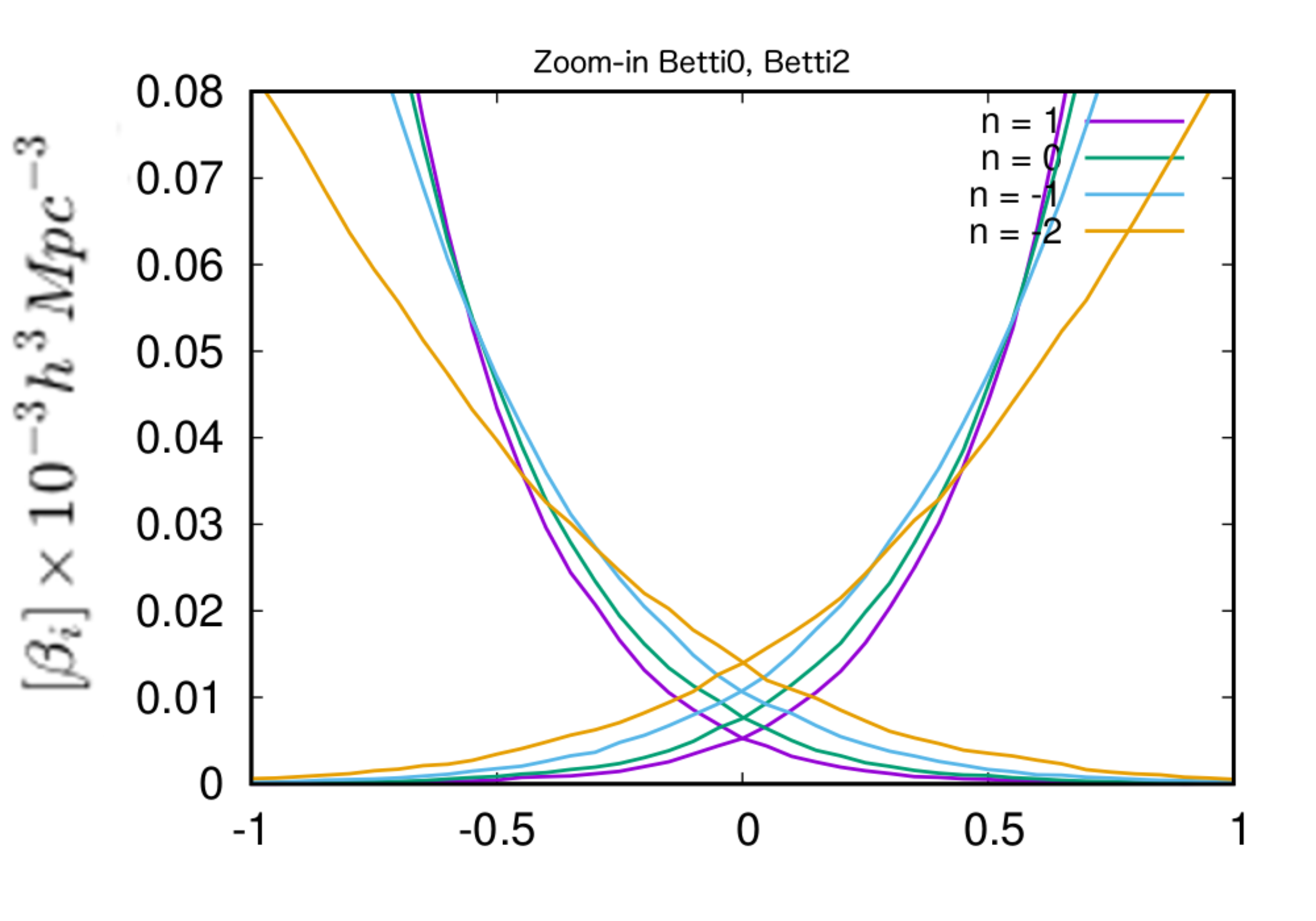}}
	{\includegraphics[width=0.45\textwidth]{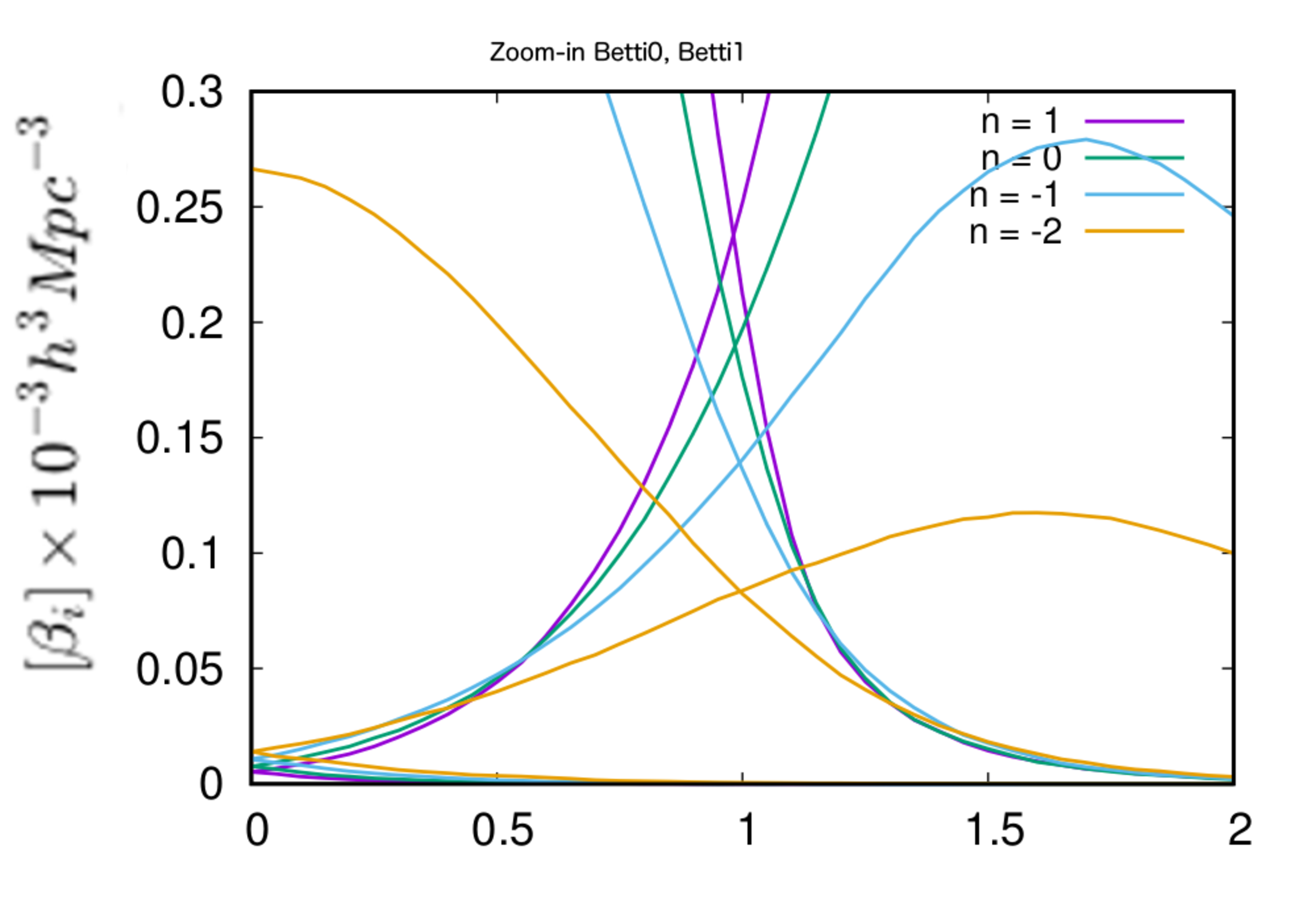}}
	\caption{Top: Unscaled Betti number curves of Gaussian random fields for the power-law models. The curves are drawn for spectral index $n = 1, 0, -1$ and $-2$, as a function of the dimensionless density threshold $\nu$. $\beta_0$ and $beta_2$ curves are symmetric to each other with respect to the median density threshold, but exhibit visible skewness, increasing with increasing spectral index. $\beta_1$ curve is symmetric to itself with respect to $\nu = 0$, and exhibits no skewness. Bottom: Zoom-in into the region around median density threshold of the top panel shows the small but non-zero contributions from $\beta_0$ and $\beta_1$, more for lower spectral indices. The manifold is not strictly \emph{Sponge-like} in general, but consists of multiple isolated objects as well as enclosed cavities, as opposed to the expectation of a single percolating overdense region intertwined with a single percolating underdense  network of tunnels, for all the models examined.}
	\label{fig:betti_unscaled}
\end{figure*}

\begin{figure*}
	\centering   
	\vspace{-0.5truecm}
	\rotatebox{-90}{\includegraphics[height=0.7\textwidth]{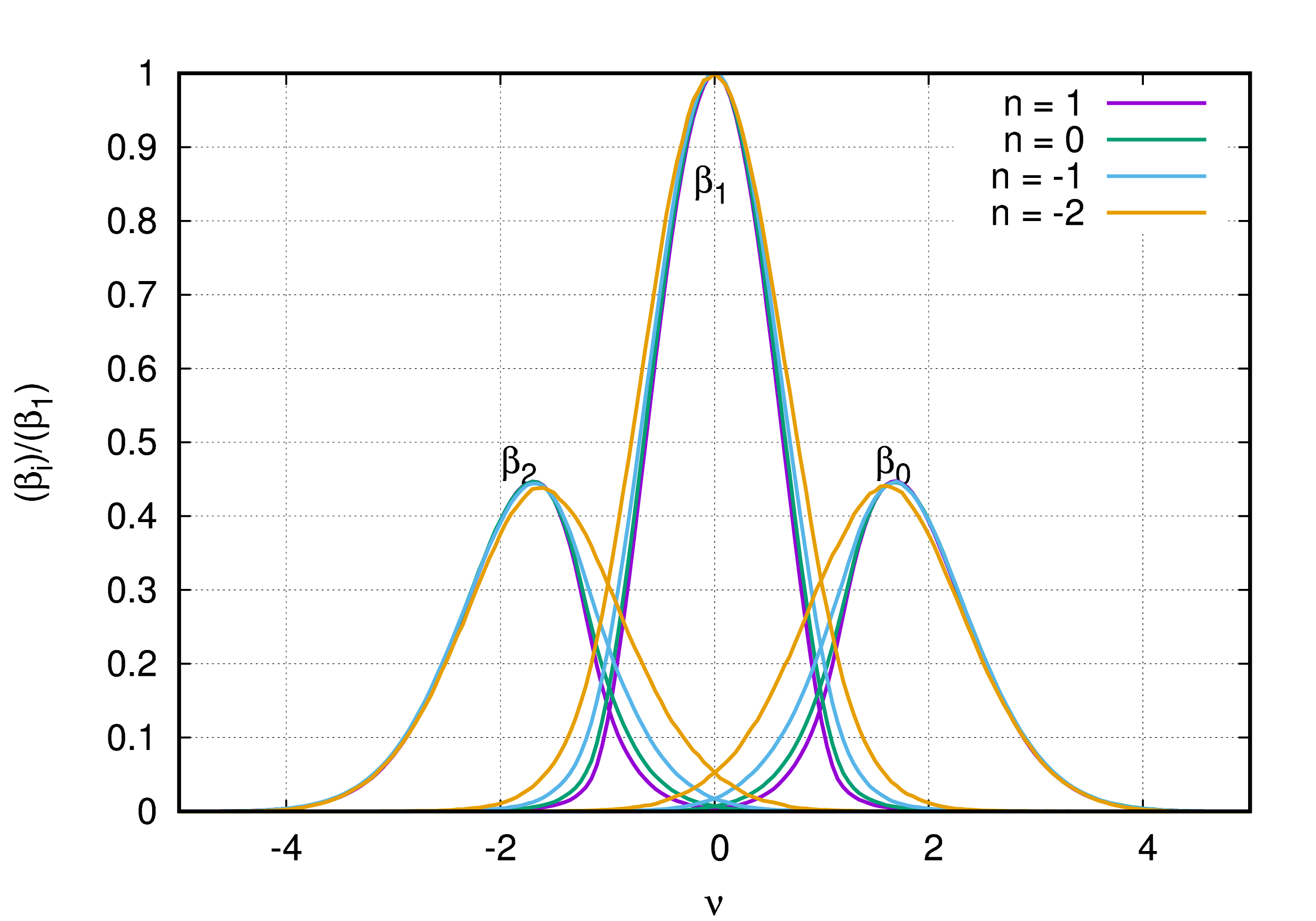}}\\
	{\includegraphics[width=0.7\textwidth]{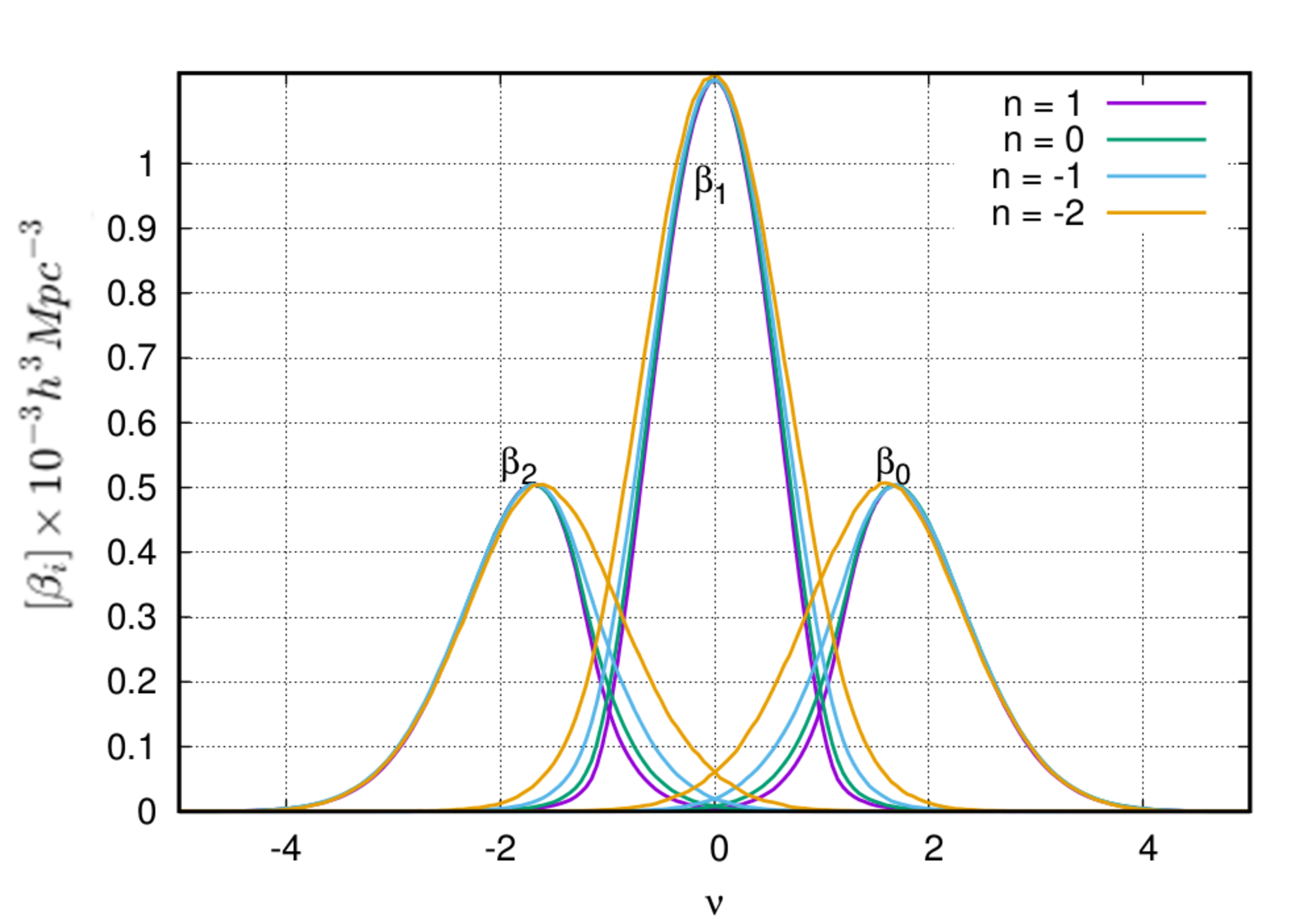}}\\
	\caption{Scaled Betti number curves of Gaussian random fields. The top panel plots Betti numbers, where each $\beta_i$ of a model is normalized by the $\beta_1$ corresponding to that model (also cf. Park et al. 2013). The bottom panel plots the Betti numbers, where  each $\beta_i$ is normalized by the amplitude of
          $\beta_i$ of the $n = 0$ model. }
	\label{fig:betti_scaled}
\end{figure*}

\begin{figure*}
	\centering   
	{\includegraphics[width=\textwidth]{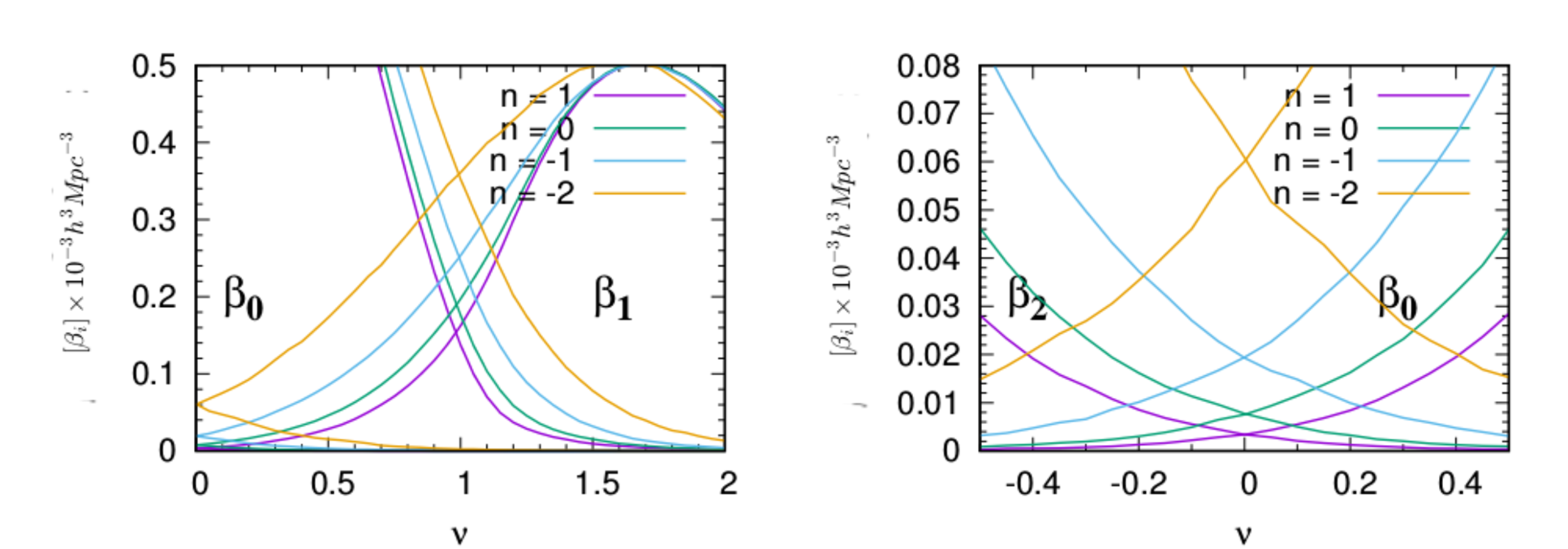}}\\
	\rotatebox{-90}{\includegraphics[height=\textwidth]{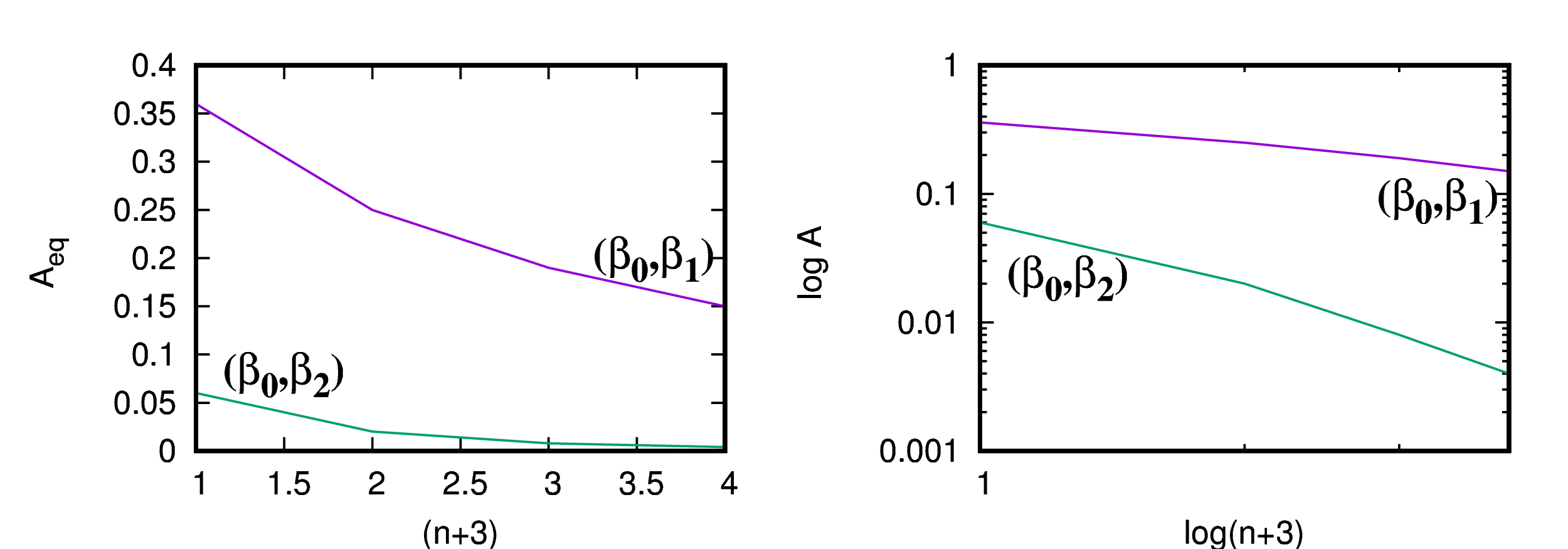}}\\
	\caption{The top-left and the top-right panels present the enlargements of the density range where scaled Betti number curves of Gaussian random fields overlap. The curves are drawn for the  power law models as a function of the dimensionless density threshold $\nu$. The bottom-left and bottom-right panels plot the value of $\nu$ at which the overlap occurs as a function of ($n+3$), where n is the spectral index, in absolute and log units respectively.}
	\label{fig:betti_zoomin}
\end{figure*}

\begin{figure*}
	\centering   
	\rotatebox{-90}{\includegraphics[height=\textwidth]{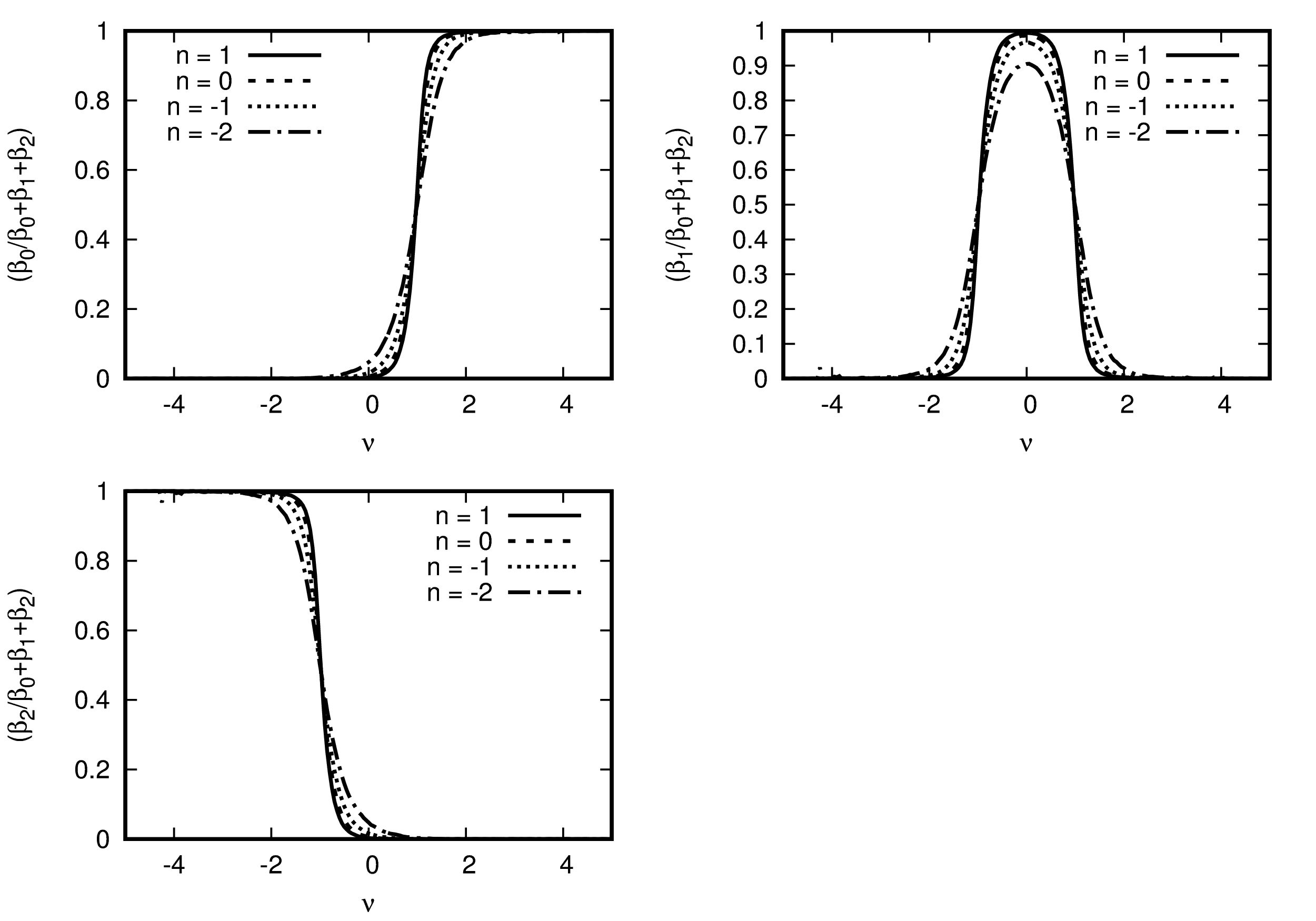}}\\
	\caption{ The fractional contribution of islands, tunnels and voids to the 
		sum of Betti numbers in all three dimensions for rms threshold range 
		between ($-5\sigma:5\sigma$) for the different powerlaw models. Top-left: $\beta_0$/($\beta_0+\beta_1+\beta_2$) 
		-- fractional contribution of islands to the total sum of Betti 
		numbers. Top-right: $\beta_1$/($\beta_0+\beta_1+\beta_2$) -- 
		fractional contribution of tunnels to the total sum of Betti 
		numbers. Bottom-left: $\beta_2$/($\beta_0+\beta_1+\beta_2$) -- 
		fractional contribution of voids to the total sum of Betti 
		numbers.}
	\label{fig:fractional_betti}
\end{figure*}


\begin{figure*}
	\centering
	{\includegraphics[width=0.75\textwidth]{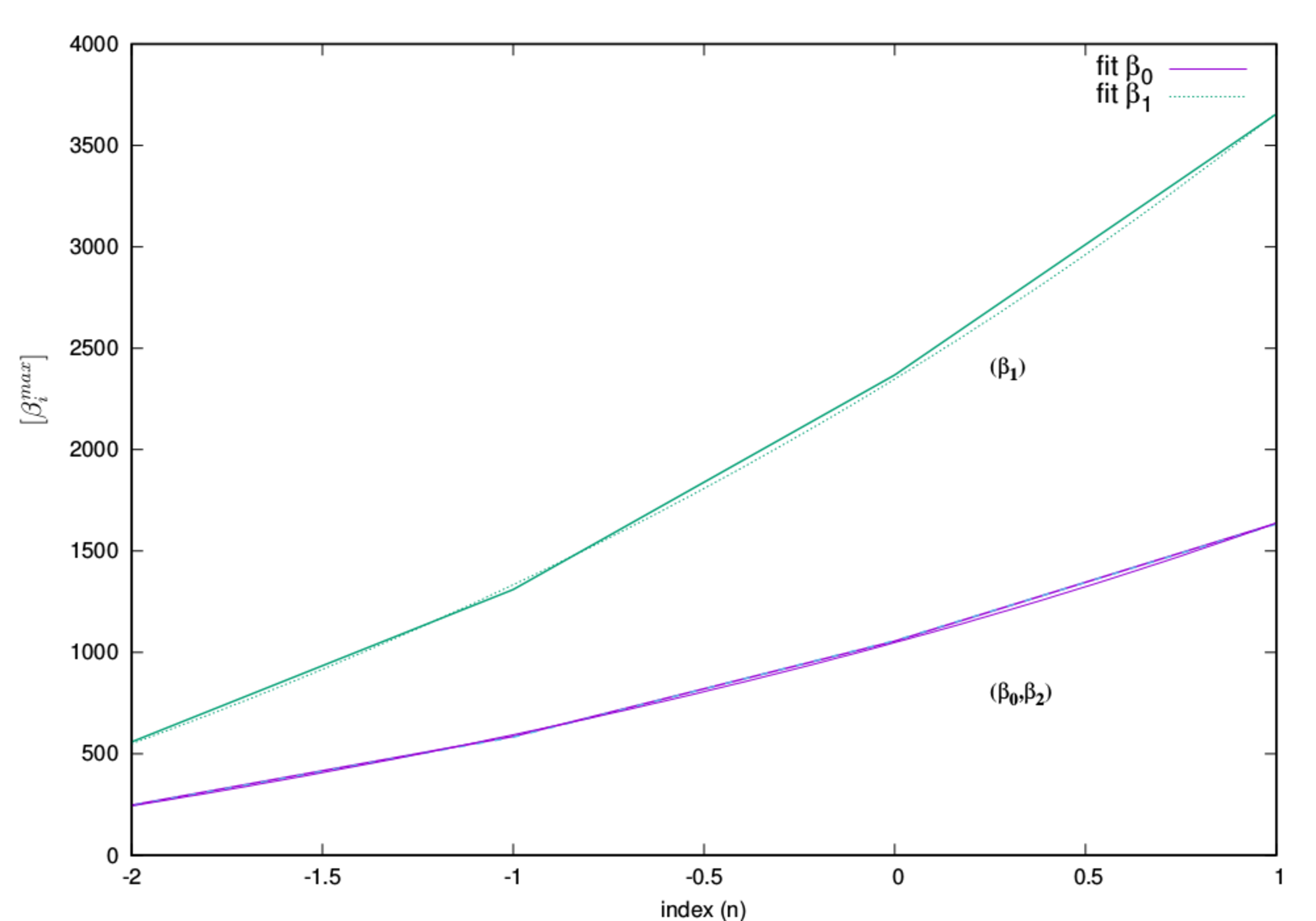}}\\
	\caption{Amplitude of Betti numbers as a function of the spectral 
	  index (also cf. Park et al. 2013). The curves are fit to an exponential of the form
          $A_0\cdot \exp{n/\tau}$, with $\tau = 2 \pm 0.19$.}
	\label{fig:betti_amp_vs_index_scale}
\end{figure*}

Having established the generic behaviour of the Betti curves of Gaussian fields, we investigate their
systematic trends and dependence on the power spectrum of the field realizations. One of the findings
that we earlier reported in \cite{PPC13} is that Betti numbers depend on the shape of the power spectrum. 
The top panel of Figure~\ref{fig:betti_unscaled} shows the unscaled Betti number curves for the various power law models.
The bottom-left and the bottom-right panels plot enlargements of the regions of overlap between ($\beta_0$, $\beta_2$) and
($\beta_0$, $\beta_1$) respectively.

The first direct observation is that there is a steep increase of all three Betti curves, over the
entire density range, as the power spectrum index $n$ increases. That is, the number of topological
features - islands, cavities and tunnels - is steeply increasing as the small-scale fluctuations in
the density field are more prominent and have a higher amplitude. This is in line with what would
be expected for Gaussian fields. 

For all power spectra, we find that the $\beta_0$ curve reaches its maximum at the characteristic
density threshold $\nu=\sqrt{3}$, while the $\beta_2$ reaches its maximum at $\nu=-\sqrt{3}$,
and $\beta_1$ at the mean density level $\nu=0$. Also, for all power spectra we find that at $\nu=0$,
the number of over-dense islands is the same as under-dense cavities, i.e. $\beta_0=\beta_2$.
Furthermore, there is an overlap of the different Betti numbers in determining the topological identity of the manifold, at different density thresholds. The bottom two panels of Figure~\ref{fig:betti_unscaled} substantiate this claim. The number of isolated islands is equal to the number of isolated voids at $\nu = 0$, and the number of isolated islands (voids) is equal to the number of isolated tunnels at $\nu = 1     (-1)$.  This symmetry is related to the fact that the simulations are realized on the 3-torus, implying that the manifold is without boundary. 

The corresponding scaled Betti curves are shown in Figure~\ref{fig:betti_scaled}. The top panel plots Betti numbers, where each $\beta_i$ of a model is normalized by the $\beta_1$ corresponding to that model. The bottom panel plots the Betti numbers, where  each $\beta_i$ is normalized by the amplitude of $\beta_i$ of $n = 0$ model. Interestingly, both the Figures present identical shapes for the curves, even though differing in the normalization procedure (see the vertical axes for values and units). The amplitudes of $\beta_0$ ($\beta_2$) curves compared to that of $\beta_1$ are also different for the different normalization procedure. The top-left and the top-right panels of Figure~\ref{fig:betti_zoomin} presents the enlargements of the relevant regions of overlap between the different scaled Betti number curves. 

\subsubsection{Overlap of Betti curves: topological identity}
\label{sec:betti_overlap}

The scaled Betti number curves provide supplementary information on systematic trends with respect to 
the relative importance and prevalence of the various topological features. It allows us to investigate in
how far the observed changes in Betti number curves affect the range over which we can speak of \emph{Meatball-like} and
\emph{Swiss-cheeselike} topology. It also allows us to assess in how far the \emph{Sponge-like} appearance at median
density range is affected.

All models retain the exclusive \emph{Meatball-like} topology at high density thresholds $\nu \gg \sqrt{3}$, i.e. an almost
exclusive presence of isolated islands, and a similar exclusive dominance by cavities for $\nu \ll -\sqrt{3}$,
outlining a typical \emph{Swiss-cheeselike} topology. At intermediate density range we notice a few interesting
trends as the spectral index $n$ decreases and large scale fluctuations attain a larger prominence in the Gaussian
fields. We see a systematically growing overlap between the various Betti curves.

The $\beta_0$ and $\beta_2$ curves become less skewed and hence more symmetric as the value of $n$ decreases. It means that the
number of these features at $|\nu|<\sqrt{3}$ is more comparable to that at the higher density levels.
Instead of a steep falloff of $\beta_0$ and $\beta_2$ for $|\nu| < \sqrt{3}$ due to the rapid merging
of objects and cavities into a single percolating island and ocean, a relatively large number of them
remain intact over a wider density range. It implies that the presence of stronger large scale
perturbations goes along with a - relatively - higher number of disconnected objects and cavities at medium
density values and that these characteristically large-scale objects and cavities remain independent down to
lower density levels. It is most likely a manifestation of the lower level of clustering in Gaussian fields with a
lower spectral index $n$.

Also interesting is the fact that as the index $n$ decreases, we start to see an increasing range of overlap
between $\beta_0$ and $\beta_2$. In other words, at $|\nu|<\sqrt{3}$ we not only see a relatively larger number
of islands or cavities, there is also the presence - in absolute numbers - of an increasing number of these
that even remain below the mean density value $\nu=0$. In terms of $\beta_0$, from Figure~\ref{fig:betti_scaled}, and zoom-ins in the concerned overlapping regions in Figure~\ref{fig:betti_zoomin},
we see there are isolated islands living in large under-dense regions. The presence of a progressively larger
number of isolated islands at density levels below the mean can also be inferred from the fractional Betti number
plots in Figure~\ref{fig:fractional_betti}. For the $n=-2$ model, we find a non-zero $\beta_0$ for thresholds as low as
$-2\sigma$. In a sense, it is reminiscent of the 
\emph{cloud-in-void} process identified by \cite{shethwey2004} in their description of the formation of
voids in the cosmic mass distribution. One aspect of this is existence of overdense isolated halos (islands)
in an overall underdense void region, which would emanate from precisely the primordial configuration
identified here by the $\beta_0$ Betti number curve being non-zero at negative density values $\nu$. The opposite
process, \emph{void-in-cloud}, is reflected in the increasing presence of $\beta_2$ at positive density thresholds
while the spectral index $n$ is lower. Such cavities still existing at positive densities $\nu$ may be compared to lakes in a mountain range. 

The presence of tunnels also changes as a function of the power spectrum. Relatively speaking, for decreasing spectral
index $n$ there is a lower number of tunnels per object at the lower density levels $|\nu|<1$ (Figure~\ref{fig:fractional_betti}).
As tunnels are often forming along with the merging of islands, the lower number of tunnels may be a consequence of the
relatively large number of islands - and cavities - that did not yet connect up at these density levels. Interestingly,
at higher density levels of $|\nu|>1$, there appear to be more tunnels per island or cavity for lower spectral indices $n$.
At lower density levels, a sizeable number of these tunnels appear to have filled up and disappeared. The bottom-left and the bottom-right panels of Figure~\ref{fig:betti_zoomin} plot the value $A_{eq}$ at which the overlap occurs, and the numbers for the concerned topological identities are equal, as a function of ($n+3$), where n is the spectral index, in absolute and log units respectively.

\bigskip
Finally, we may answer the question in how far we can still describe the topology around the mean density threshold as
\emph{Sponge-like}. Clearly, as the spectral index $n$ decreases, we see an increasing presence of islands and cavities at the
mean density, going along with a decreased presence of tunnels. Instead of one connected and percolating overdense region
intertwined with a similarly connected underdense region, the topological identity of the density field at median
density is one in which more than one disconnected overdense region is filling up space with several similarly shaped
underdense regions. These regions are also marked by irregular, intricately shaped boundaries, marking a complex
intertwining of each other, involving a large number of tunnels and loops. The number of disjunct overdense and underdense
complexes increases with decreasing spectral index $n$. Strictly speaking this is not a \emph{Sponge-like} topology,
although practically speaking it shares a similar morphology of a complex convoluted structural pattern. See Figure~\ref{fig:contour_surfaces} for a visualization of the contour surfaces for differtent density thresholds for the $n = 1$ and $n = -2$ models.


\begin{figure*}
	\centering
	\subfloat[]{{\includegraphics[width=0.48\textwidth]{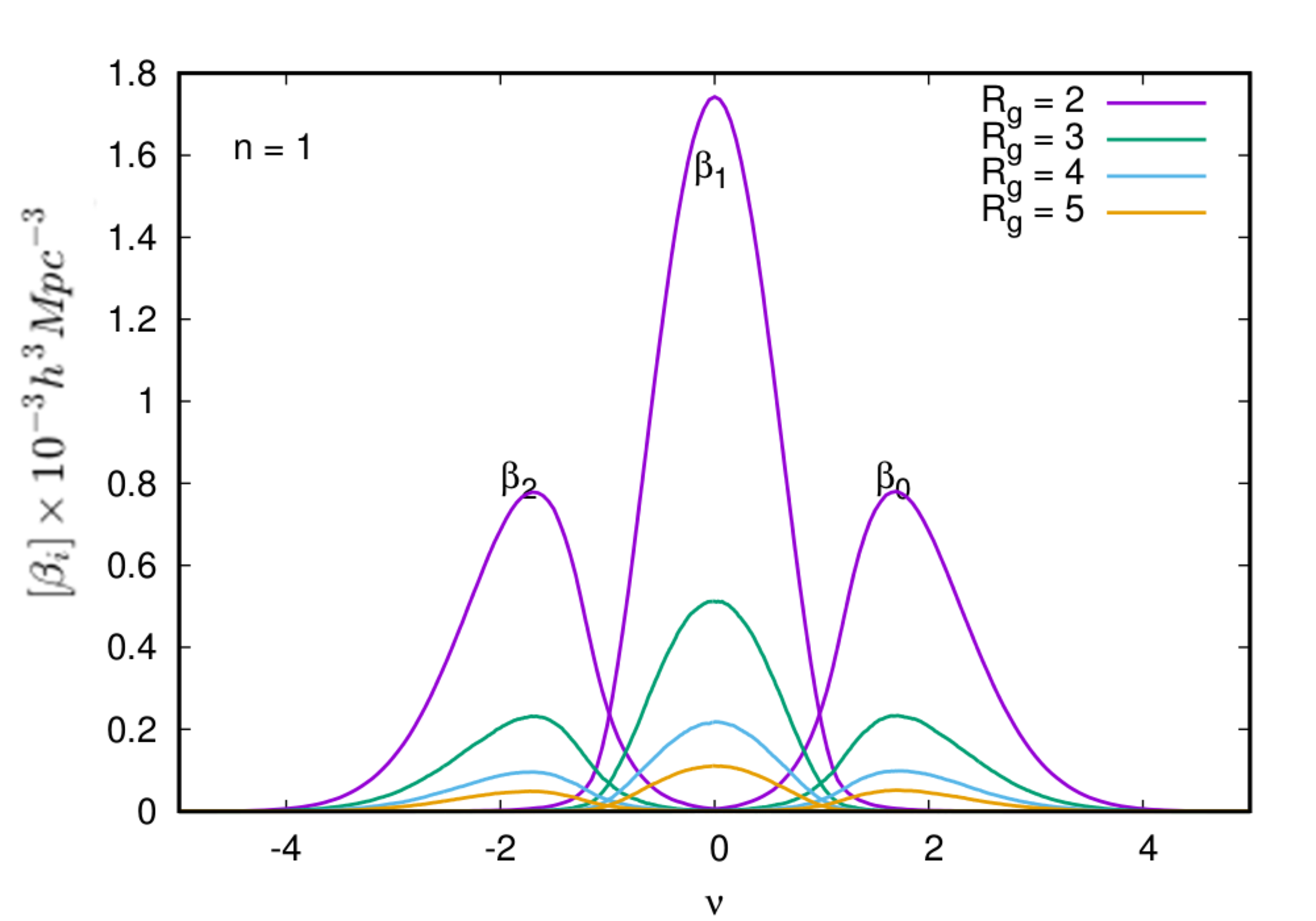}}}
	\hspace{0.01\textwidth}
	\subfloat[]{{\includegraphics[width=0.48\textwidth]{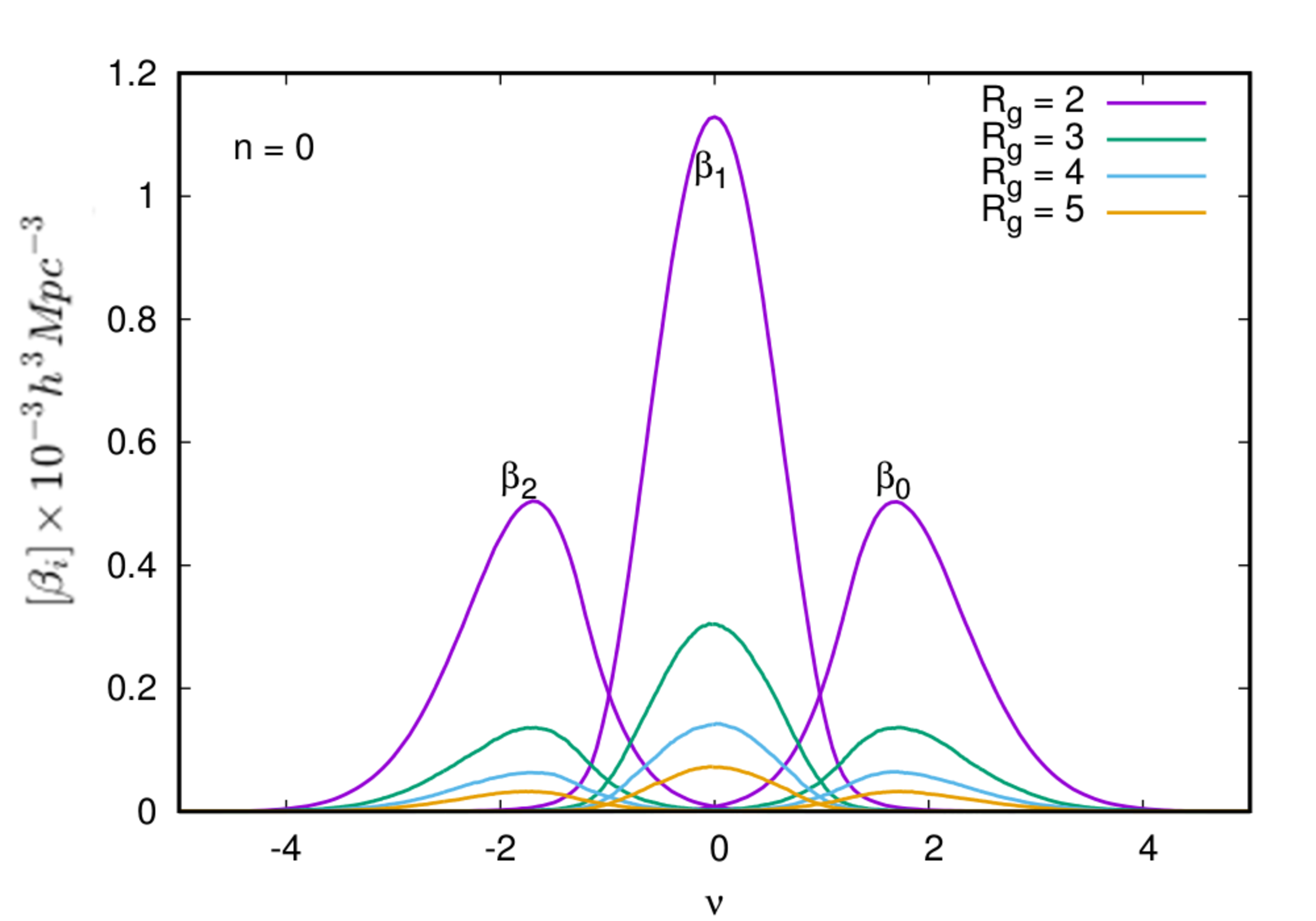}}}\\
	\subfloat[]{{\includegraphics[width=0.48\textwidth]{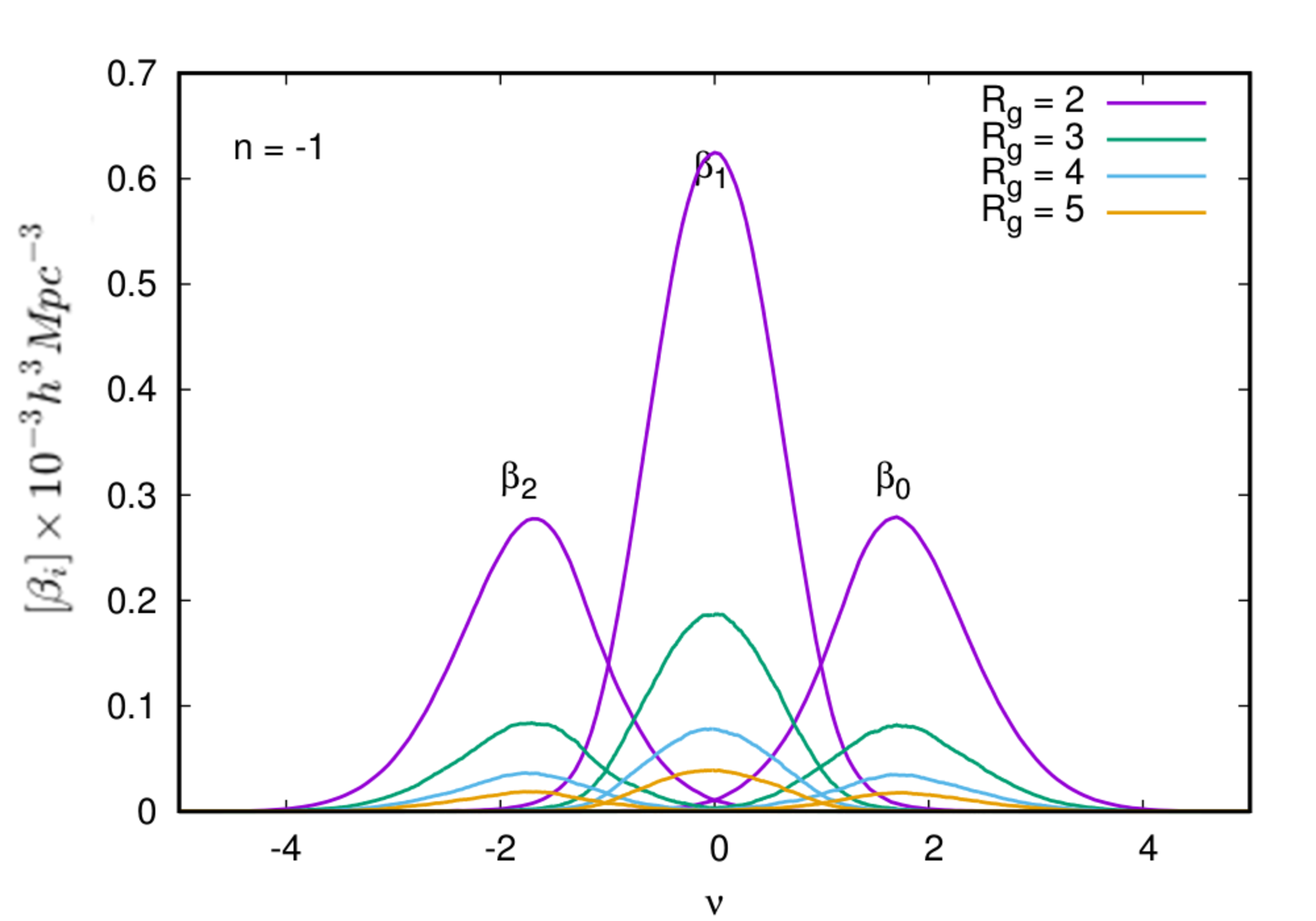}}}
	\hspace{0.01\textwidth}
	\subfloat[]{{\includegraphics[width=0.48\textwidth]{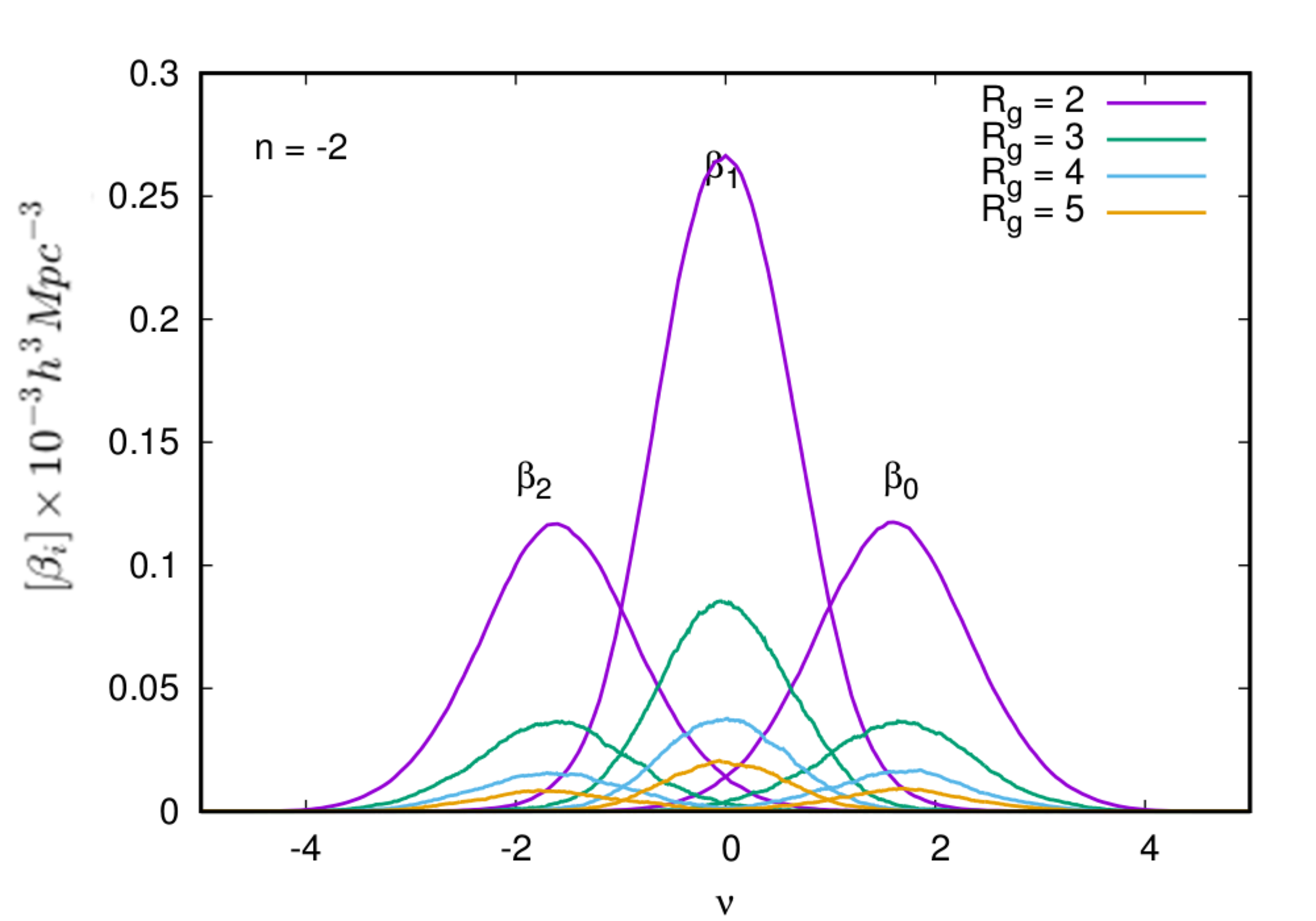}}}\\
	\subfloat[]{\rotatebox{-90}{\includegraphics[height=0.48\textwidth]{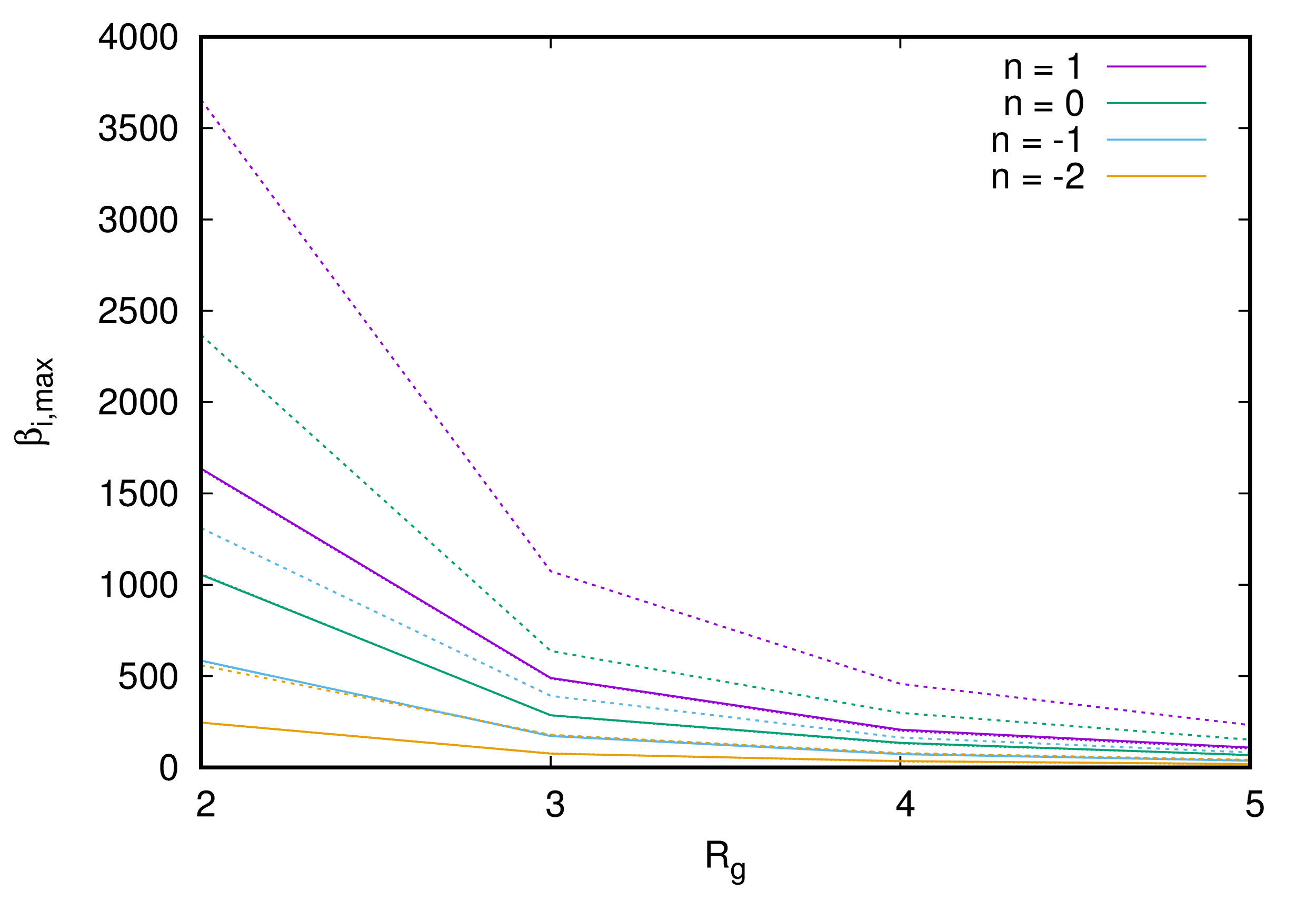}}}
	\hspace{0.01\textwidth}
	\subfloat[]{\rotatebox{-90}{\includegraphics[height=0.48\textwidth]{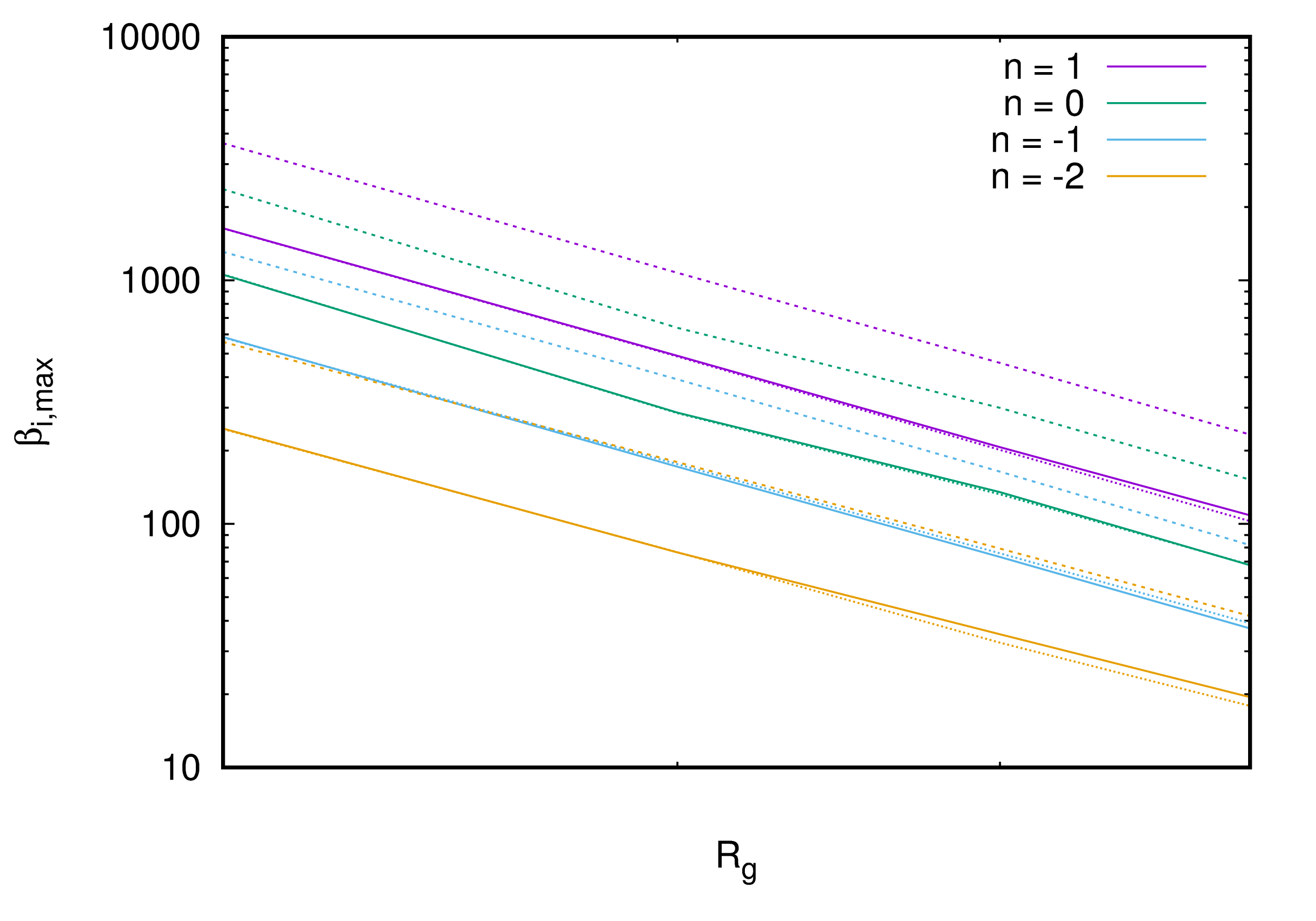}}}\\
	\caption{Dependence of the amplitude of Betti number curves on the filter radius $R_g$. Panels (a) to (d): Betti number curves for the $n = 1, 0, -1$ and  $-2$ respectively, where for each model, the field is smoothed by a Gaussian kernel of varying radii $R_g = 2, 3, 4$ and $5$. Panel (e): Graph presenting the amplitude of the Betti curves as a function of smoothing radius. Curves with the same colors represent the same model. The dotted curves are for $\beta_1$, while the solid and the dashed curves represent $\beta_0$ and $\beta_2$ respectively. Panel (f): Same as panel (e), but on a logarithmic scale. }
	\label{fig:grf_plaw_rg_scaling}
\end{figure*}

\subsection{Spectral scaling properties of Betti numbers}
\label{sec:betti_scaling}

Given the clear dependence of the Betti number curves on the power spectrum of the
Gaussian field, it would be insightful to evaluate the scaling of descriptive
parameters of the Betti curves. We find that the systematic changes are entirely
equivalent for the $\beta_0$ and $\beta_2$ curves. Two aspects that we investigate are
the amplitude and shapes of the Betti number curves. The shapes are investigated in terms of the skewness and curtosis of the curves. 

\bigskip

\begin{figure*}
	\centering   
	\includegraphics[width=0.8\textwidth]{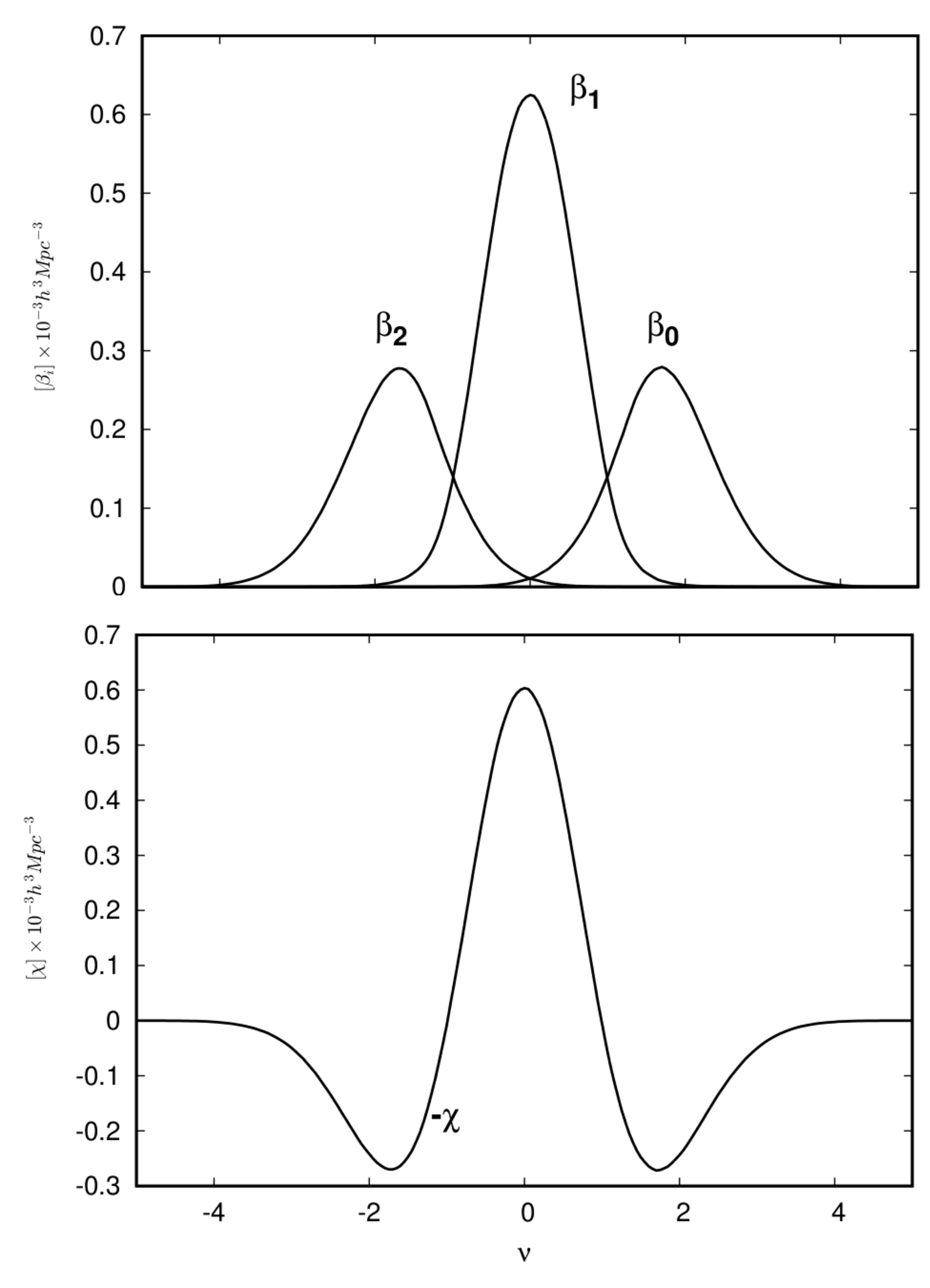}\\
  	\caption{Betti numbers and Euler characteristic for a single model. Different topological entities dominate the different density threshold ranges in the Euler characteristic curve.}
  	\label{fig:betti_EC_singleN}
\end{figure*}

\subsubsection{Scaling of Betti number amplitude with spectral index}

The amplitudes of the unscaled Betti numbers depend on the value of the spectral index $n$. The trend 
of the dependence of the maximum of the Betti number curves on the value of the spectral index is 
shown in Figure~\ref{fig:betti_amp_vs_index_scale}. The amplitude of the Betti numbers, defined
as the maximum of the Betti number curves, approximately follows an exponential. By fitting 
\begin{equation}
f(n)\,=\,A_0 \,\exp{\left\{\frac{n}{\tau}\right\}}\,,
\end{equation}
we find a decay parameter $\tau \approx 2$. Amongst others, it implies that the amplitude of the Betti number curves decreases
roughly exponentially as the value of spectral index $n$ decreases.

\begin{figure*}
  \centering
    	\mbox{\hskip -0.75truecm\includegraphics[width = \textwidth]{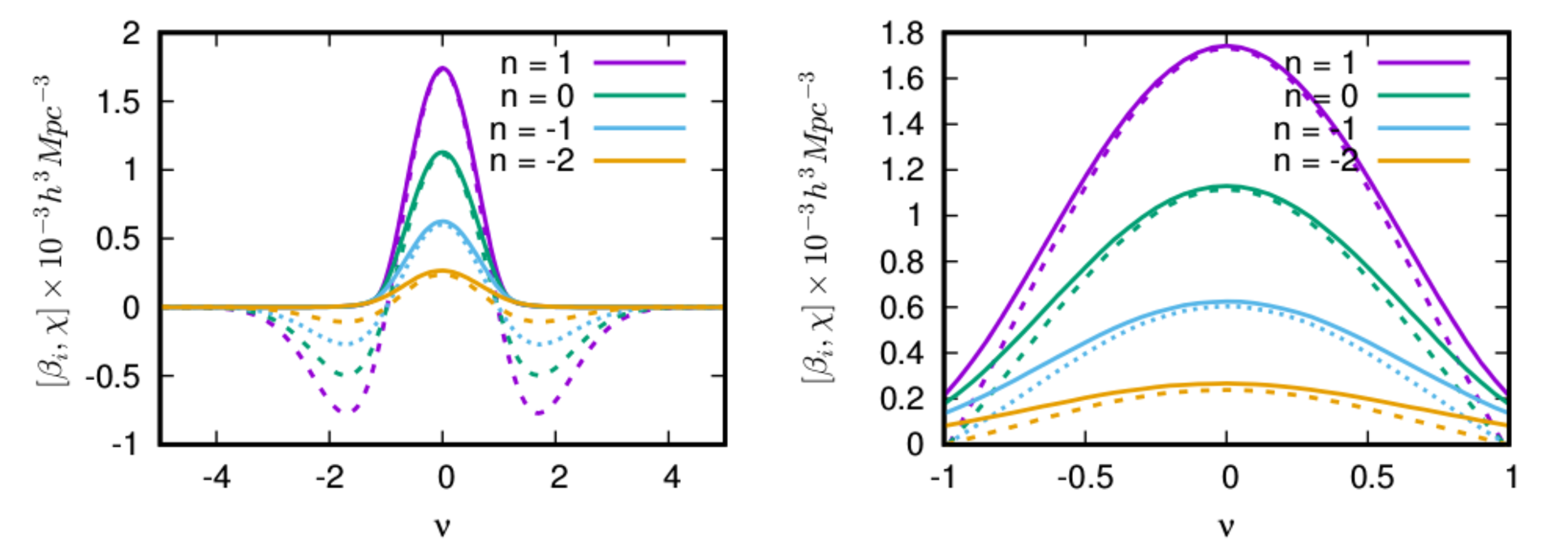}}
  	\caption{Comparison of relative contributions of $\beta_1$ vs. Euler characteristic.}
  	\label{fig:betti_EC_rel}
   \end{figure*}
        \vskip 0.5truecm
\begin{figure*}
	\centering   
	{\includegraphics[width=\textwidth]{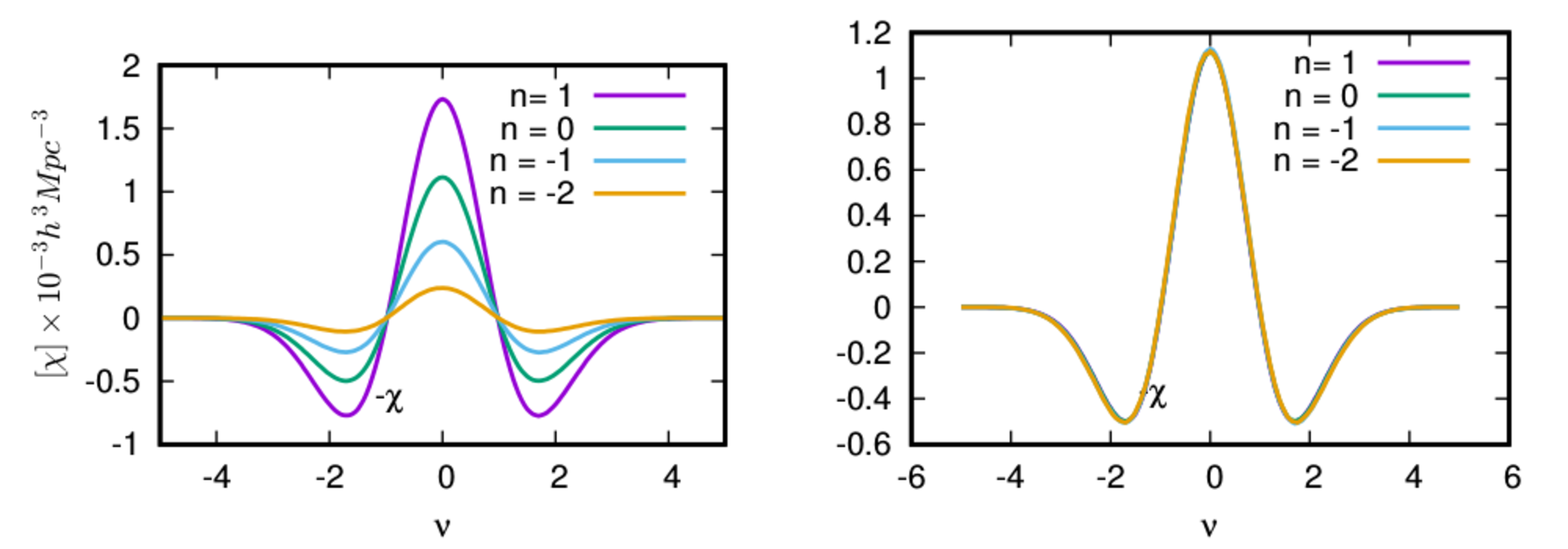}}\\
  	\caption{Scaling behaviour of the Euler characteristic curves for the power law models. Left panel: unscaled Euler characteristic curves for the power law models. Right panel: Scaled curves for the same. The curves for the different models are scaled so that their amplitudes coincide. The scaled curves are indistinguishable from each other. }
  	\label{fig:betti_EC_varyN}
\end{figure*}

\subsubsection{Shape of the Betti curves: skewness \& curtosis}

\begin{table*}
 \begin{center}
	\begin{tabular}{|r||rrr|rrr|} \hline
		
		& \multicolumn{3}{|c|}{Skewness}  & \multicolumn{3}{|c|}{Kurtosis} \\
		
		model & \multicolumn{1}{c}{$\beta_0$} & \multicolumn{1}{c}{$\beta_1$}
		
		& \multicolumn{1}{c}{$\beta_2$}
		
		& \multicolumn{1}{c}{$\beta_0$} & \multicolumn{1}{c}{$\beta_1$}
		
		& \multicolumn{1}{c}{$\beta_2$} \\ \hline \hline
		
		n = 1      & -4.583	& 5.727e-05 &4.584 & 9.134 & 0.047 & 9.136 \\
		
		n = 0       & -4.737	& -7.33e-05&4.736 & 9.546 & 0.055 & 9.541 \\
		
		n = -1       & -4.977	& -2.117e-05 & 4.974 & 10.260 & 0.090 &10.260 \\
		
		n = -2       & -5.015	& -3.789e-05 & 5.029 &  10.764 & 0.280 &10.803 \\
		
		 \hline\hline	
	\end{tabular}
	\caption{Table listing the skewness and kurtosis values for the Betti curves of the various models.} 
\end{center}
	\label{tab:betti_stats}	
\end{table*}

The shape of the Betti number curves  show a dependence on the choice of the power spectrum:
the Betti number curves become broader as $n$ decreases,  the $\beta_0$ and $\beta_2$
curves also appear more symmetric. To appreciate this optimally, we scale the Betti number curves.
Figure~\ref{fig:betti_scaled} nicely illustrates the systematic changes in width and
asymmetry of the curves as a function of $n$. The Figure also shows that changes
of the Betti number curves on $n$ are nearly exclusively confined to the range  
$|\nu|\le \sqrt{3}$. 

Quantities that characterize the shape of the Betti number curves are the skewness $\mu_3$ and curtosis $\mu_4$, quantifying properties such as asymmetry and
narrowness of the curves. The systematic relation between these quantities and power law index is presented in Table~2.
All the quantities exhibit non-zero skewness. Interestingly, this is also the case for $\beta_1$, even though examining by eye they look symmetric for all the models. This is perhaps due to the influence of the tail of the distribution, where, even though, the numbers are small, they are not the same for positive and negative thresholds. 


\subsubsection{Scaling of Betti number amplitude with Gaussian filter radius $R_g$}

Figure~\ref{fig:grf_plaw_rg_scaling} presents the graphs of the Betti number curves for the various power-law models, where they are smoothed by a Gaussian filter radius $R_g = 2, 3, 4$ and $5 \Mpch$. Panels (a) to (d) present the curves for $n = 1, 0, -1$ and $-2$ models respectively. Within each panel, the different colors of the curves represent the different smoothing radii. Panel (e) shows the graph of the amplitude of the Betti curves as a function of smoothing radius. Curves with the same colors represent the same model. The dotted curves are for $\beta_1$, while the solid and the dashed curves represent $\beta_0$ and $\beta_2$ respectively. The curves for $\beta_0$ and $\beta_2$ coincide with each other, indicating that their amplitudes are identical. Panel (f) presents the same curves as panel (e), but on a logarithmic scale. 

The amplitudes for $\beta_0$, $\beta_1$ and $\beta_2$ 
scale the same as a function of the smoothing length $R_g$. We fit the 
maximum of the peaks to the function $\beta_i^{max} = A_0 R_g^{-\tau}$. The value of the power-law index of the fit is $\tau = 3$, irrespective of the model.

  \begin{figure*}
  \centering   
   {\includegraphics[height=0.5\textwidth]{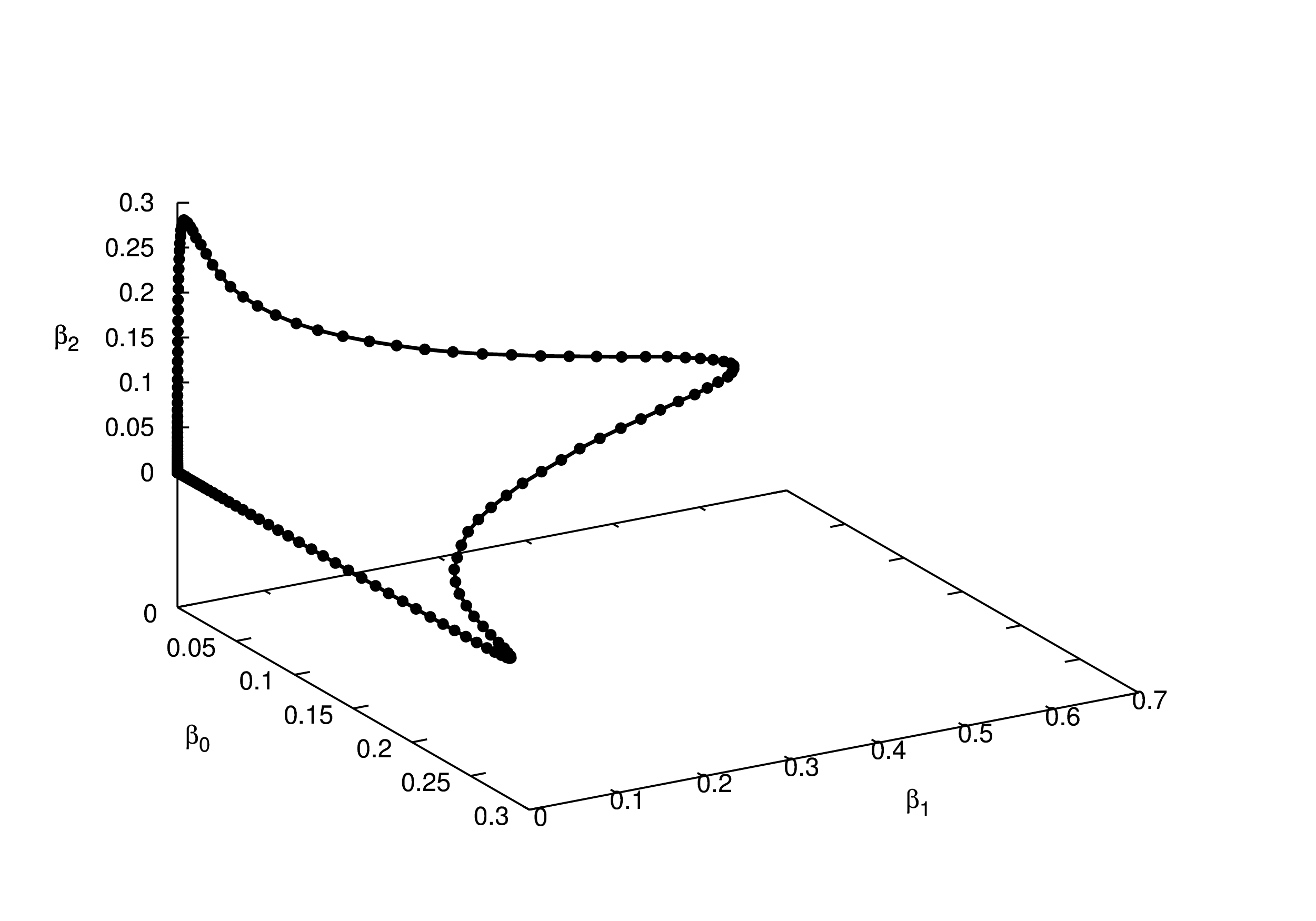}}\\
    \rotatebox{-90}{\includegraphics[height=0.5\textwidth]{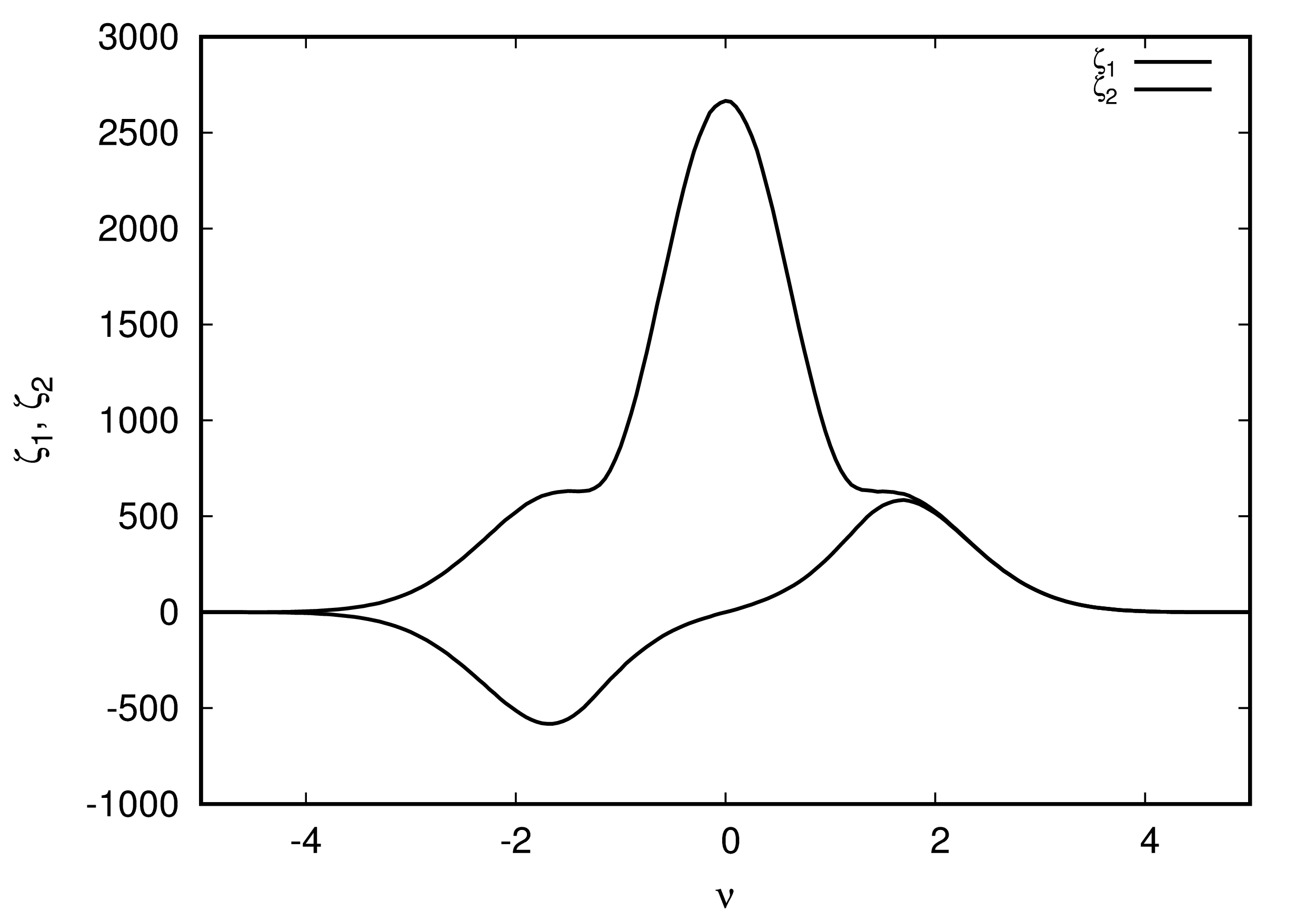}}
    \caption{$\beta$-track (top) and $\zeta$-track (bottom) for for a typical Gaussian field model. They represent orthogonal information to the Euler characteristic, see text for details..}
    \label{fig:betti_track}
   \end{figure*}
\begin{figure*}
\centering   
   \rotatebox{-90}{\includegraphics[height=0.75\textwidth]{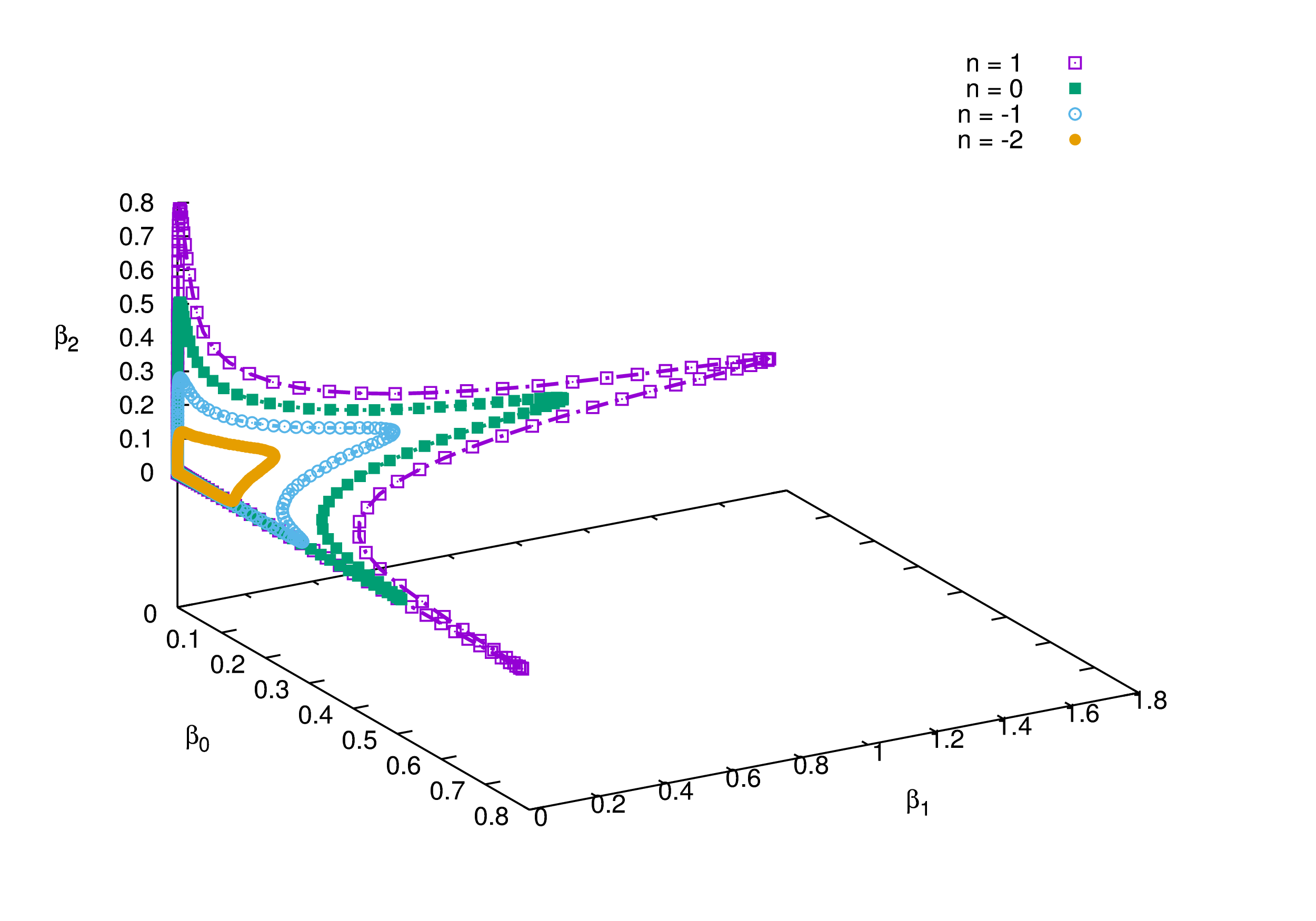}}\\
  \rotatebox{-90}{\includegraphics[height=0.75\textwidth]{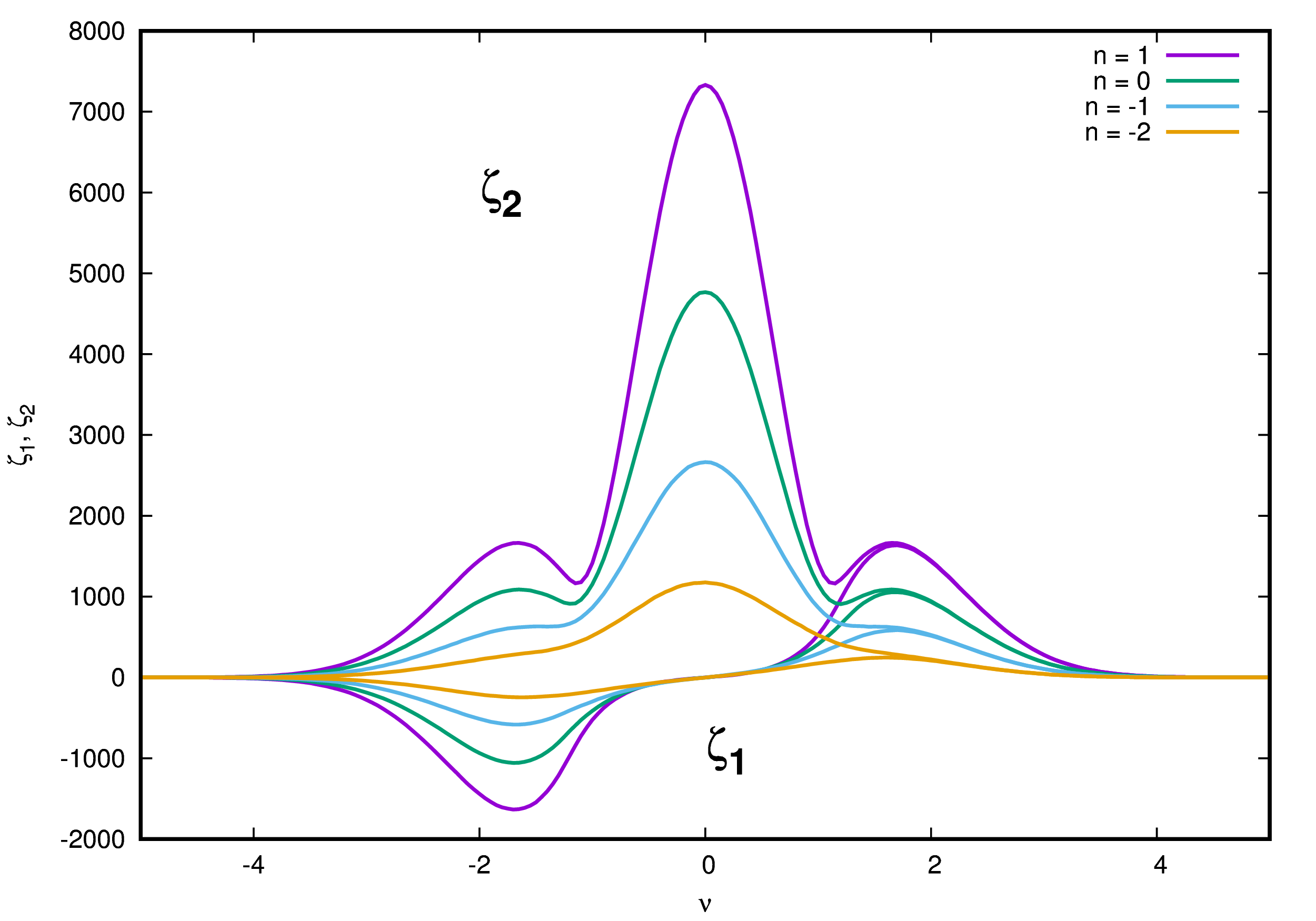}}
  \caption{  Betti tracks and ($\zeta_1, \zeta_2$) for the various power law models, as a function of the spectral index. There is a clear dependence on the power spectrum for the quantities.}
  \label{fig:betti_track_varyN}
  \end{figure*}

\subsection{Betti numbers and Euler characteristic}
\label{sec:betti_EC}

The top panel of the Figure~\ref{fig:betti_EC_singleN} presents the Betti curves for the $n = -1$ model. The bottom panel of the same Figure presents the Euler characteristic. The extrema of the three 3D Betti numbers 
correspond to the three extrema of the Euler 
characteristic curve. Only for 
large thresholds of $|\nu|> 3$, $\beta_0$ and $\beta_2$ are almost equal 
to $-\chi$. This is because the absolute value of the Euler characteristic is 
very close to the 
number of excursion sets or peaks in the asymptotic limit of high 
density thresholds \citep{Adl81,bbks,PPC13}. For thresholds as large as $\nu 
\sim 2$, there is a
significant contribution from $\beta_1$ to $\chi$. 

The left panel of the Figure~\ref{fig:betti_EC_rel} presents the Betti numbers as well as the Euler characteristic for the different power law models. 
Note that the different Betti numbers dominate different regions of the 
Euler characteristic curve. The magnitude of the amplitude of the Euler characteristic curve is lower in general compared to the Betti curves
As an example, for the $n = -2$ 
model, this is even as large as
$10\% - 15\%$. This can be confirmed independently from the right panel 
of Figure~\ref{fig:fractional_betti}. Similarly, in the right panel of  
Figure~\ref{fig:betti_EC_rel} we see that the  
amplitude of $\chi$ is lower than the amplitude of $\beta_1$. The 
difference becomes larger as the spectral index decreases.
It is an indication of the presence of a 
significant number of islands and voids at $\nu = 0$ for lower 
spectral indices. 

The above observations can be related to the 
nature of the density fluctuation field as a function of spectral 
index. For higher spectral indices, there is significant power only 
at smaller scales. This results in high density peaks connected by 
low density saddles, giving the field a distinctly spiky appearance. 
These peaks get connected before they start forming tunnels 
and voids, resulting in a clear cut demarcation of \emph{Meatball-like}, \emph{Sponge-like}
or \emph{Swiss-cheeselike} topology. As the spectral index decreases, the
demarcation diffuses. As the spectral index decreases, progressively 
more and more isolated islands contain additional topological holes of 
higher dimensions, at thresholds well before the manifold becomes a 
singly connected entity. This is reflected in the broadening and 
increased overlap of the Betti number curves, indicating an increase in 
the mixture of topology as the spectral index decreases. In contrast, 
the Euler characteristic curve does not have this 
dependence. As a result, this additional information about the 
inherent differences in the topological structure of the various power 
law models is not available from the Euler characteristic curves. 

The left panel of Figure~\ref{fig:betti_EC_varyN} shows 
the unscaled Euler characteristic curves for the power law models. 
The right panel presents the scaled Euler 
characteristic curves for the same models. The scaled 
curves fall on top of each other, indicating that the shape of the 
Euler characteristic curve is insensitive to the choice of power 
spectrum. This is unlike the Betti numbers, whose shapes show a 
characteristic dependence on the choice of the power spectrum. 
The dependence of Euler characteristic on the choice of the 
power spectrum is restricted to the expression for amplitude, through 
the variance term.  

The above remarks lead us to conclude the 
following. In general, only for positive spectral indices, it is 
feasible to describe the topology of the field as either \emph{Meatball-like}, 
or \emph{Sponge-like} or \emph{Swiss-cheeselike}. For negative 
spectral indices, the demarcation is not clear, except near the tails 
of the density distribution. The topology is an increasing mixture of 
the three types as the spectral index decreases. It is clear that the 
Betti numbers add extra information to the description of topology than 
that by the Euler characteristic.

\subsection{$\beta$- and $\zeta$-tracks}
\label{sec:betti_track_single}

Another means of gleaning the topological information content from the Betti numbers is to visualize them assuming they define coordinates in a 3D space of ($\beta_0$,$\beta_1$,$\beta_2$). The left panel of Figure~\ref{fig:betti_track} presents such \emph{Betti tracks} for a typical Gaussian field realization. There are more ways to exploit the additional information content of the Betti numbers. Since the Euler characteristic is the alternating sum of Betti numbers,
$\chi = \beta_0 - \beta_1 + \beta_2,$ this can be interpreted as the projection of ($\beta_0,\beta_1,\beta_2$) on to the line in direction (1,-1,1). Following this geometrical interpretation, we may identify combinations of the Betti numbers that are orthogonal to the line (1,-1,1), and thus provide independent additional topological information to the
Euler characteristic:

\begin{align}
	\zeta_1 &= \beta_0 - \beta_2\nonumber \\
	\zeta_2 &= \beta_0 + 2 \beta_1 + \beta_2\nonumber \\
			&= (\beta_0 + \beta_1) + (\beta_1+\beta_2),
\end{align}
\noindent i.e., the vectors
(1,0,-1), (1,2,1) and (1,-1,1) form an orthogonal system.
By looking at the distribution of ($\zeta_1,\zeta_2$) ($\zeta$-tracks from now on) in the plane defined by axes (1,0,-1) and (1,2,1) we get an appreciation for the supplementary topological information yielded by Betti numbers, in addition to the Euler characteristic. The right panel of Figure~\ref{fig:betti_track} presents the curves for $\zeta$-track for a typical field realization. We notice that for high density thresholds up to $\nu \sim 2$, $\zeta_1$ and $\zeta_2$ coincide with each other, but show different trends for lower thresholds. 
Note that the $\beta$- and the $\zeta$-tracks  may provide additional tools when model discrimination is the primary focus. 

\subsubsection{Spectral dependence of $\beta$- and $\zeta$-tracks}
\label{sec:betti_track_varyN}
Figure~\ref{fig:betti_track_varyN} presents the $\beta$- and $\zeta$-tracks for the various models. The left panel plots the $\beta$-tracks and the right panel plots the $\zeta$-tracks for the models. Recall that the $\beta$-tracks trace the quantity ($\beta_0,\beta_1,\beta_2$) in a coordinate where the axis represent the individual Betti numbers. Similarly, $\zeta_1$ and $\zeta_2$ (see Section~\ref{sec:betti_track_single} for definitions) are various combinations of the Betti numbers and provide orthogonal information compared to the Euler characteristic. It is evident from the left and the right panels of the Figure that both the quantities show a dependence on the spectral index.

\begin{figure*}
\centering   
\subfloat[]{\includegraphics[width=0.31\textwidth]{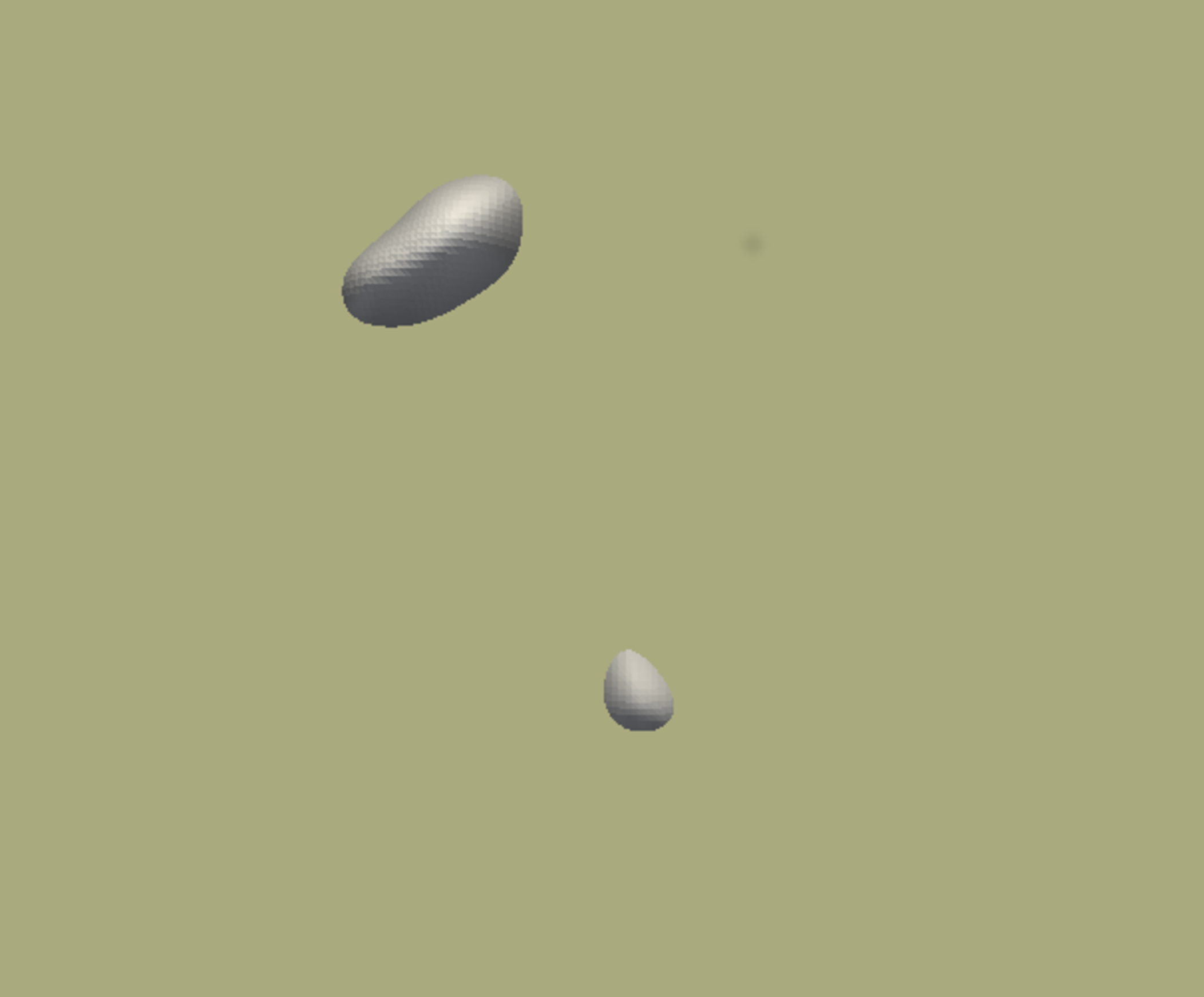}}
   \hspace{0.01\textwidth}
   \subfloat[]{\includegraphics[width=0.31\textwidth]{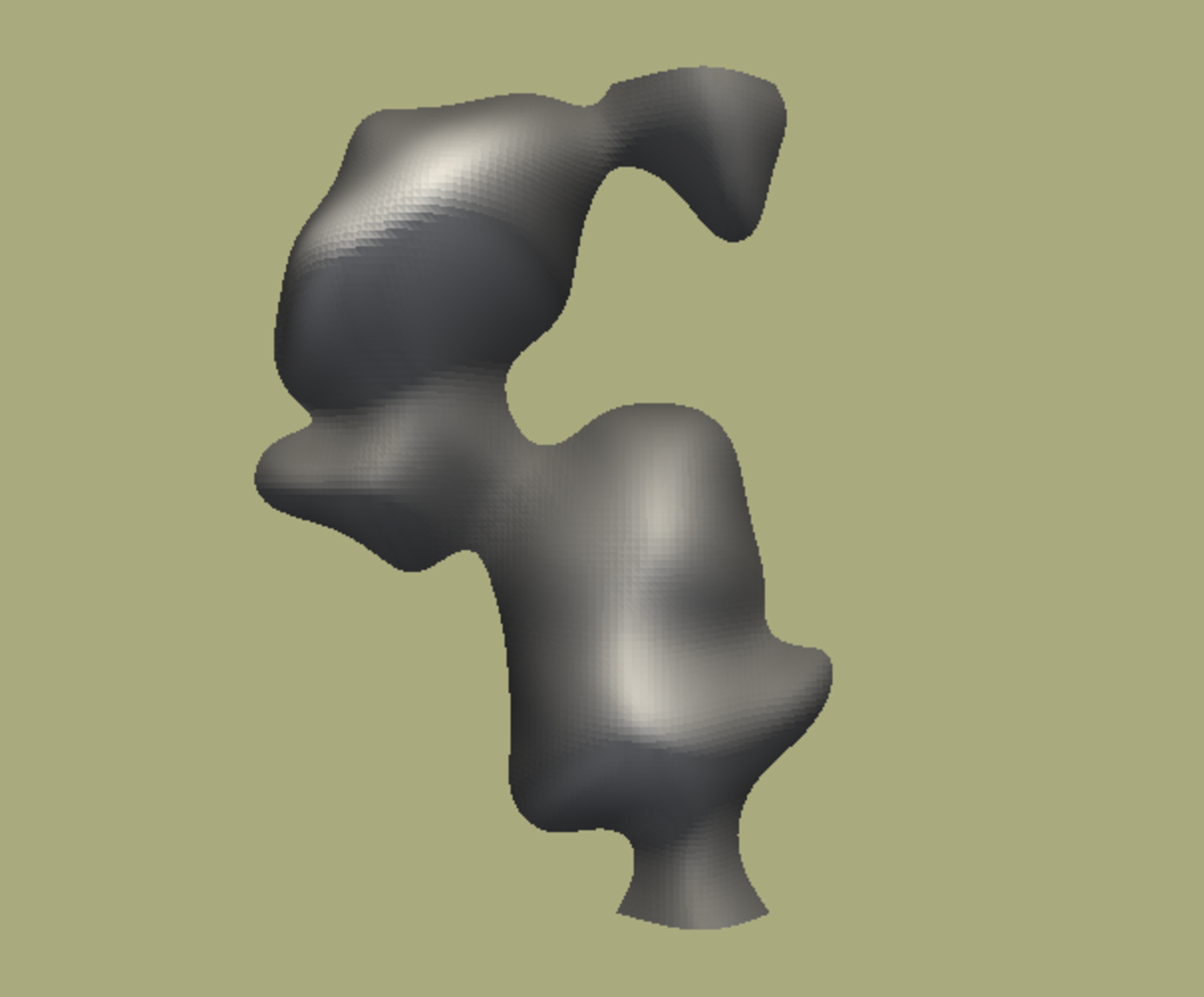}}
   \hspace{0.01\textwidth}
   \subfloat[]{\includegraphics[width=0.31\textwidth]{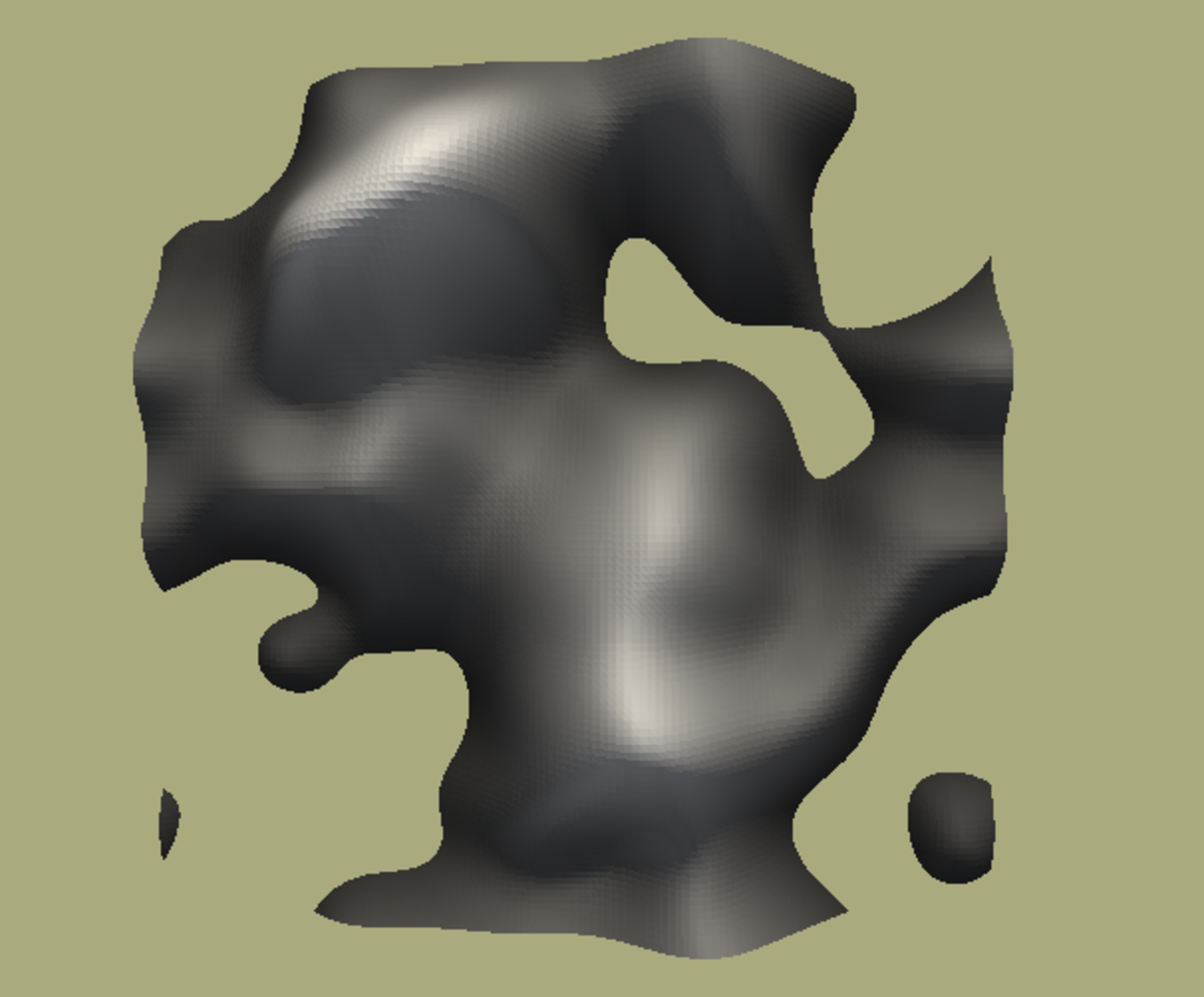}}
  \caption{  Figure illustrating the difference between peaks and 
  islands. The left panel illustrates two peaks. They are composed of 
  a single maximum. However, since they are also trivially connected 
  and isolated objects, they double up as islands also. The middle and 
  the right panels illustrate islands with a more complex topology. In 
  the middle panel the island is a connected object, and contains many 
  peaks. In the right panel, the island encloses a loop as well. }
  \label{fig:peaks_and_islands}
\end{figure*}

\section{Peaks vs. Islands}
\label{sec:peaks_vs_islands}

There is a telling distinction between 
\emph{peaks} such as described by \cite{bbks}, and the 
\emph{islands} of our definition. 
A peak is the location of a 
local maximum of the function. An island is a single connected object. In general, an island may be marked by 
many peaks. However, at the higher density 
thresholds, when no saddle points have yet been introduced in the 
manifold, there will be necessarily one peak per island. As the 
threshold is lowered, the number of peaks per island increases. 
As this happens, the manifold starts developing complex connectivity.
This happens because the peaks merge 
through saddles, such that they are a part of a single connected component. As the density threshold lowers still further, such a connected component may exhibit more topological features, like a hole.
 
Figure~\ref{fig:peaks_and_islands} illustrates the above mentioned phenomenon in 2D.  The left panel illustrates two separate peaks. They are composed of 
a single maximum. However, since they are trivially connected 
and isolated objects, they can also be classified as islands. The middle and 
the right panels illustrate islands with a more complex topology. In 
the middle panel, the island is a connected object, and contains many 
peaks. This is due to the fact that as the density threshold lowers, saddle points are also introduced in the manifold, which connect two disjoint maxima to form a single connected component. In general, at sufficiently low thresholds, one may identify islands composed of multiple maxima and saddles, as is the case in the middle panel. In this 
context, we point out that the number of peaks per island, as a 
function of the density threshold, is a topological quantification of the 
strength of clustering of a model. 

In the right panel, the island encloses a loop as well. This is due to the introduction of a saddle point that connects the boundary of an already connected component, forming a closed loop.  Below, we investigate the  
model dependent variation of the number of peaks and islands for the 3D 
Gaussian field models.

\subsection{Peaks vs. Islands: the Gaussian case}
\label{sec:peaks_vs_islands_grf}

\begin{figure*}
  \vskip 0.5truecm
\centering   
  \mbox{\hskip -0.0truecm\includegraphics[width = \textwidth]{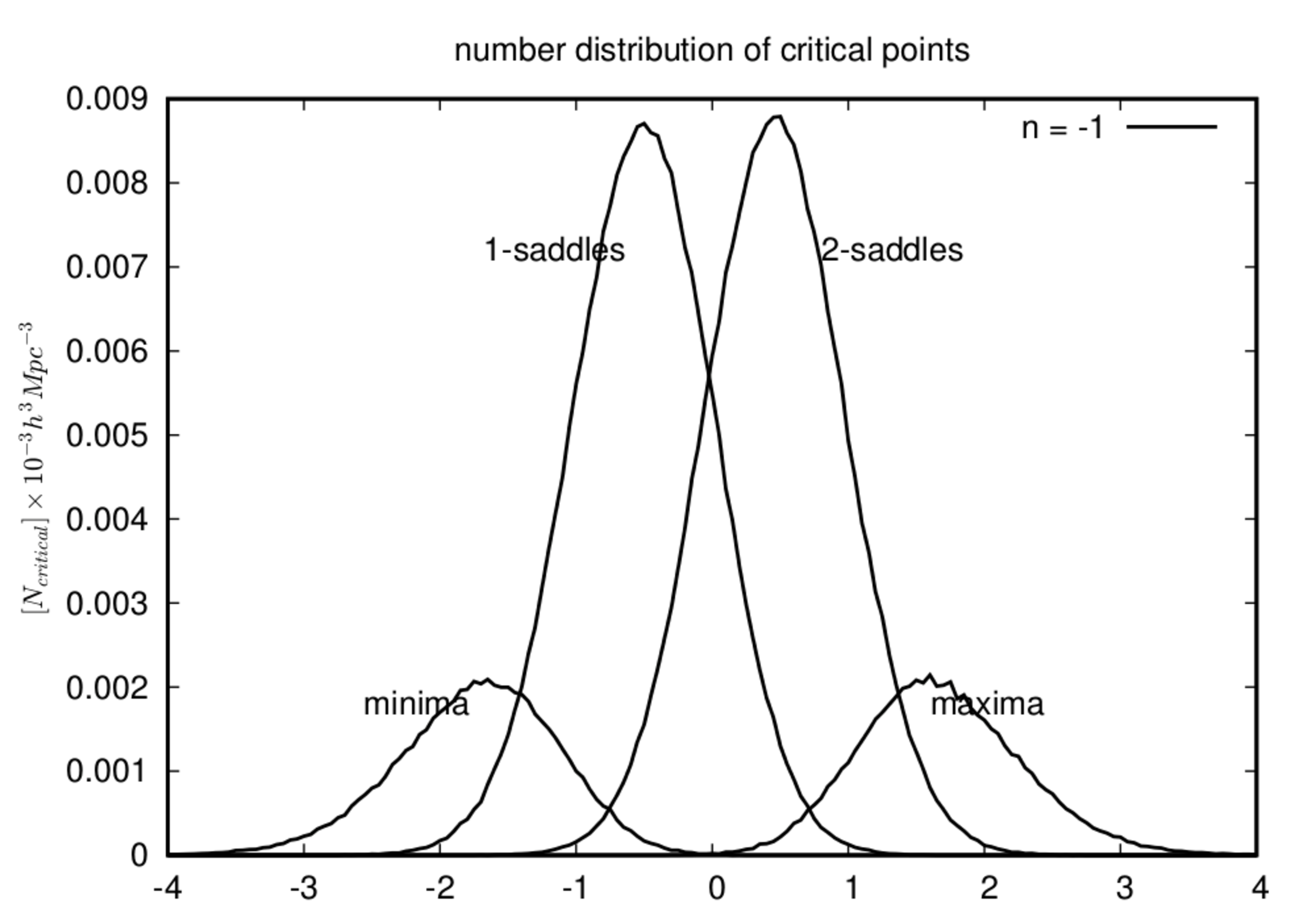}}
  \caption{  Number distribution of critical points for the $n = -1$ model. Critical points with different indices are dominant over different ranges, but there is a strong overlap between them, as a function of the density threshold.}
  \label{fig:cps}
  \vskip 1.0truecm
  \centering
   \mbox{\hskip -0.75truecm\includegraphics[width = 1.1\textwidth]{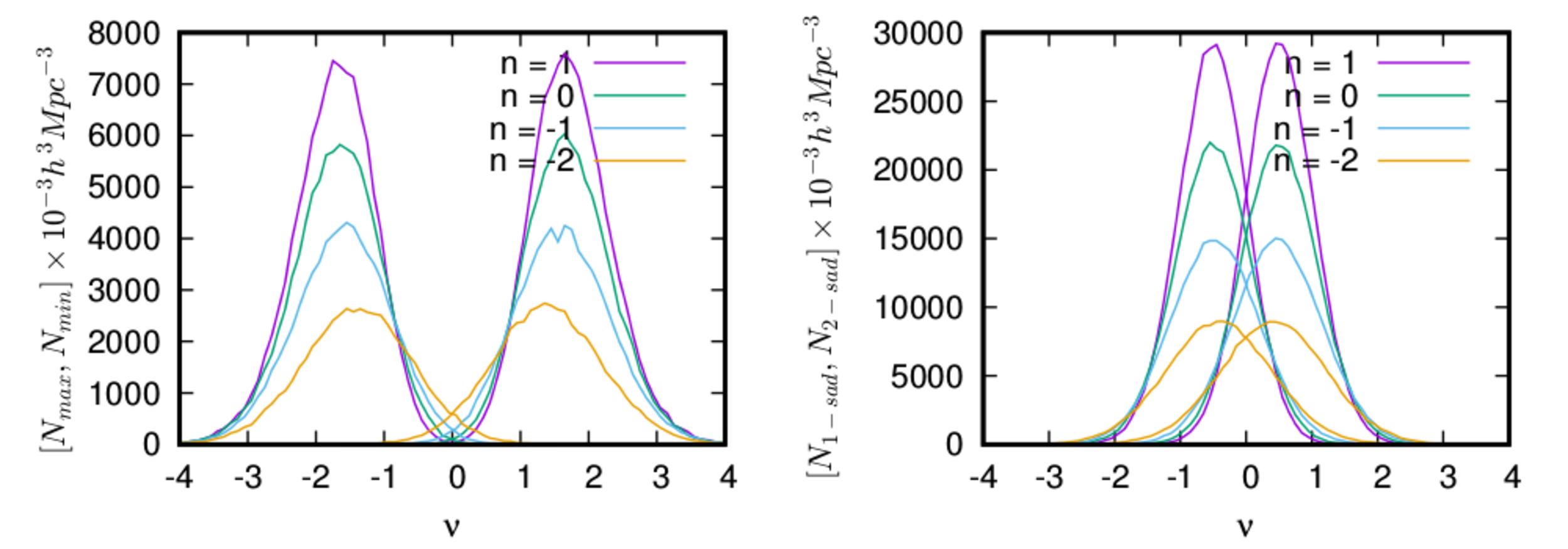}}
  \caption{  Number distribution of critical points as a function of the spectral index. The left panel plots the distribution of the maxima/minima as a function of the spectral index, and density threshold. The right panel plots the same for $1$- and $2$-saddles.}
  \label{fig:cps_varyN}
\end{figure*}

As we noted earlier, \emph{peaks} and \emph{islands} (see definition above) are related yet different topological entities. In particular, peaks are a submanifold of islands. The former are the
local maxima, while the latter grow depending on the density threshold 
$\nu$. An 
island may contain multiple peaks.  In fact, as an island grows it 
might get arbitrarily complicated, acquiring tunnels and even voids, while 
always staying connected. But in the asymptotic limit of high $\nu$, 
every island will contain only one peak. In this connection, it is instructive to examine the number distribution of critical points in general, and peaks in particular.

\subsubsection{Number distribution of peaks (critical points)}
\cite{bbks} derive the differential number distribution of peaks for Gaussian random fields, as a function of the
dimensionless density threshold $\nu$  (see appendix~\ref{app:peakdens} for details):
\begin{equation}
	\Nspace_{pk}(\nu) \,d\nu = \frac{1}{(2\pi)^{2}R_{\star}^{3}} e^{-\nu^{2}} G(\gamma,\gamma\nu).
	\label{eqn:diff_peak_density}
\end{equation}
\noindent The function $G(\gamma,\gamma\nu)$ depends on the spectral parameter $\gamma$ and height $\nu$ of the
peaks. \cite{bbks} derived and specified a highly accurate fitting function for $G(\gamma,\gamma\nu)$, whose
details are specified in appendix~\ref{app:peakdens}). The parameters $\gamma$ and $R_{\star}$ are combinations
of sevreal moments $\sigma_j$ of the (filtered) power spectrum $P_s(k)$ (see appendix~\ref{app:peakdens}.
The spectral scale $R_{\star}$ is proportional to the smoothing scale $R_s$ of the field, ie. $R_{\star} \propto R_s$. For
power-law power spectra, $\gamma$ and $R_{\star}/R_s$ depend on the spectral index $n$
(see appendix~\ref{app:peakdens}). 

\begin{figure*}
\centering   
  {\includegraphics[width=\textwidth]{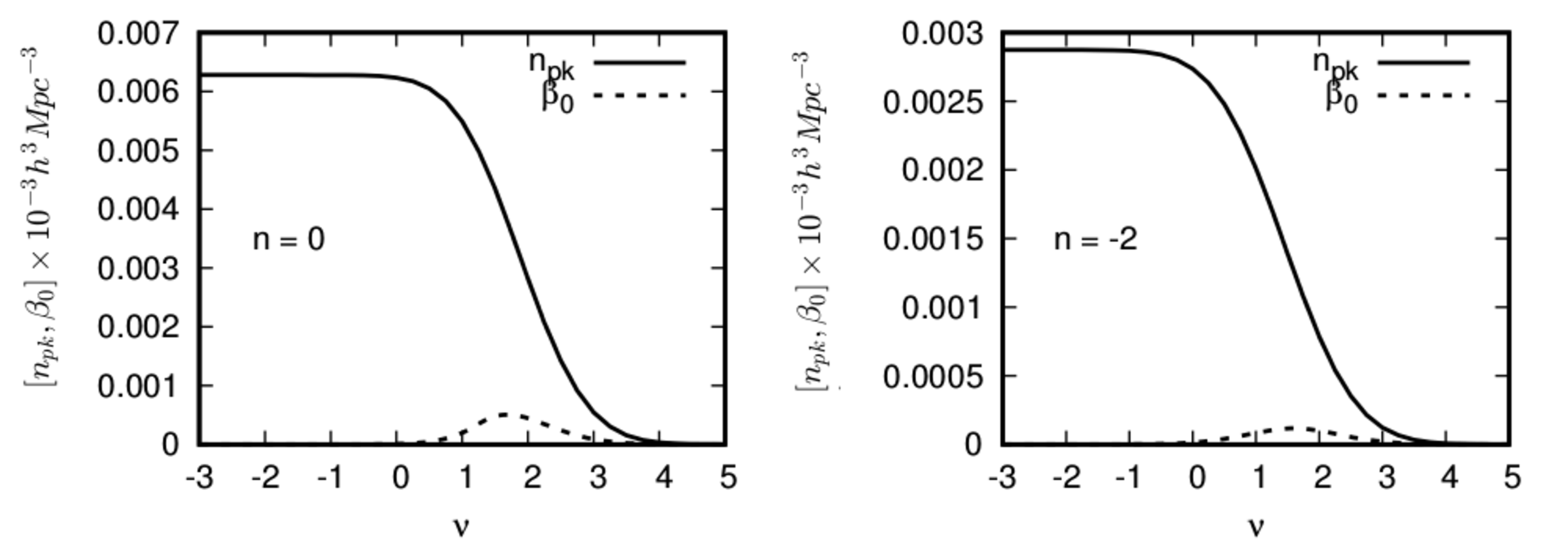}}\\
  \caption{  Cumulative number distribution of peaks and the distribution of islands (zeroth Betti number) for comparison, as a function of density threshold. The left panel presents the graphs for the $n = 0$ model, and the right panel plots it for the $n = -2$ model.}
  \label{fig:cum_peaks_vs_islands}
\vskip 0.5truecm

\centering   
{\includegraphics[width=\textwidth]{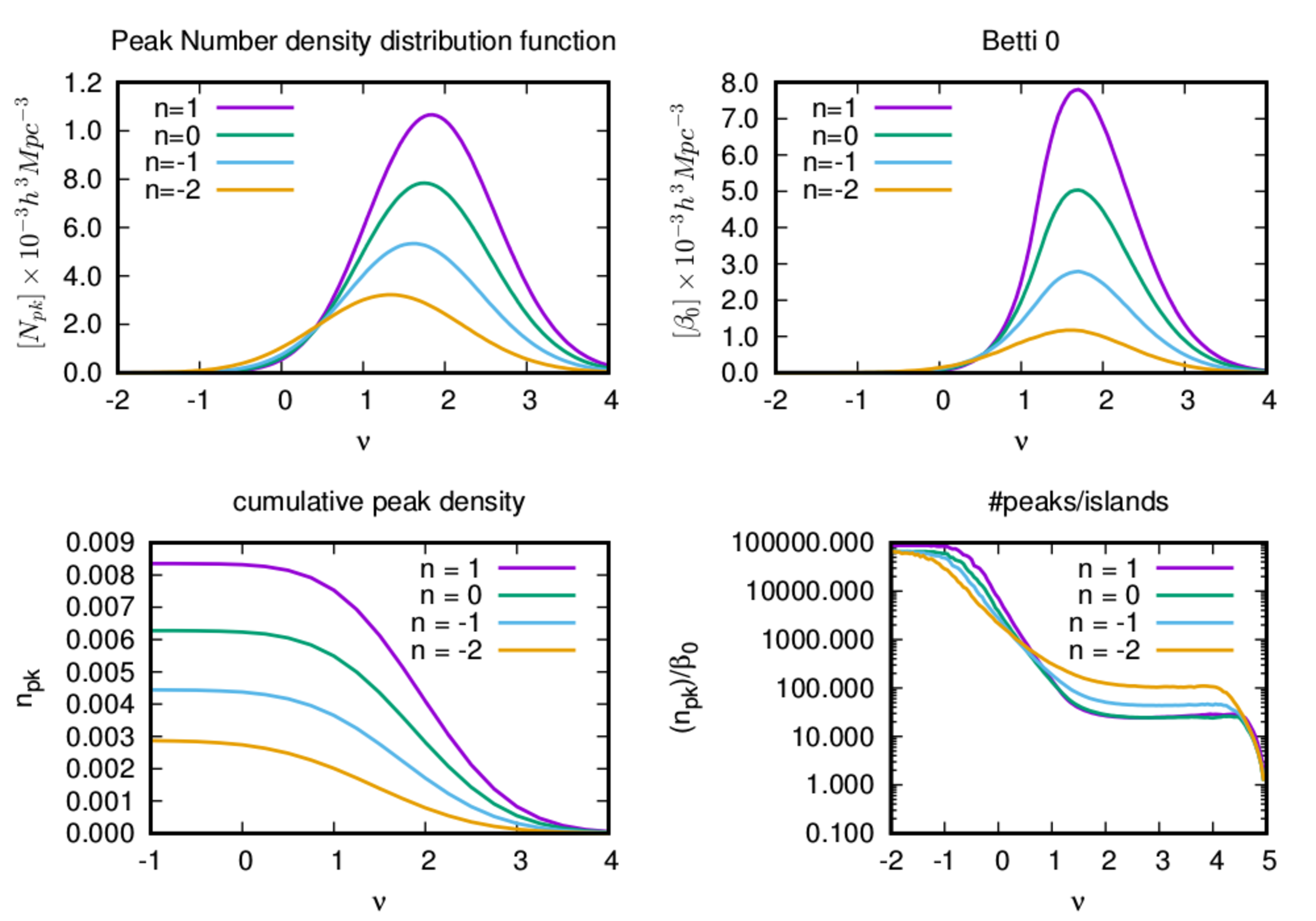}}\\
  \caption{ Figure illustrating the relation between peaks and islands. The top-left panel plots the number distribution of peaks, $N_{pk}$.  The top-right panel plots the number of independent islands as denoted by $\beta_0$. The zeroth Betti number counts the number of isolated islands at a particular 	density threshold. The curves for both the quantities show a 
  	characteristic dependence on the index of the power spectrum. The location of the maximum shifts
  	towards lower density thresholds considerably for the number distribution of peaks as the spectral index decreases. The location of the maximum for the $\beta_0$ curve remains unaffected as the spectral index changes. The bottom-left panel plots the cumulative number density of peaks as a function of the spectral index. It increases with increasing spectral index, which is related to the fact that as the power shifts to smaller scales for larger spectral index, it is accompanied by larger number small scale peaks packed in a given volume. The bottom-right panel plots the cumulative number of peaks per island as a function of power spectrum, and density threshold $\nu$. For very large values of $\nu$ there is one peak per island irrespective of the spectral index. As the density threshold decreases, this number has a characteristic dependence on the spectral index.}
  \label{fig:diff_peak_density}
\end{figure*}

For the total number of peaks $n_{pk}(\nu)$ in excess of a density threshold $\nu$ in the smoothed density field, we may compute
the cumulative peak density, 
\begin{equation}
n_{pk}(\nu)\,=\,\int_{\nu}^{\infty}\,\Nspace_{pk}(\nu) \,d\nu\,.   
  \end{equation}
From this, we may infer - analytically - that the total density of peaks $n_{pk}(-\infty)$ is given by 
\begin{equation}
    n_{pk}\,=\,0.016\,R_{\star}^{-3}\,.
  \end{equation}
In other words, in a Gaussian field one expects in total some 62 to 63 peaks per cubic volume $R_{\star}^3$. 

Figure~\ref{fig:cps} plots the number distribution of the critical points for the $n = -1$ model, as computed from a single realization. One may notice that the distribution is symmetric about the mean density threshold. In particular, the distribution of maxima is symmetric with respect to the minima. Similarly, the distribution of the 2-saddles is symmetric with respect to the 1-saddles. To arrive at the expressions for the spatial density of all critical points - maxima, minima and saddle points - one may follow a similar calculation as that for maxima (peaks). Along these lines, \cite{pogosyan2009} arrives at approximations for the
distribution function of the critical points, while \cite{codis2013} even managed to obtain expressions for mildly non-Gaussian fields. More recently,
\cite{cheng2015} specified the exact formula for the number distribution of critical points for the Gaussian case.

Figure~\ref{fig:cps_varyN} presents the distribution of critical points for the different power law models. The left panel plots the curves for maxima/minima, and the right panel plots the curves for 2-saddles/1-saddles. As the spectral index decreases, the amplitude of the curves drop, accompanied by a broadening of the curves. This indicates that, as the spectral index decreases, there is a bigger overlap between the distribution of the various critical points as a function of the density threshold. The lower amplitude of the curves can be seen as the effect of a generally smoother field, as the spectral index decreases, which results in lesser number of critical points in the same given volume. As we will see shortly, these characteristics of the critical point distribution have a repercussion on the overall distribution of peaks and islands, as well as their ratio.

\subsubsection{Asymptotic behavior of peaks and islands}
For very high values of rms density 
threshold, as long as the peaks do not start merging, we expect  
the cumulative number density of peaks to be equal to the number density of 
islands. This is confirmed in Figure~\ref{fig:cum_peaks_vs_islands} 
where we 
present the cumulative number density of peaks $n_{pk}$ and the 
number density of islands $\beta_0$ per unit volume, as a function of $\nu$, for the $n = 0$ and $n = 2$ models in the left an the right panels respectively. The cumulative number density of peaks equals the number density of islands asymptotically for very large rms density 
thresholds. The equivalence starts breaking down rapidly at thresholds 
even as high as 
$\nu \sim 4$. This is attributed to the fact that for high 
thresholds all the peaks represent disconnected regions almost surely 
\citep{bbks}, while they 
start connecting up and forming complex topology as the threshold 
decreases. In general, one can also notice from both the panels that the distribution of peaks and islands show a characteristic dependence on the choice of the power spectrum.

For high values of spectral index, the small scales are dominant. In terms of the 
structures in the density 
fluctuation field, this means that the number of small scale peaks of 
high amplitude is large. They are also separated by low-density saddles. 
There is no discernible large scale feature in the density field. 
As the index of the power spectrum decreases, the power shifts to large 
scales. The small scale peaks are 
separated by saddles occurring at relatively high density thresholds. 
It is also accompanied by a decrease in the amplitude of the 
global maximum of the field. As noted earlier, this is because the variance of the density 
field in the box decreases with decreasing spectral index: $\sigma_0 
\propto R^{-(n+3)}$. This phenomenon is reflected 
in the curves of the number distribution of peaks in the 
top-left panel and top-right panels of the Figure~\ref{fig:diff_peak_density}. For a larger $n$, the number 
distribution of peaks attains its maximum at a higher density threshold 
compared to a smaller $n$. For the $n = 1$ model, the maxima is located as 
high as $\nu = 2$. In contrast, for 
the $n = -2$ model, the maximum is located at $\nu \sim 1$, and there 
are significant number of peaks even below $\nu = 0$. 
As noted earlier, this is a direct reflection of the fact that there 
are progressively more number of peaks for lower thresholds, as the 
spectral index decreases \citep{bbks}. 
In contrast, in the same Figure, the location of the maxima of $\beta_{0}$ curves shows a negligible dependence on the value of spectral index. 

\subsection{Distribution of peaks and islands: a comparison}
The top left and top right panels of Figure~\ref{fig:diff_peak_density} plots the number distribution of 
peaks and the Betti numbers for the 3D Gaussian fields. The left and the right panels plot the specific number
distribution of peaks and the zeroth Betti number $\beta_0$ respectively. The curves show the behaviour of these quantities
as a function of the dimensionless density threshold $\nu$. The curves for both quantities reveal a largely
similar behaviour as a function of the density threshold $\nu$, and in their dependence on the spectral index
of the power spectrum.

Nonetheless, we identify subtle properties that manifest themselves in distinct
differences when we assess the number of peaks - or saddles and minima - that populate a given island complex. 
From the top panels of Figure~\ref{fig:diff_peak_density}, it is evident that as the spectral index decreases, the location
of peaks shifts towards lower density thresholds for both the quantities, as the spectral index decreases. This effect is strong for the
cumulative distribution of peaks, but small for $\beta_0$. We also note a substantial difference in amplitudes of the
distribution functions. The (cumulative) number density of peaks therefore differs substantially from that for the zeroth Betti number,
indicating that they measure different features associated with the topology of the density distribution. 

The bottom left panel of Figure~\ref{fig:diff_peak_density}, presents the cumulative number density of peaks for the different models.
The bottom right panel plots the ratio of the cumulative peak density and the number of islands $\beta_0$. This quantity is an indicator
of the average number of peaks of height $\nu$ and higher that populate an island at $\nu$.  The average number of peaks per island shows 
a characteristic dependence on the power spectrum. As expected, for high density thresholds the number of peaks per island approaches unity.
While the density thresholds have a positive value, lower spectral indices correspond to a higher number of peaks per island. A major reason
for this is that Gaussian fields with a lower spectral index contain larger coherent features. Net, this lower number of large islands
contains a higher number of peaks (and other singularities).

In this context, we may also identify a subtle complementary effect. The interior structure of each island
is marked and largely determined by the spatial distribution of peaks and 2-saddles. In general, the number density of
these behave differently \citep{bbks,pogosyan2009}. This can be immediately inferred from their distribution functions in Figure~\ref{fig:cps_varyN}.
Not all 2-saddles at a given density threshold are therefore responsible for bridging two previously disconnected peaks. Moreover, 
the fraction of 2-saddles that join two isolated objects is, in general, a function of the density threshold (also 
see~\cite{feldbrugge2012} for a semi-analytic approximation describing this). At low densities we therefore see an increasing fraction
of them involved in establishing connections between two or more already connected peaks, thereby forming loops or tunnels \citep{EdHa10,PEW16},
and the crackled appearance of the island.

The spectral dependence of the peak population of islands reverses at median and low field densities. As borne out by
Figure~\ref{fig:diff_peak_density}, at underdense field values we observe a steep rise in the number of peaks per island as the
density decreases. It reflects the merging of an increasingly larger fraction of the volume into an ever larger
connected and percolating complex as individual disconnected overdense islands start to connect. They merge into a single or a few volume
percolating regions, leading to a field topology attaining a \emph{Sponge-like} character (see Section~\ref{sec:spongetopology}). As we descend to lower
densities, we therefore see the absorption of the remaining peaks into the remaining percolating region(s). It results in the observed steep
rise of the number of peaks per island. 

As we described extensively in section~\ref{sec:betti_plaw}, Gaussian fields with a lower spectral index retain a slightly higher number of disconnected islands
at low density thresholds than those with higher spectral index (also see Figure~\ref{fig:betti_unscaled}). Meanwhile, as a consequence of the dominance of
high frequency modes in high spectral index Gaussian fields, they are marked by a considerably higher number of peaks (see Figure~\ref{fig:diff_peak_density},
lower left-hand panel). It translates into the steeper rise of the ratio $n_{pk}/\beta_0$ for Gaussian fields that have a higher spectral index seen in 
Figure~\ref{fig:diff_peak_density} (right-hand panel). 

\begin{figure*}
\centering   
   \rotatebox{-90}{\includegraphics[height=0.9\textwidth]{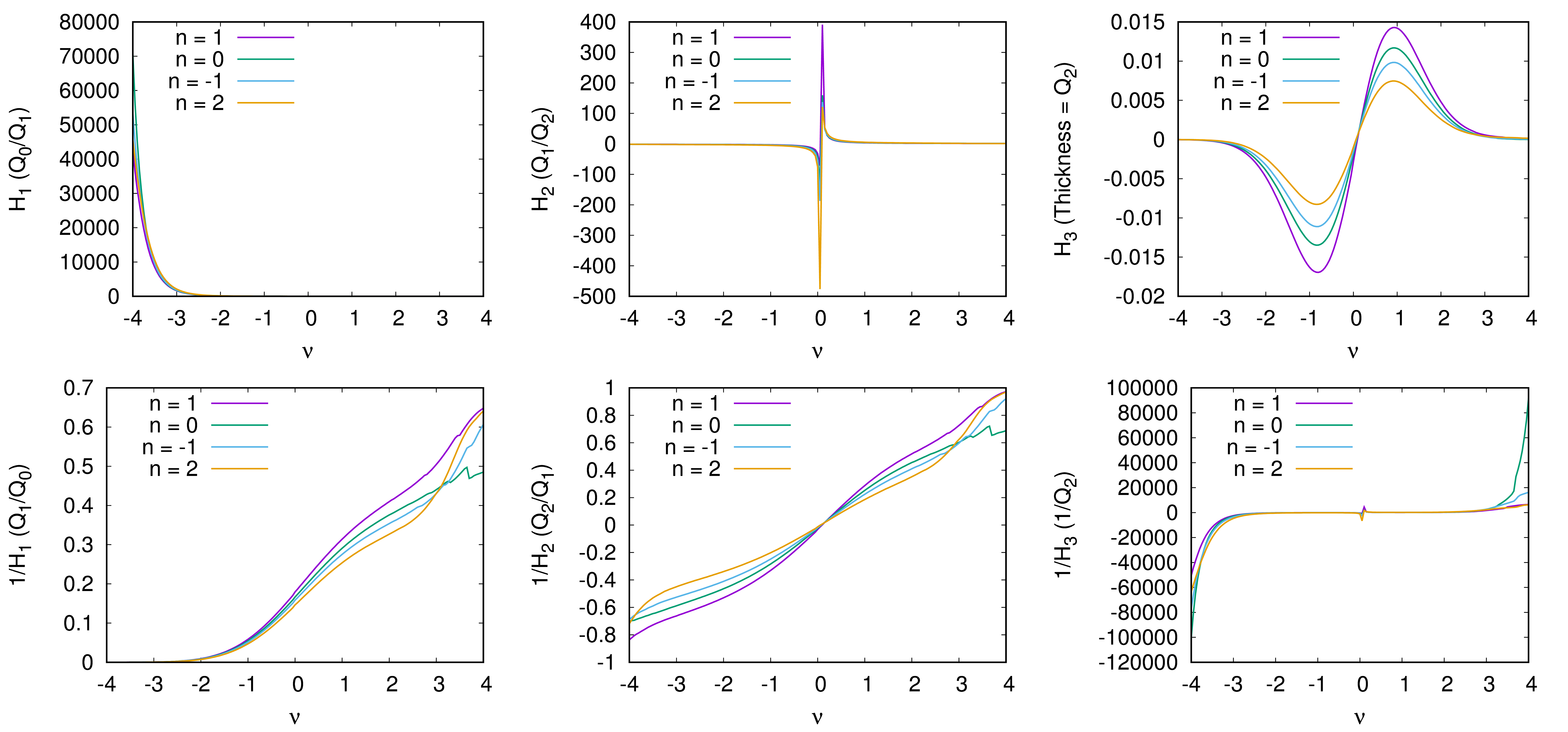}}\\
  \caption{  Minkowski functionals of the power law
  models. From the top to the bottom, we present the volume, area, integrated curvature, and Gaussain curvature, or the Euler characteristic. The left column presents the unscaled version, while the right panel present the scaled version.
  All the functionals are normalized by the 
  total volume of the simulation box. The volume functional is 
  invariant with respect to the choice of the power spectrum. The 
  amplitude of the area, contour length and Euler characteristic shows 
  a dependence on the choice of the power spectrum. The shape of the scaled curves is indistinguishable across the models. This implies the shape of the Minkowski functionals has no dependence on the spectral index.}
  \label{fig:minkowski_grf}
\end{figure*}

\begin{figure*}
\centering   
   \rotatebox{-90}{\includegraphics[height=0.9\textwidth]{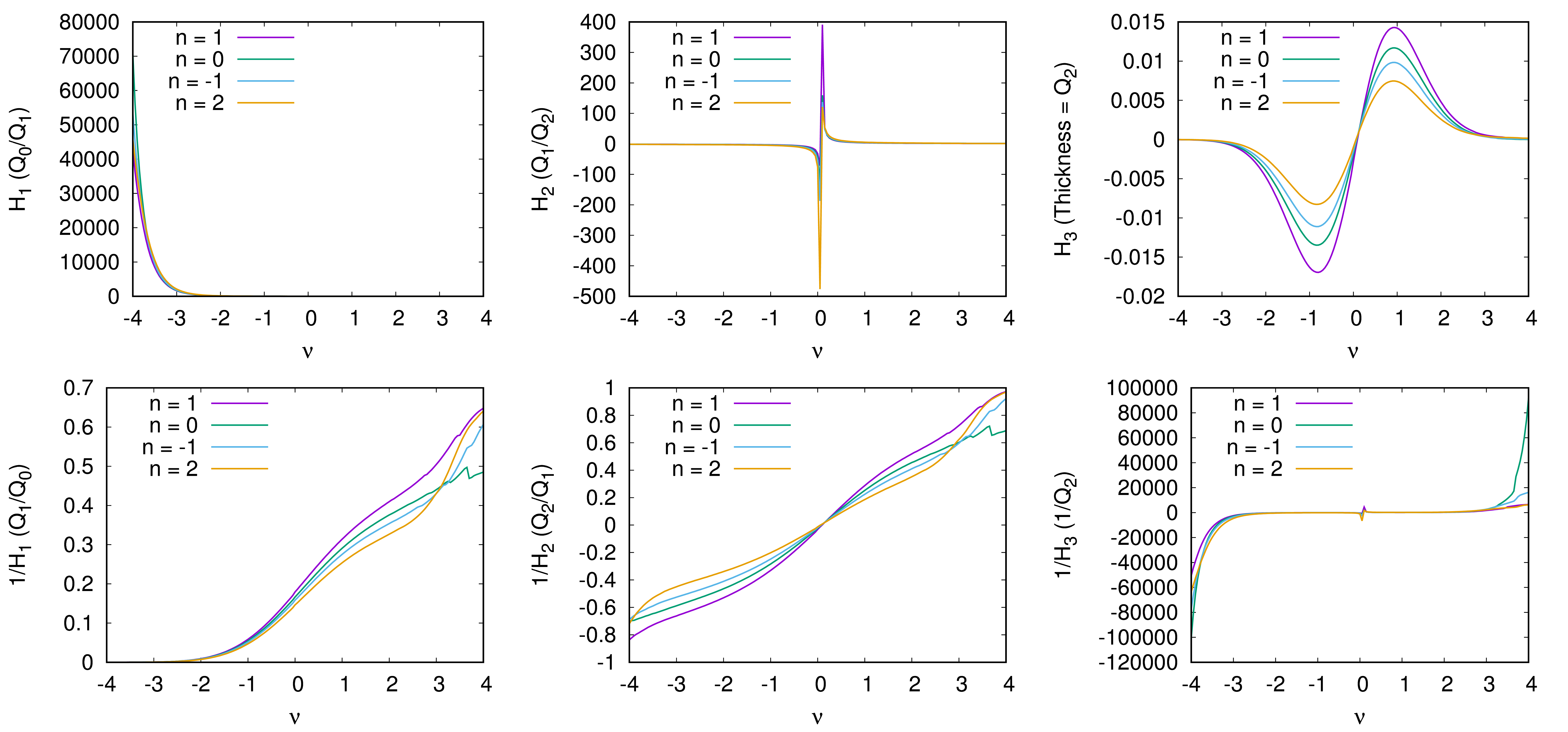}}\\
  \caption{  Ratios of Minkowski functionals. Left: The ratio $Q_1/Q_0$, 
  denoting the ratio between the total occupied volume and the total 
  surface area corresponding to the occupied volume. Middle: The ratio 
  $Q_2/Q_1$, denoting the ratio between the total length of contours  
  and the total surface area. Right: The ratio $Q_2/Q_0$, denoting the 
  total length of contours per unit volume occupied. The curves are 
  drawn with respect to the dimensionless density threshold $\nu$. For the bottom row, the numbers for high density thresholds, are not reliable, due to division by small numbers.}
  \label{fig:minkowski_ratio}
\end{figure*}

\begin{figure*}
\centering   
  \rotatebox{-90}{\includegraphics[height=0.9\textwidth]{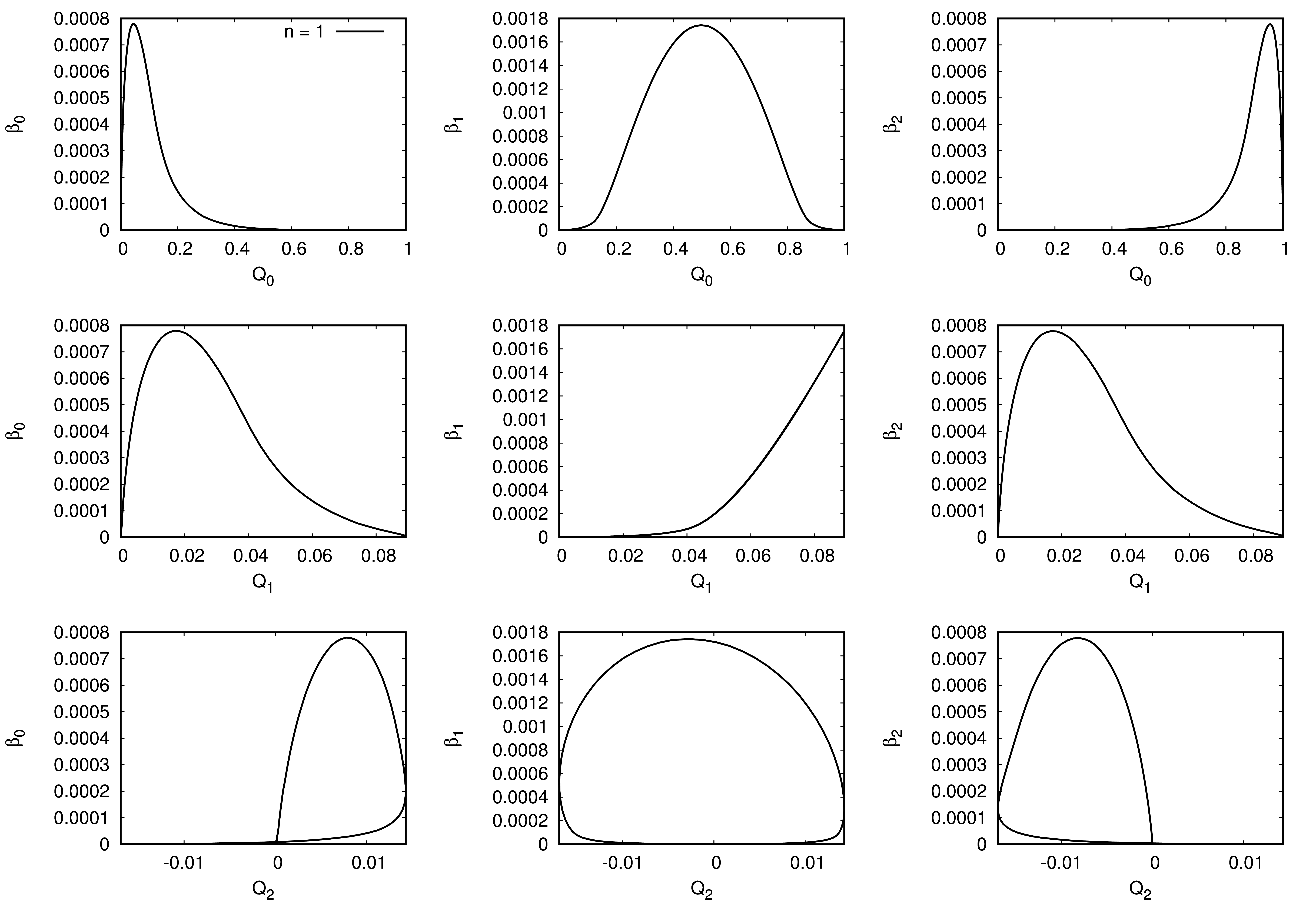}}\\
  \caption{  Betti numbers vs. the Minkowski functionals. }
  \label{fig:betti_vs_minkowski}
\end{figure*}
  \begin{figure*}
\centering   
  \rotatebox{-90}{\includegraphics[height=0.9\textwidth]{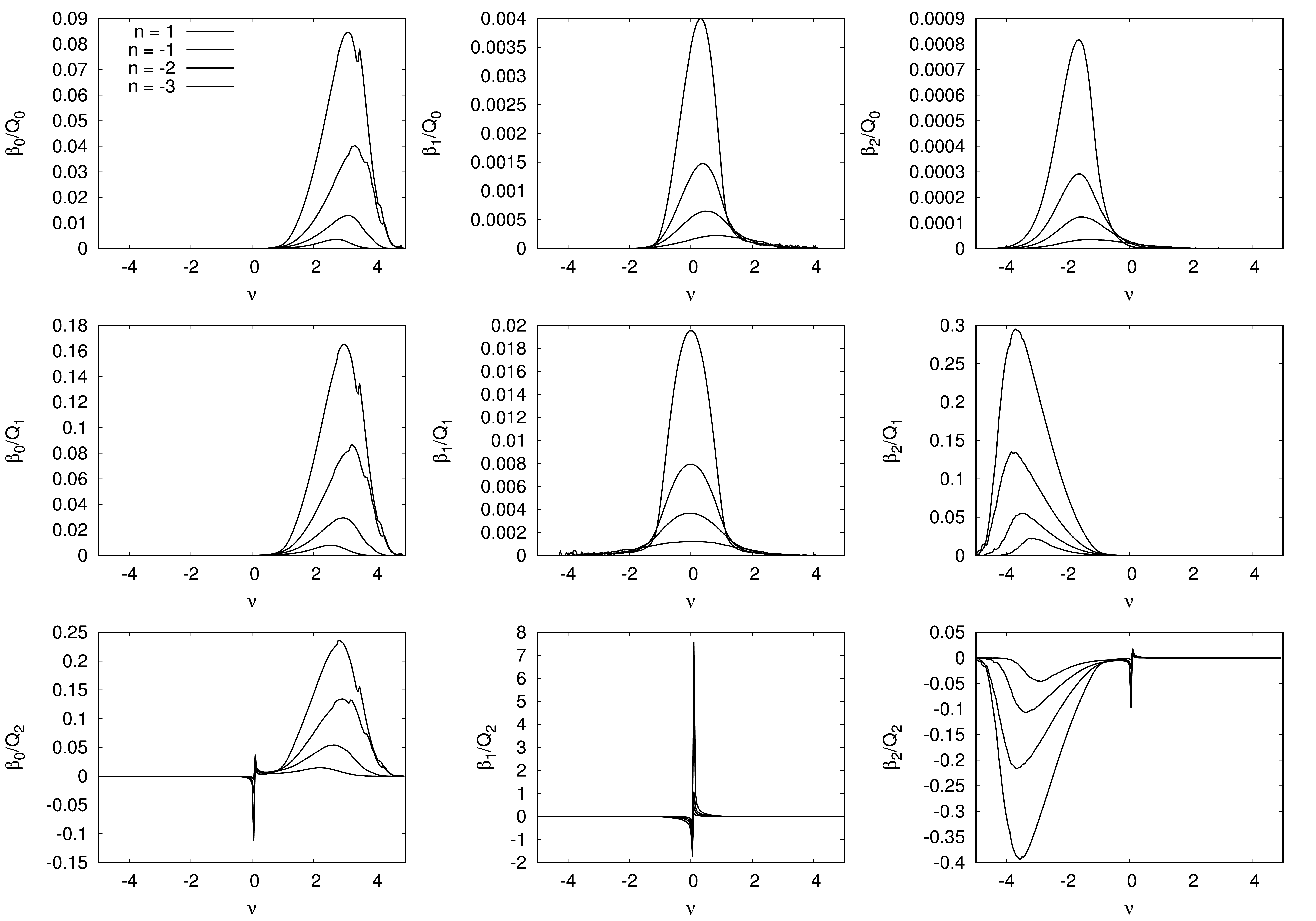}}\\
  \caption{  Ratio of Betti numbers to the various 
  Minkowski functionals. The graphs as plotted for a range of density 
  threshold.}
  \label{fig:betti_and_minkowski_ratio}
\end{figure*}

\section{Minkowski functionals of the models}
\label{sec:minkowski_result}

The left column of Figure~\ref{fig:minkowski_grf} presents the un-normalized (original) graph of the Minkowski 
functionals for the power law models. The graphs are averaged over 100 
realizations. The quantities are plotted as a function of the density 
threshold $\nu$. 
The fractional volume $Q_0$ is 
invariant with respect to the choice of the power spectrum. This is to be expected, because the first Minkowski functional simply takes the
form of the error function. All the 
other functionals show a systematic dependence on the choice of the 
power spectrum. The amplitude of the graphs 
of the area functional, the integrated mean curvature functional, and 
the Euler characteristic decreases monotonically with the decrease in 
the index of the power spectrum. 

The right column of the Figure~\ref{fig:minkowski_grf} presents the rescaled Minkowski 
functional curves. The graphs for the rest of the power law models 
have been scaled to the amplitude of the curve of the $n = 0$ model. The 
shape of the rescaled graphs falls neatly on top of 
each other. This indicates that the 
shape of the Minkowski functional curves is independent of the choice of 
the power spectrum. This observation is in line with equation~(\ref{eqn:minkowski_ch2}). 
The dependence on the choice of the power spectrum 
comes in only through the amplitude term. This dependence is parametrized in terms 
of $\lambda$, which is a function of the correlation function, or 
equivalently, the power spectrum. 

That the shape of the Minkowski functional curves is independent of the 
choice of the power spectrum is an important observation, when seen in 
comparison to the shape of the Betti number curves, which show a 
characteristic dependence on the choice of the power spectrum. We 
present a detailed analysis of the Betti numbers with respect to the 
Euler characteristic in Section~\ref{sec:betti_and_genus_result}.
This indicates that the Betti numbers are potentially more discriminatory 
than the Minkowski functionals, an observation we have already established in \cite{PPC13}. 

\subsection{Minkowski functionals as shapefinders}

Recall that the ratio of the Minkowski functionals are simplified
indicator of the morphological 
properties of manifold, given by \citep{sahni1998,sheth2003,shandarin2004} (also see \cite{schmalzing1999} for the relation to iso-perimetric
inequalities and Blaschke diagrams):
\begin{equation}
H_1 = Q_0/Q_1; \quad H_2 = Q_1/Q_2; \quad H_3 = Q_2.
\end{equation}

For example, a high surface 
area to volume ratio indicates a more pancake like morphology of 
structures. The reverse indicates a more filamentary morphology.
 Figure~\ref{fig:minkowski_ratio} presents presents the shapefinders for the models in the top panel, and their inverse quantities in the bottom panel. In the left column, we 
present $H_1$ and $H_1^{-1}$, in the top and the bottom panels respectively. The middle and the right columns present $H_2 (H_2^{-1})$, and $H_3 (H_3^{-1})$ respectively.
The curves are drawn with respect to the dimensionless density threshold $\nu$. 

All three quantities show a characteristic dependence on the 
choice of the power spectrum, more clearly in the inverse quantities for $H_1$ and $H_2$, and directly for $H_3$. The curves for the lower
power spectra increase more steeply towards the extremes of the density 
threshold, and flatten out as the threshold moves to further extremes.
For all the 
models, the surface 
area to volume ratio is high for high density thresholds. It indicates that 
the structures are more flattened for high 
thresholds. Interesting is the sharp rise in the value for the negative 
spectra. This indicates that at very high thresholds, the structures in 
the $n = -2$ model are the most flat. This ties in with the observation 
that for the $n = -2$ model, the large scale structures have 
significant power, giving rise to the overall flattened characteristics 
of the density field. The large structures are a consequence of significant powers at those scales. 

In summary, the Minkowski functionals characterize the geometric 
properties of the manifold predominantly. The connection to topology 
comes through the Euler characteristic. Hence, the 
Minkowski functionals maybe seen as complimentary to the topological 
descriptors such as the Betti numbers. The Minkowski 
functionals, together with the information on the homology of a manifold, provide a richer and more 
comprehensive morphological and topological information about the 
manifold. 

\begin{figure*}
	\centering   
	{\includegraphics[width=0.99\textwidth]{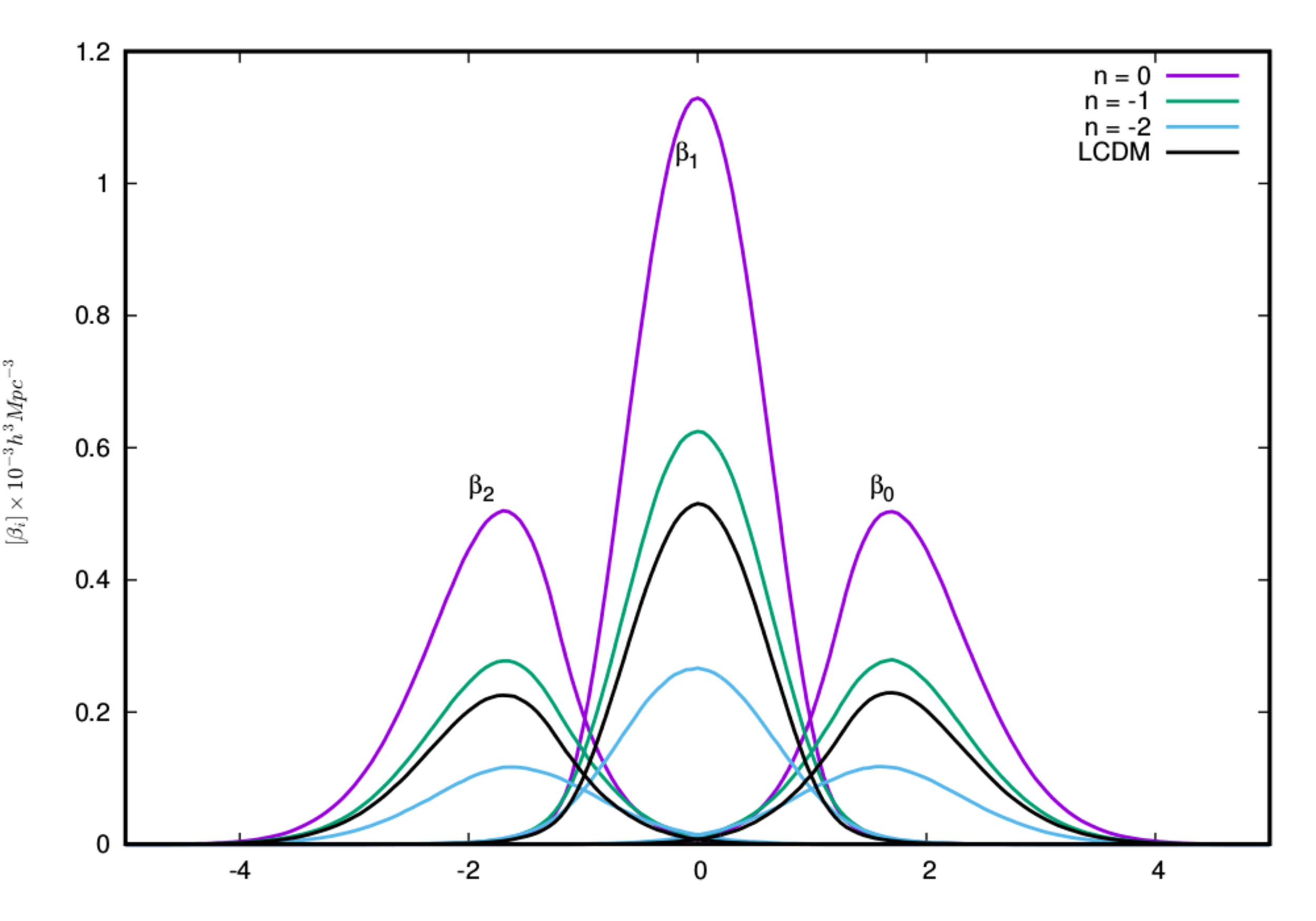}}\\
	\caption{Betti numbers of the LCDM model (in black). Alongside are plotted the curves for the power law models,
          $n = 0, -1 \text{and} -2$. The curves for the LCDM model are closest to the $n = -1$ model.}
	\label{fig:betti_LCDM}
\end{figure*}

\subsection{Betti numbers vs Minkowski functionals}

As we learnt in the previous sections, the Betti numbers are
topological quantities. They measure topology by assessing the number 
of independent holes in the different dimensions. On the other hand, the 
Minkowski functionals are primarily morphological measures, the 
exception being the Minkowski functional $Q_3$, or the Euler 
characteristic, $\chi$. The first three Minkowski functionals are 
associated with the volume ($Q_0$), surface area ($Q_1$) and the 
integrated mean curvature length ($Q_2$) of the manifold. 
However, an important question one may ask is if the Betti 
numbers and the Minkowski functionals convey different information 
about the manifold characteristics. With a view to investigate this, we 
assess the correspondence between the Betti numbers and the Minkowski 
functionals. 

Figure~\ref{fig:betti_vs_minkowski} presents the Betti numbers 
plotted against the various Minkowski functionals. The top-left panel 
of the Figure plots $\beta_0$ on the vertical axis against $Q_0$ on the 
horizontal axis, and so on. We notice that almost all the pairs of 
quantities exhibit a degeneracy. For example, in the top-left panel, we 
notice that there are two values of $Q_0$ for which the value of 
$\beta_0$ is the same. The exception is the peak of the curve, at which 
$\beta_0$ is associated with a unique value of $Q_0$. The only 
exception to this trend of degeneracy is the middle 
panel of the middle row, where we plot $\beta_1$ against $Q_1$. The 
curve is monotonic, indicating that $\beta_1$ and $Q_1$ behave in a 
similar fashion. In general, a monotonic curve between any two plotted 
quantities indicates a similar behaviour of the quantities.

The ratio of the Betti numbers to the 
Minkowski functionals is another interesting quantity to analyse, particularly 
in certain cases,  where they can be readily associated with a particular 
features of the manifold. For example, the ratio $\beta_0 /Q_0$ 
represents the number of isolated objects per unit occupied volume. 
Note that the occupied volume is different from the total volume of the 
manifold. The total volume is a constant, while the occupied volume is 
a function of the density threshold. Similarly, the ratio $\beta_1 / 
Q_1$ indicates the number of independent tunnels per unit surface area. 
This may be regarded equivalent to the information on the genus density of the 
manifold. 

Figure~\ref{fig:betti_and_minkowski_ratio} presents the ratio 
of the various Betti numbers to the various Minkowski functionals, as a 
function of the density threshold $\nu$. The plots are presented for 
the different power law models. We notice a dependence of the 
quantities on the choice of the spectral index. It is important to note 
that a constant, or a monotonically increasing or decreasing curve 
indicates quantities have a simple dependence on each other. We notice that none of the 
pair of quantities exhibit a monotonic ratio. This indicates crudely 
that the Betti numbers and the Minkowski functionals behave differently 
from each other in general.

\section{Topology of the LCDM model}
\label{sec:lcdm_result}
In this section, we briefly discuss the topology of the LCDM model. This is pertinent, since the LCDM model is the standard model of
cosmology. However, recall that the LCDM model is also characterized in terms of a spectral index, which is a running function of the
wavelength. This is unlike the power law models, where the spectral index is constant. Due to a corresponding description through the index
of the power spectrum, for both the LCDM and the power law models, it makes sense to compare the characteristics of the power law and
the LCDM models.

Figure~\ref{fig:betti_LCDM} presents the Betti curves for the LCDM model (drawn in black). We also present the curves for the power law
models with spectral index $n = 0, -1,\text{and} -2$ models.  Here we compare the spectra at a scale
of $R_f=2 \Mpch$, with an effective index $n_{eff} \approx -2$. Our topological analysis appears to yield Betti number curves that
lie in between those for power law spectra with $n=-1$ and $n=-2$, tending more towards the first.

For a complete comparison, we  need to take into account that over the range of the simulation box, the LCDM spectrum is
represented by frequencies running from Nyquist frequency to fundamental frequency (see Figure~\ref{fig:plaw_spectrum},
Section~\ref{sec:model_lcdm}). Hence, the effective index at the frequencies respresented in the realized LCDM Gaussian density fluctuation
field varies from $n_{eff} \sim -2.5$ to $n_{eff} \sim -0.5$. The combination of these fluctuations appear to lead to a Betti number
topology that resembles best that of a power-law power spectrum close to $n=-1$. In order to get more insight in how these
topological synergy between different modes works, we need to invoke the concept of persistence. For this we refer to the
upcoming accompanying study \cite{pranavb2018}.

\section{Discussion and Summary}
\label{sec:discussion_ch2}

In this study we present a largely numerical study of the topology of Gaussian random fields on the basis of homology,  
specifically in terms of Betti numbers. Homology describes the topological structure of a manifold in terms of the topological
features - or topological holes - it contains, whereby it concentrates on their boundaries. These and other concepts from
algebraic topology provide a fundamental and rich framework for a quantitative characterization of the shapes and connectivity of
structures in the cosmological mass distribution. An important aspect is the intimate relationship with the Euler characteristic,
which is equal to the alternating sum of the Betti numbers. Their individual assessment therefore enables us to understand the role
and contribution of the various structural components in establishing the overall topology of the cosmic mass distribution
encapsulated in the Euler characteristic. This can be obtained from the decomposition of the well-known curve for the genus
or Euler characteristic, as a function of superlevel threshold, into separate curves for the individual Betti numbers.

The topology of Gaussian random fields functions as key reference against which topological measures of the cosmic mass and
galaxy distribution in more advanced evolutionary stages should be assessed. In cosmology, the topological and geometric structure
of two-dimensional and three-dimensional Gaussian random fields has been extensively analysed in terms of genus, the Euler characteristic
and Minkowski functionals. An aspect of importance for the use of Gaussian random fields as reference point is the existence of closed
analytical expressions for the statistical distribution of their Euler characteristic and Minkowski functionals
\citep{Adl81,bbks,WGM87,tomita1993,schmalzing1997}. In fact, the analytic expressions for the mean of the Euler characteristic and
Minkowski functionals of Gaussian random fields belong to an extensive family of such formulae, all emanating from the so called
{\it Gaussian kinematic formula} or GKF \citep{Adl10,ATSF,ARF}.  While hardly known in the cosmological literature, given its
central role for our assessment and understanding of the topology and morphology of Gaussian fields, we devote an extensive
discussion on its formulation and ramifications in section~\ref{sec:gkf}. Most relevant for our purpose is the observation that it is
not possible via this GKF route to establish similar closed analytical expressions for Betti numbers and persistence. While other routes 
might be feasible, it establishes the principal motivation for the numerical approach used here. 

The present study is part of a series of papers that seeks to extend this to the richer language of homology, with
the second part \citep{pranavb2018} discussing the description of the multi-scale topological aspects of Gaussian fields in terms of
persistence \citep{EdHa10,robinsa2015,robinsb2015,PEW16}. An accompanying upcoming article \citep{feldbrugge2018} contains a
mathematical analysis and formalism for the description of Gaussian random field homology. 

\subsection{Gaussian field Betti Numbers: properties}
Our statistical study of three-dimensional Gaussian field homology consists of Betti number curves $\beta_0$, $\beta_1$ and $\beta_2$
as a function of density field threshold $\nu$ of the corresponding superlevel set of a Gaussian field realization. The curves are averaged over
100 field realizations for each separate power spectrum.

In section 5 we computed the Betti numbers using the \cite{BEK10} algorithm, an optimal and exact
algorithm for computing all Betti numbers of a discretely sampled image on a cubic grid. It starts 
with a slight deformation of the grid, the calculation of the corresponding unique simplicial complex
defined by the grid-point distribution, and the subsequent computation of Betti numbers of superlevel
sets of the field - for a range of density thresholds - via the construction and simplification of
a boundary matrix of the simplicial elements.

The Gaussian fields for which we evaluate the homology measures are three-dimensional in nature, and consist of two classes. The first family of
realizations is a series of pure power law power spectra, with the intention to assess systematic trends as the power shifts from
dominant high frequency modes to low frequency modes. We also measure the Betti numbers in Gaussian fields with a LCDM power spectrum. 

\subsubsection{Gaussian fields: topological components}
At extreme low and high density values, the topology of Gaussian fields is dominated by a single class of features. 
At very high density levels these are islands and thus fully specified by $\beta_0$, defining a predominantly 
\emph{Meatball-like} topology. The same is true at very low density levels, where the topology is distinctly \emph{Swiss-cheeselike}, 
exclusively dominated by cavities and thus entirely specified by $\beta_2$.

At more moderate levels, for $|\nu| \lesssim 2$, the topology attains an increasingly \emph{Sponge-like} character. In these regimes at least two Betti numbers are
needed to describe the topology of the superlevel manifolds. On the lower density side, the topology is dominated by $\beta_2$ and $\beta_1$. It reflects
a pattern of isolated cavities, and agglomerates of density troughs interspersed by tunnels and loops. On the higher density side, the topological
signal consists mostly of $\beta_0$ and $\beta_1$.

The corresponding spatial pattern is that of isolated islands and large agglomerates of
islands, connected by bridges/loops, and infused and punctured by numerous tunnels. In a relatively narrow density range around the mean density,
for $|\nu| \lesssim 0.1-0.2$, we even observe the simultaneous existence of all three topological features, cavities, islands, and tunnels. In that
regime, all three Betti numbers are needed to quantify the Gaussian field topology.

\bigskip
Also interesting is the finding that the topological identity at the median density level, at $\nu=0$ for Gaussian fields, is
not exactly - ie. ideally - sponge-like. At that level, we see the presence of an equal number of islands and cavities as $\beta_0=\beta_2$ at
$\nu=0$. Conventionally, it is assumed that all overdense regions have merged into one percolating complex, interlocking with
an one equivalent underdense ``ocean''. While this is the definition for a pure sponge-like topology, we find that in general this
it not even the case for Gaussian fields. The topology at median levels is determined by a few - disconnected - overdense complexes,
intertwined with a few underdense ones. The dissection of the genus curve into the contributing Betti curves leads to new insights
on to this issue. 

\subsubsection{Betti numbers vs. Power spectrum}
An important aspect of Betti numbers of Gaussian fields is they are sensitive to their power spectra. In other words, they reflect 
the nature of Gaussian random fields. The shifting prominence from large wavelength modes to that of short wavelength modes
has a significant impact on the resulting topology. This is entirely different from the behaviour of the Euler characteristic,
whose curve is known to be entirely independent of the nature of the underlying power spectrum (see Equation~\ref{eqn:genusgauss})
\citep{Adl81,bbks,WGM87}.

This aspect of Gaussian field topology is most clearly revealed in the scaled Betti number curves - scaled with respect to the maximum of the $\beta_1$ curves,
ie. with respect to the maximum number of tunnels. While the spectral insensitivity of the Euler characteristic makes it a highly robust measure for
testing the level of Gaussianity of a field, it also implies that it yields only a rather limited amount of topological information. This
concerns key aspects such as the topological composition of the cosmic mass distribution, and the connectivity of the various topological
elements. The implication is that considering homology, in terms of Betti numbers and even more that of topological persistence \citep{PEW16,pranavb2018},
represent a major advance in understanding Gaussian fields.

The Betti number curves reveal a systematic dependence of relative populations of isolated mass concentrations, tunnels and enclosed voids as a
function of the power spectrum of a Gaussian random field. Our study finds a monotonic increase of the width of the $\beta_1$ curves as the
power spectrum index $n$ decreases': the number of tunnels increases steeply as the large scale wavelength modes become more prominent. 
In addition, we find that there is a considerably larger density range over which the topology resembles a sponge-like morphology. In other
words, configurations marked by the simultaneous presence of tunnels and cavities at the low density regime, and of tunnels and islands at the
high density regime, exist over a wider density range as the spectral index $n$ is lower. This also concerns the narrow density regime around the
median density where all three topological features exist simultaneously. 

The implications for the topology of the resulting spatial pattern of the evolved mass distribution are substantial. Gravitational
evolution amplifies the topological differences in the initial conditions. A sponge-like topology evolves into a mass distribution
resembling a connected network, while one that only involves isolated islands would merely produce a field of isolated collapsed
density clumps. Given that primordial Betti numbers already elucidate and highlight such fundamental topological differences,
suggests they have the potential of quantifying crucial aspects of the connectivity of the evolved cosmic matter distribution.

For true insight into the hierarchical evolution and development of the topology of the mass field, we will need to
characterize in more detail how the various features connect up with each other. This is the subject of the second part of our
investigation, to be reported in \cite{pranavb2018}, where we will present and discuss the persistent homology of Gaussian fields. 

\subsubsection{Singularities and Betti Numbers}
As a prelude to our study of the persistent topology of Gaussian random fields \citep{pranavb2018}, we also evaluate the
relation between minima, saddle points and maxima in the density field and Betti numbers. We focussed in
particular on the relationship between the zeroth Betti number and the number of maxima, ie. peaks, in the
density field. Assisted by the useful analytical expressions for the number density of peaks and minima \citep{bbks}
as consistency check, we have subsequently assessed the growth of peak number per density island in Gaussian fields.
It reveals the subtle dependence of number of peaks per island as a function of power spectrum. While the number of
islands and peaks are similar at very high density levels, we see that the convergence towards unity of the ratio of peaks
to islands is very slow. This shifts strongly and systematically towards higher density levels as the index $n$ decreases. 

\subsubsection{Betti numbers and Minkowski Functionals}
We also study in detail the extent to which Betti numbers contain topological information that is complementary to
Minkowski functionals. One immediate observation is that of the power spectrum dependence of Betti numbers,
which is a fundamental difference with the power spectrum independence of the Minkowski functional curves. 

One major difference concerns the power spectrum dependence of Betti numbers. By contrast, the shape of Minkowski functional
curves,  ie. the value of Minkowski functionals as a function of density level $\nu$, is independent of the Gaussian field's
power spectrum. Nonetheless, ratios of Minkowski functionals do reveal a dependence on power spectrum, a fact that was exploited
in a series of studies by Sahni and collaborators \citep[see e.g.][]{sahni1998,sheth2003,shandarin2004} in the morphological analysis of
structural Megaparsec features in evolved cosmic density fields. As we show here, it manifests itself in features in the
primordial Gaussian field having a more flattened shape as the spectral index is lower,

We present a comprehensive visual assessment of the differences between Minkowski functionals and Betti numbers. In combination
with the difference in power spectrum dependence, the systematic comparison demonstrates that Betti numbers, and
persistence \citep[see][]{pranavb2018}, contain a considerable amount of topological information that is complementary
to that contained in Minkowski functionals. 

\subsection{Homology \& Cosmology: Potential}
The potential for exploiting the rich topological language of homology to the observational reality
of the Universe is substantial. It opens the path towards a richer, more powerful and insightful
analysis of the connectivity and organization characteristics of emerging cosmic mass distribution in the form of the topologically
complex and intricate structure of the cosmic web. Also, it allows a better understanding of structural aspects
of the Gaussian primordial density field, and might even shed new light on the nature of the primordial
CMB perturbations. 

One particular intriguing example where topological signatures might reveal
yet unknown cosmological features concerns the detection of possible non-Gaussianities in the primordial perturbation
field. The discovery of such primordial non-Gaussianities in the temperature fluctuations in the Cosmic Microwave Background
would provide unique insights on the physical processes that determined the nature of our Universe during the inflationary epoch.
\citep{bartolo2004,baumann2009,chen2010}.

Even while the Planck satellite has set stringent upper limits on the amplitude of primordial non-Gaussian fluctuations, one may not
exclude that these had a more intricate and elusive character than suggested by most multi-field inflation theories. It may
reflect itself in subtle topological markings for which the rich language of homology, in terms of Betti numbers and even
moreso of topological persistence, may provide a means of uncovering. There are theoretical indications, such as those
presented by \cite{feldbrugge2018}, that demonstrate the potential of persistence to find non-Gaussian signatures. 
Even only Betti numbers have the ability to detect non-Gaussian signatures, as was discussed in considerable detail by \cite{chingangbam2012} \citep[for a recent contribution, see also][]{cole2018}. In fact, the recent study by \cite{pranav2019} of the homology of CMB measurements by the Planck satellite have uncovered some interesting effects, when comparing the observed CMB maps with respect to simulations based on Gaussian prescriptions. However, it remains to be ascertained if these signals are a genuine cosmological signal, and not arising from yet unknown data systematic or foregrounds.

\bigskip
Notwithstanding such uncertainties, as the present study has argued in detail, homology considerably enriches the 
language for exploring the nature and describing the spatial patterns of cosmological structures. 

 
 
\section*{Acknowledgements} 
We are indebted to Herbert Edelsbrunner for encouragement, numerous discussions
and incisive comments. We also thank Keimpe Nevenzeel, Matti van Engelen and Mathijs Wintraecken
for many discussions and insights. 

Parts of this work have been supported by the 7th Framework Programme for Research  
of the European Commission, under FET-Open grant number 255827 (CGL Computational  
Geometry Learning), ERC advanced grant, URSAT (Understanding Random Systems via
Algebraic Topology, PI: Robert Adler) number 320422, and ERC advanced grant ARTHUS (Adavances in the research on Theories of the Dark Universe; PI: Thomas Buchert), number 740021. RvdW also acknowledges support  from the New Frontiers of Astronomy and Cosmology program at the Sir John Templeton Foundation.

\bibliographystyle{mn2e}

\bibliography{references}

\appendix

\section{Gaussian Field Peak Density}
\label{app:peakdens}
For an evaluation of the number density of peaks in a Gaussian field, we use the
expressions derived by \cite{bbks} for the (comoving) differential peak density
$\Nspace_{pk}(\nu)$ for peaks of (normalized) density $\nu=f_{pk}/\sigma$ in a density
field $f_s({\vec x})$ filtered on a spatial scale $R_s$.

\bigskip
\noindent A Gaussian field $f({\vec x})$ filtered on a scale $R_s$ with filter kernel $W_s({\vec r}; R_s)$,
  \begin{equation}
  f_s (\vec x)  \,=\, \int f(\vec y)\, W_s(\vec y -\vec x; R_s)\, d\vec y \,,
  \end{equation}
can be written in terms of the Fourier integral, following Parseval's theorem,    
\begin{equation}
  f_s (\vec x)  \,=\,  \int_{\mathbb R^3} \frac{d^3\vec k}{(2\pi)^3}\, \,{\hat f}(\vec k)\,{\hat W}(kR_s)\, \exp(-i\vec k\cdot \vec x)\,, 
\end{equation}
in which ${\hat W}(kR_s)$ is the Fourier transform of the filter kernel. From this, it is straightforward to see that
the corresponding power spectrum $P_s(k)$ of the filtered field is the product of the unfiltered power spectrum $P(k)$
and the square of the filter kernel ${\hat W}(kR_s)$
\begin{equation}
  P_s(k; R_s)\,=\,P(k)\,{\hat W}^2(kR_f)\,.
\end{equation}
\noindent Two spectral parameters are instrumental for assessing the number density of peaks - and dips and other singularities - in the
filtered density field. The spectral parameter $\gamma$ and spectral scale $R_{\star}$ are combinations of various moments of the
filtered power spectrum $P_s(k)$,  
\begin{equation}
  \gamma\,=\,\frac{\sigma_1^2}{\sigma_2\sigma_0} \,,\qquad   R_{\star}\,=\,\sqrt{3}\,\frac{\sigma_1}{\sigma_2}\,,
 \end{equation}
in which the spectral moments $\sigma_j$ are defined as 
\begin{eqnarray}
\sigma_j^2\,=\, \int_0^{\infty} \frac{{d^3\vec k}}{(2\pi)^3}\ k^{2j} P_s(k)\,, 
  \end{eqnarray}
(and thus $\sigma=\sigma_0$). To appreciate the dependence of the spectral parameters $\gamma$ and $R_{\star}$ on the power spectrum
and filter scale $R_s$ of the field, a useful refererence are the values for a field with a power law power spectrum of index $n$ and filter
scale $R_s$ \citep{bbks},
\begin{equation}
  \gamma^2\,=\,\frac{(n+3)}{(n+5)}\,,\qquad R_{\star}\,=\,\left(\frac{6}{n+5}\right)^{1/2}\,R_s\,.
  \end{equation}

\bigskip
\noindent The influence of the power spectrum and smoothing scale $R_s$ on the cumulative and differential peak number densities propagates 
via the values and behaviour of the spectral parameters $\gamma$ and $R_{\star}$.  The differential number
density of peaks $\Nspace_{pk}(\nu)$ in the filtered field $f_s({\vec x})$ at normalized (dimensionless) density level $\nu=f_s/\sigma$ is
given by \citep{bbks}, 
\begin{equation}
	\Nspace_{pk}(\nu) \,d\nu \,=\,\frac{1}{(2\pi)^{2}R_{\star}^{3}} e^{-\nu^{2}} G(\gamma,\gamma\nu)\,. 
	\label{eqn:diff_peak_density_1}
\end{equation}
While an analytical expression for the function $G(\gamma,w)$ is not availalbe, \cite{bbks} provide a fitting formula for the function
$G(\gamma,w)$ that is accurate to better than $1 \%$ over the range $0.3 < \gamma < 0.7$ and $-1 < w < \infty$, and even better than
1 in 10000 for $w>1$,
\begin{equation}
G(\gamma,w)\ =\ \frac{w^3-3\gamma^2 w\,+\,[B(\gamma)w^2+C_1(\gamma)] \exp{[-A(\gamma)w^2 }}{1+C_2(\gamma)\,\exp[-C_3(\gamma)w]}
\end{equation}
\noindent The coefficients $A$ and $B$ can be inferred by assuring the fitting formula to agree with the asymptotic behaviour for
peak at high $\nu$, while $C_1$, $C_2$ and $C_3$ follow from the fitting procedure \citep[see][, Section 4]{bbks},

\begin{eqnarray}
  A &=&\frac{5/2}{(9-5\gamma^2)}\,,\qquad B\,=\,\frac{432}{(10\pi)^{1/2}\,(9-5\gamma^2)^{5/2}}\,,\nonumber\\
  \ \\
  C_1 &=& 1.84+1.13(1-\gamma^2)^{5.72}\,,\nonumber\\
  C_2 &=& 8.91+1.27\exp{(6.51\gamma^2)}\,,\nonumber\\
  C_3 &=& 2.58 \exp{(1.05\gamma^2)}\,.\nonumber
  \end{eqnarray}

\medskip
\noindent To find the cumulative number density $n_{pk}(\nu)$ of peaks in the filtered field $f_s({\vec x})$ with a height of
$\nu$ or higher, we need to evaluate the integral,
\begin{equation}
  n_{pk}(\nu)\,=\,\int_{\nu}^{\infty}\,\Nspace_{pk}(\nu) \,d\nu\,. 
  \end{equation}
\noindent While in general the integral has to be evaluated numerically, its asymptotic value for the full peak
density $n_{pk}$ can be inferred analytically \citep{bbks},
\begin{equation}
  n_{pk}\,=\,n_{pk}(-\infty)\,=\,\frac{29-6\sqrt{6}}{5^{3/2}2(2\pi)^2\,R_{\star}^3}\,=\,0.016\,R_{\star}^{-3}\,.
  \end{equation}
\noindent We refer to the diagram in Figure~2 of \cite{bbks} for a representative sample of differential
number density $\Nspace_{pk}(\nu)$ curves for different values of spectral parameters $\gamma$. Likewise,
Figure~3 in the same study shows the dependence of the cumulative peak density as function of normalized
density threshold $\nu$. 

\end{document}